\documentclass[aps,prd,amsmath,amssymb,nofootinbib,superscriptaddress,floatfix]{revtex4-2}
\usepackage{graphicx}
\usepackage{dcolumn}
\usepackage[utf8]{inputenc}
\usepackage{xcolor}
\usepackage[unicode=true,colorlinks,linkcolor=blue,anchorcolor=blue,urlcolor=blue,citecolor=blue,breaklinks=true]{hyperref}
\usepackage{epstopdf}
\usepackage{orcidlink}
\newcommand{\ii}{\mathrm{i}\,}
\newcommand{\ee}{\mathrm{e}}
\newcommand{\pararrow}{\mathord{\buildrel{\lower3pt\hbox{$\scriptscriptstyle\leftrightarrow$}}\over {\partial}}} % partial leftrightarrow operation
\newcommand{\pararrowk}[1]{\mathord{\buildrel{\lower3pt\hbox{$\scriptscriptstyle\leftrightarrow$}}\over {\partial}\hspace*{-0.18em}{}^#1}\hspace*{-0.18em} \,} % partial
\usepackage{bm}
\usepackage{slashed}
\usepackage{ulem}

\newcommand{\qfnu}{\affiliation{College of Physics and Engineering, Qufu Normal University, Qufu 273165, China}}

\newcommand{\hnnu}{\affiliation{Institute of Particle and Nuclear Physics, Henan Normal University, Xinxiang 453007, China}}

\begin{document}
	
	\title{Dipionic transitions of $Y(4500)$ to $J/\psi$}
	
	\author{Shi-Dong Liu\,\orcidlink{0000-0001-9404-5418}} \email{liusd@qfnu.edu.cn}\qfnu
	%\author{Ju-Jun Xie} \email{xiejujun@impcas.ac.cn} \imp \snst \scnt
	%\author{Xiao-Hai Liu\,\orcidlink{0000-0002-6159-3140}}\email{xiaohai.liu@tju.edu.cn}\tju
	\author{Qi Wu\,\orcidlink{0000-0002-5979-8569}}\email{wuqi@htu.edu.cn} \hnnu
         \author{Gang Li\,\orcidlink{0000-0002-5227-8296}} \email{gli@qfnu.edu.cn} \qfnu	
	
	\begin{abstract}
		We, using an effective Lagrangian approach, investigated the dipionic transition of the $Y(4500)$ newly well established in the process $e^+e^-\to K^+K^- J/\psi$. In this study, the $Y(4500)$ was considered as a mixture of the $\psi(5S)$ and $\psi(4\,{^3D_1})$, and its dipionic decay to the $J/\psi$ was assumed to occur via charmed meson loops including box and two kinds of triangle loops. The calculated invariant mass spectra due to the different loop mechanisms exhibit distinct differences. These spectrum patterns are found to be nearly independent of the cutoff parameter, especially for the box and $\mathcal{R}$ loops. By means of comparing our calculated results with the future experimental measurements of the $\pi^+\pi^-$ and $\pi^\pm J/\psi$ invariant mass distributions, we could judge which loop mechanism is of more importance in this dipionic transition. The interference among the different kinds of loops is also exhibited. Despite the unknown phase angles between different kinds of loop diagrams, the partial decay width for the process $Y(4500)\to\pi^+\pi^- J/\psi$ is estimated to be 250--700 keV, aligning with the estimation obtained by combining BESIII Collaboration measurements and theoretical predictions for $Y(4500)\to K^+K^- J/\psi$. We hope that the present calculations would be tested by the future BESIII or Belle II experiments.
	\end{abstract}
	
	\date{\today}
	
	\maketitle
	\section{Introduction}\label{sec:intro}
	
	With the advancement of experimental techniques, many new resonant structures, whose masses greatly exceed the threshold of open heavy-flavor meson pairs, especially in the $c\bar{c}$ sector, have been identified in recent years \cite{Liu:2023hhl,Gershon:2022xnn,ParticleDataGroup:2024cfk}.
	These resonances usually show different properties and behaviors compared to the conventional heavy quarkonia. For the higher states of the heavy quarkonia, they would mainly decay into the open heavy-flavor meson pairs, while the resonances that cannot fit into the quark model usually exhibit strong coupling to the hidden heavy-flavor channels \cite{Mo:2006ss,Wang:2019mhs,Wang:2018rjg,ParticleDataGroup:2024cfk,BESIII:2016bnd}. 
	In terms of their special features (e.g., the near-threshold effect) or unexpected decay modes, various models have been proposed to interpret their natures, including molecules, multiquarks, hybrids, cusp effects, and so on (see reviews \cite{Chen:2024eaq,Liu:2024uxn,Chen:2016qju,Chen:2022asf,Meng:2022ozq,BESIII:2020nme,Guo:2019twa,Brambilla:2019esw,Liu:2019zoy,Guo:2017jvc,Liu:2013waa,Wang:2025sic} and references therein).
	These novel resonances not only broaden our perspective of the hadron spectrum in the heavy quarkonium energy region, but also provide us with additional platforms to understand the nonperturbative properties of strong interactions.
	
	In the $c\bar{c}$ sector, the celebrated vector charmoniumlike state with energy larger than 4.0 GeV is the $Y(4260)$, first discovered in the processes $e^+e^-\to \pi^+\pi^- J/\psi$ by the \textit{BABAR} experiment in 2005 \cite{BaBar:2005hhc}, and soon confirmed in the same process by the CLEO \cite{CLEO:2006ike} and Belle \cite{Belle:2007dxy} Collaborations. 
	Higher-statistics data of the $e^+e^-\to\pi^+\pi^- J/\psi$ from the BESIII Collaboration \cite{BESIII:2016bnd} revealed a shift in the peak position to a lower mass and a reduction in the width, thereby leading to the renaming of the state $Y(4260)$ as $Y(4230)$ or $\psi(4230)$ with a mass of $(4222.1\pm 2.3)~\mathrm{MeV}$ and a width of $(49\pm 7) ~\mathrm{MeV}$ \cite{ParticleDataGroup:2024cfk}. 
	A series of experimental measurements demonstrate that the $Y(4230)$ decays predominantly through hidden charm modes \cite{BESIII:2022joj,ParticleDataGroup:2024cfk,BESIII:2022qal} but has a minimal probability to the open charmed final states (plus one pion) \cite{ParticleDataGroup:2024cfk,BESIII:2023cmv}, despite its mass being well above the $D\bar{D}$ threshold.
	This indicates that the $Y(4230)$ is likely to be an exotic state, such as the hybrid charmonium \cite{Zhu:2005hp}, the tetraquark \cite{Dubnicka:2020xoh,Wang:2018ejf,Wang:2018ntv}, the mesonic molecule \cite{Chiu:2005ey,Yuan:2005dr,Albuquerque:2008up,Ding:2008gr,Close:2010wq,Guo:2013zbw,Li:2013yla,Liu:2013vfa,Wang:2013cya,Cleven:2013mka,Peng:2022nrj,Dong:2019ofp,Cleven:2016qbn,Li:2014gxa}, and baryonium \cite{Qiao:2005av}, but not a conventional charmonium \cite{Zhu:2005hp,CLEO:2006ike,Chen:2019mgp}.
    In Refs. \cite{Wang:2019mhs,Fu:2018yxq}, the authors interpreted the $Y(4230)$ as a $4S$-$3D$ mixture.
	Moreover, it was also interpreted as a nonresonant structure \cite{Chen:2010nv} or cusp effect \cite{Liu:2014spa}, namely not a genuine resonance.
	
	Recent measurements by the BESIII Collaboration from 2022 to 2024 reveal plentiful massive charmoniumlike states \cite{Liu:2023hhl,BESIII:2022joj,BESIII:2022qal,BESIII:2023cmv,BESIII:2023wqy,BESIII:2023wsc,BESIII:2024jzg}. Among these, some states are consistent with previously established resonances, while others, such as the $Y(4500)$, $Y(4710)$, and $Y(4790)$, are observed for the first time. 
	So far, the $Y(4710)$ and $Y(4790)$ were observed in only one process: $Y(4710)$ in $e^+ e^- \to K^+ K^- J/ \psi$ \cite{BESIII:2023wqy} and $Y(4790)$ in $e^+ e^- \to D_s^{\ast +} D_{s}^{\ast -}$ \cite{BESIII:2023wsc}, while the $Y(4500)$ seems to have been observed in three different processes: $e^+ e^- \to K^+ K^- J/\psi$ \cite{BESIII:2022joj}, $\pi^+\pi^- J/\psi$ \cite{BESIII:2022qal}, and $D^{\ast 0}D^{\ast -} \pi^-$ \cite{BESIII:2023cmv}.
	The $Y(4500)$ was well constructed in the $e^+ e^-\to K^+ K^- J/\psi$ with a more than $8\sigma$ significance \cite{BESIII:2022joj}, whose mass and width were determined to be $(4484.7\pm 13.3\pm 24.1)~\mathrm{MeV}$ and $(111.1\pm 30.1 \pm 15.2)~\mathrm{MeV}$, respectively. 
	It is noted that the signal of the $Y(4500)$ was also hinted in the dipion process $e^+ e^-\to \pi^+ \pi^- J/\psi$, exhibiting similar mass and width, although the significance was reported to be only $\sim$$3\sigma$ \cite{BESIII:2022qal}.
	The process $e^+ e^-\to D^{\ast 0}D^{\ast -} \pi^-$ provides possible evidence of the $Y(4500)$, but this channel shows a lower mass of $(4469.1\pm 26.2 \pm 3.6)~\mathrm{MeV}$ and larger width of $(246.3\pm 36.7\pm 9.4)~\mathrm{MeV}$ \cite{BESIII:2023cmv}. Very recently, the $e^+e^-\to \pi^+\pi^-h_c$ cross section measured by the BESIII Collaboration also signals the $Y(4500)$, giving a mass of $(4467.4^{+7.2+3.2}_{-5.4-2.7})~\mathrm{MeV}$ and a width of $(62.8^{+19.2+9.9}_{-14.4-7.0})~\mathrm{MeV}$ with the statistical significance greater than $5\sigma$ \cite{BESIII:2025bce}.

    The nature of these newly observed resonances remains an open question. Not only is the experimental status of the $Y(4500)$, particularly in its mass and width, difficult to reach a unanimous conclusion, but the theoretical interpretations also remain debated.
	The $Y(4500)$ is compatible with the $D$-wave charmonium $\psi(4{}^3D_1)$ having mass of $4484~\mathrm{MeV}$ predicted in the unquenched potential model \cite{Wang:2019mhs} but also accords with the $\psi(3{}^3D_1)$ having mass of $4486 (4463)~\mathrm{MeV} $ predicted by the quenched (unquenched) quark model \cite{Deng:2023mza}.
	Particularly, in the latter work \cite{Deng:2023mza}, the total width of the $\psi(3{}^3D_1)$ is predicted to be about $119~\mathrm{MeV}$, agreeing well with the experimental results \cite{BESIII:2022joj,BESIII:2022qal}. 
	If the $Y(4500)$ is the $D$-wave charmonium $\psi(3{}^3D_1) $, it would decay dominantly via the open charmed channels $DD_0(2550)$, $DD_2^\ast(2460)$, and $D^{(\ast)}D_1(2420)$.
	Moreover, it is also consistent with the heavy-antiheavy hadronic molecule $D_s\bar{D}_{s1}(2536)$ with mass of $4503~\mathrm{MeV}$ \cite{Dong:2021juy}.
	The interpretation of $Y(4500)$ as a tetraquark state with quark content $c\bar{c}s\bar{s}$ \cite{Chiu:2005ey} or $c\bar{c} q\bar{q}$ ($q=u,\,d$) \cite{Wang:2023hsc,Wang:2024qqa} appears feasible as well. 
	In the case of $c\bar{c} q\bar{q}$ components, the QCD sum rules give a total width of about $200~\mathrm{MeV}$ \cite{Wang:2024qqa}, in accordance with the BESIII data \cite{BESIII:2023cmv}.
	The primary decay modes include the open-charm channels $D\bar{D}$, $D^\ast\bar{D}^\ast$, and $DD_1(2420)$ and the hidden ones $\chi_{c0(c1)}\omega$ and $J/\psi f_0(500)$.
	In Ref. \cite{Wang:2022jxj}, the authors explained the $Y(4500)$ as the $5S$-$4D$ mixture and refitted the BESIII data, giving a mass of $(4504\pm 4)~\mathrm{MeV}$ and a width of about $50~\mathrm{MeV}$. 
	This predicted width is 2--5 times smaller than the BESIII measurements \cite{BESIII:2023cmv,BESIII:2022joj,BESIII:2022qal}, which could be attributed to the cross-section enhancement by the charmoniumlike structures around $4.6~\mathrm{GeV}$ \cite{Wang:2022jxj}.
	Additionally, the branching fraction of the $Y(4500)\to K^+K^- J/\psi$ was predicted to be $(1.3$--$2.0)\times 10^{-3}$ using the charmed meson loop mechanism, where the $K^+K^-$ were assumed to be produced via the scalar meson $f_0(980)$. 
	So far, experimental and theoretical studies on the $Y(4500)$ are still scarce.
	To deepen understanding of the $Y(4500)$ nature, more effort is needed.
	
	Although the process $e^+e^-\to \pi^+\pi^- J/\psi$ signals weakly the $Y(4500)$ resonance \cite{BESIII:2022qal}, the ratio of the Born cross sections for the $e^+e^-\to \pi^+\pi^- J/\psi$ to $e^+e^-\to K^+K^- J/\psi$ is approximately $2.5$ near the center-of-mass energy of $4.5~\mathrm{GeV}$ \cite{BESIII:2018iop,BESIII:2022joj,BESIII:2022qal}.
	If the $Y(4500)$ is the only contribution to the $\pi^+\pi^- J/\psi$ and $K^+K^- J/\psi$,
	this naive cross section ratio implies that the decay process $Y(4500)\to \pi^+\pi^- J/\psi$ might occur with a higher rate around $4.5~\mathrm{GeV}$. 
	%
	%The fact that the significance of the $Y(4500)$ signal in the $e^+e^-\to \pi^+\pi^- J/\psi$ is much smaller might be attributed to the highly strong productions $Y(4230)$ and $Y(4320)$ \cite{BESIII:2022qal}, while in the $e^+e^-\to K^+K^- J/\psi$ the $Y(4230)$ and $Y(4500)$ productions are about of equal rate \cite{BESIII:2022joj}.

	In the present work, we shall study the dipionic transition of the $Y(4500)$ to $J/\psi$ using an effective Lagrangian approach. 
	Following the pioneering theoretical studies \cite{Wang:2022jxj,Wang:2019mhs}, we also regard the $Y(4500)$ as a mixture of the $\psi(5S)$ and $\psi(4\,{}^3D_1)$.
    QCD multipole expansion can deal well with the hadronic transitions between low-lying heavyquarkonium~\cite{Yan:1980uh,Kuang:1981se}. However, coupled-channel effects become important for higher charmonia~\cite{Wang:2022jxj,Wang:2015xsa,Jia:2023pud,Liu:2024ogo}.
    We assume the decay to occur via the charmed meson loops, including the triangle and box loops. 
    This work contributes to testing the $5S$-$4D$ mixing scheme of $Y(4500)$ and offers deeper insights into the role of coupled-channel effects in the decay of higher charmonia.
	In the following, we first give in Sec \ref{sec:lags} the Lagrangians. Then, in Sec. \ref{sec:results} the numerical results and discussion are described in detail. Finally, a summary is given in Sec. \ref{sec:summary}.

\section{Theoretical Consideration}\label{sec:lags}
\begin{figure*}
	\centering
	\includegraphics[width=0.82\linewidth]{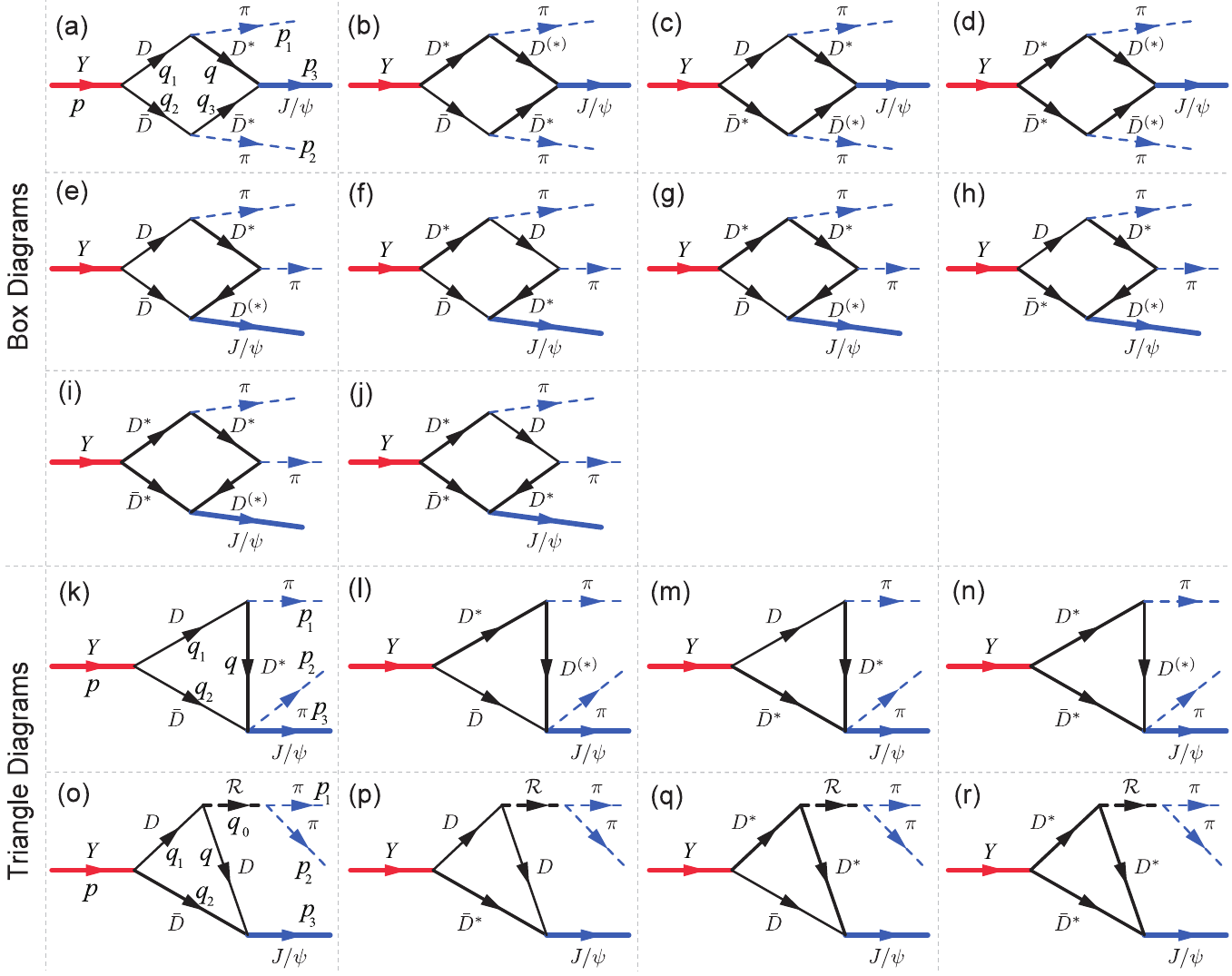}
	\caption{Feynman diagrams for the $Y(4500)\to\pi^+\pi^- J/\psi$ decay process via the intermediate meson loops. As indicated, the diagrams (a)--(j) describe the box loops, while the diagrams (k)--(r) are the triangle loops. The $Y$ represents the $Y(4500)$, and the $\mathcal{R}$ stands for the $f_0(500),\,f_0(980)$, and $f_2(1270)$. The relevant kinematics [$p,\,p_{1(2,3)},\,q,\,q_{0(1,2,3)}$] are explicitly indicated in the graphs. The charge conjugated loops are not shown here but included in the calculations.}
	\label{fig:feyndiags}
   \end{figure*}

   The dipionic processes $Y(4500)\to \pi^+\pi^- J/\psi$ are assumed to occur possibly via the intermediate meson loops, including the box and triangle ones \cite{Chen:2019gfp,Chen:2016mjn} as shown in Fig. \ref{fig:feyndiags}. In view of the experimental observations that the invariant mass distributions for the $K^+K^-$ \cite{BESIII:2018iop,BESIII:2022joj} exhibit clear signals of the mesons $f_0(980)$ and $f_2(1270)$, along with the appreciable decay rate of $Y(4500)\to J/\psi f_0(500)$ predicted by the tetraquark model \cite{Wang:2024qqa}, we incorporate these intermediate mesons into the triangle loops as depicted in Figs. \ref{fig:feyndiags}(o)--(r). 
   %It should be pointed out that some box and triangle loops might be double-considered. However, we cannot identify which ones are equivalent. Thus, 
   The contributions from the different types of loops will be calculated separately in the following, as we did in our previous work \cite{Jia:2023pud,Liu:2024ogo}. The present results show that the $\pi\pi$ invariant mass distributions due to different loops vary significantly. The great difference might be easily recognized by the experimental measurements. Furthermore, the interference effects of the three different types of loop mechanisms are also considered, showing significant influence on the invariant mass distributions of the final states.

   Following the treatment in Refs. \cite{Wang:2019mhs,Wang:2022jxj}, here we also consider the $Y(4500)$ as the mixed state of the $\psi(5S)$ and $\psi(4\,{}^3D_1)$. The $\psi(5S)$ has a mass of 4443 MeV and the $\psi(4\,{}^3D_1)$ 4484 MeV, predicted in the unquenched potential model \cite{Wang:2019mhs}. Within the $S$-$D$ mixing framework, the wave function of the $Y(4500)$ has the following form \cite{Wang:2019mhs,Wang:2022jxj}:
   \begin{equation}\label{eq:wavefunction}
	\tilde{Y}(4500) = -\tilde{\psi}(5S) \sin\theta  + \tilde{\psi}(4\,{}^3D_1) \cos\theta\,,
   \end{equation}
   where the mixing angle $\theta$ is determined to be $30^\circ$ \cite{Wang:2022jxj}, agreeing well with the widely accepted mixing angle for the higher charmonium $S$-$D$ \cite{Badalian:2008dv}. The mixing angle can be determined by either the mass-mixing formula or fitting the experimental dielectron width \cite{Wang:2019mhs}. Here, it should be pointed out that, in some quark models \cite{Barnes:2005pb,Ebert:2011jc,Gui:2018rvv,Deng:2016stx,Deng:2023mza}, the masses of $\psi(4S)$ and $\psi(3\,{}^3D_1)$ are predicted between 4400 and 4500 MeV. Hence, the $4S$-$3D$ mixing scenario for the $Y(4500)$ appears also possible, which corresponds to a bit smaller mixing angle of $\lesssim 25^\circ$. In this work, we consider only the $5S$-$4D$ mixing assumption, regardless of the possibility of the $4S$-$3D$ mixing. The dipionic transitions of the $\psi(4415)$, another mixed state of the $\psi(5S)$ and $\psi(4\,{}^3D_1)$, have been studied in Refs. \cite{Chen:2011xk,Chen:2010nv,Chen:2013wca}, in which the $\psi(4415)$ is regarded as the pure $S$-wave charmonium. Moreover, Ref. \cite{Anwar:2016mxo} shows that for the $S$-wave dominant state with a mixing angle around $30^\circ$, its hidden-charm decay deviates mildly from that of the pure $S$-wave charmonium. Thus, in our calculations, we focus on only the dipionic transition of the $Y(4500)$.
	
   According to the heavy quark effective theory, the interactions of the $S$-wave charmonium $\psi(nS)$ and $D$-wave $\psi(n\,{}^3D_1)$ with the charmed and anticharmed mesons are described by the Lagrangians \cite{Casalbuoni:1996pg,Wang:2022jxj}
  \begin{subequations}\label{eq:LagSD}
	\begin{align}
	\mathcal{L}_S =& -\ii g_{\psi \mathcal{D}\mathcal{D}}\psi_\mu \mathcal{D}^\dagger \pararrowk{\mu}\mathcal{D}\nonumber\\
	&+ g_{\psi \mathcal{D}\mathcal{D}^\ast} \varepsilon_{\mu\nu\alpha\beta}\partial^\mu \psi^\nu (\mathcal{D}^\dagger \pararrowk{\alpha} \mathcal{D}^{\ast\beta} - \mathcal{D}^{\ast\dagger\beta} \pararrowk{\alpha} \mathcal{D})	\nonumber\\
	&+\ii g_{\psi \mathcal{D}^\ast\mathcal{D}^\ast}\psi_\mu(\partial^\nu \mathcal{D}^{\ast\dagger\mu}\mathcal{D}^\ast_\nu-\mathcal{D}^{\ast\dagger}_\nu\partial^\nu D^{\ast\mu}\nonumber\\
	&+ \mathcal{D}^{\ast\dagger\nu}\pararrowk{\mu}\mathcal{D}^{\ast}_\nu)\,,\\
	\mathcal{L}_D =&\, \ii g_{\psi_1 \mathcal{D}\mathcal{D}}\psi_{1\mu} \mathcal{D}^\dagger \pararrowk{\mu}\mathcal{D}\nonumber\\
	&- g_{\psi_1 \mathcal{D}\mathcal{D}^\ast} \varepsilon_{\mu\nu\alpha\beta}\partial^\mu \psi_1^\nu (\mathcal{D}^\dagger \pararrowk{\alpha} \mathcal{D}^{\ast\beta} - \mathcal{D}^{\ast\dagger\beta} \pararrowk{\alpha} \mathcal{D})	\nonumber\\
	&+\ii g_{\psi_1 \mathcal{D}^\ast\mathcal{D}^\ast}\psi_{1\mu}(\partial^\nu \mathcal{D}^{\ast\dagger\mu}\mathcal{D}^\ast_\nu-\mathcal{D}^{\ast\dagger}_\nu\partial^\nu D^{\ast\mu}\nonumber\\
	&+ 4\mathcal{D}^{\ast\dagger\nu}\pararrowk{\mu}\mathcal{D}^{\ast}_\nu)\,.
   \end{align}
  \end{subequations}
   Here $\mathcal{D}^{(\ast)\dagger} = (\bar{D}^{(\ast)0},\,D^{(\ast)-},\,D_s^{(\ast)-})$ is the charmed meson triplet. In the case of the $Y(4500)$ as the $\psi(5S)$-$\psi(4\,{}^3D_1)$ mixture, we need the coupling constants $g_{\psi\mathcal{D}^{(\ast)}\mathcal{D}^{(\ast)}}$ and $g_{\psi_{1}\mathcal{D}^{(\ast)}\mathcal{D}^{(\ast)}}$, which are estimated in the Appendix. For the $J/\psi$, the couplings $g_{J\mathcal{D}\mathcal{D}},\,g_{J\mathcal{D}\mathcal{D}^\ast}$, and $g_{J\mathcal{D}^\ast\mathcal{D}^\ast}$ are linked to each other by a global coupling strength $g_J$ and the masses of the involved mesons as
  \begin{subequations}\label{eq:gJs}
	\begin{align}
		g_{J\mathcal{D}\mathcal{D}} &= 2 g_J \sqrt{m_{J}}m_D\,,\\
		g_{J\mathcal{D}\mathcal{D}^\ast} &= 2 g_J \sqrt{m_Dm_{D^\ast}/m_{J}}\,,\\
		g_{J\mathcal{D}^\ast\mathcal{D}^\ast} &= 2 g_J \sqrt{m_{J}}m_{D^\ast}\,.
	\end{align}
  \end{subequations}
   In terms of the vector meson dominance \cite{Colangelo:2003sa,Deandrea:2003pv}, $g_J = \sqrt{m_{J}}/(2m_Df_{J})$. The $J/\psi$ decay constant $f_J$ can be extracted from $J/\psi\to e^+e^-$ \cite{Badalian:2009bu,Li:2012as,Liu:2023gtx}:
  \begin{equation}
	\Gamma_{ee} = \frac{16\pi\alpha_\mathrm{EM}^2}{27m_J}f_J^2\,,
  \end{equation}
   where $\alpha_{\mathrm{EM}}=1/137$ is the fine-structure constant. Using the newly updated PDG data \cite{ParticleDataGroup:2024cfk}: $m_J=(3096.900\pm 0.006)~\mathrm{MeV}$ and $ \Gamma_{ee}=(5.53\pm 0.10)~\mathrm{keV}$, we find $f_{J} = (416\pm 4)~\mathrm{MeV}$ and, thereby, $g_J=(1.13\pm0.01)~\mathrm{GeV^{-3/2}}$.

   In the heavy quark limit and chiral symmetry, the interactions of the light pseudoscalar mesons with the heavy charmed mesons read \cite{Wang:2022qxe,Li:2013yla,Wu:2021udi}
	\begin{align}\label{eq:LVP}
		\mathcal{L}_\pi =&-\ii g_{\mathcal{D}\mathcal{D}^\ast \mathcal{P}}\big(\mathcal{D}^{i \dagger}\partial^{\mu} \mathcal{P}_{ij}^\dagger \mathcal{D}_\mu^{\ast j} - \mathcal{D}_\mu^{\ast i\dagger}\partial^\mu \mathcal{P}_{ij}^\dagger \mathcal{D}^j\big) \nonumber\\
		&+ \frac{1}{2} g_{\mathcal{D}^\ast\mathcal{D}^\ast\mathcal{P}}\epsilon_{\mu\nu\alpha\beta} \mathcal{D}_i^{*\mu\dagger}\partial^\nu \mathcal{P}^{ij\dagger}\pararrowk{\alpha} \mathcal{D}_j^{*\beta}\,,
	\end{align}
	where the $\mathcal{P}$ represents the pseudoscalar mesons in the matrix form
	\begin{equation}
		\mathcal{P} = \begin{pmatrix}
			\frac{\pi^0}{\sqrt{2}} + \frac{\eta}{\sqrt{6}} & \pi^+ & K^+\\
			\pi^- & -\frac{\pi^0}{\sqrt{2}} + \frac{ \eta }{\sqrt{6}} & K^0\\
			K^- & \bar{K}_0 & - \sqrt{\frac{2}{3}}\eta \label{eq:P}
		\end{pmatrix}\,.
	\end{equation}
The coupling constants $g_{\mathcal{D}^{(\ast)}\mathcal{D}^\ast\mathcal{P}} $ are linked to each other by the following relation:
\begin{equation}
g_{\mathcal{D}^{\ast}\mathcal{D}^\ast\mathcal{P}}=\frac{g_{\mathcal{D}\mathcal{D}^\ast\mathcal{P}}}{\sqrt{m_Dm_{D^\ast}}} = \frac{2g}{f_\pi}
\end{equation}
with the pion decay constant $f_\pi=(130.2\pm 1.2)~\mathrm{MeV}$ \cite{ParticleDataGroup:2024cfk}. Using the Lagrangian in Eq. \eqref{eq:LVP}, we get 
\begin{subequations}\label{eq:gddp}
	\begin{align}
		\Gamma(D^{\ast +}\to D^0\pi^+) &= \frac{g^2}{6\pi f_\pi^2} \frac{m_D}{m_{D^\ast}}\left| \bm{p}_\pi \right|^3\,,\\
		\Gamma(D^{\ast +}\to D^+\pi^0) &= \frac{g^2}{12\pi f_\pi^2} \frac{m_D}{m_{D^\ast}}\left| \bm{p}_\pi \right|^3\,.
	\end{align}
\end{subequations} 
According to the world average data \cite{ParticleDataGroup:2024cfk}, it yields $g=0.57\pm 0.01$, in good agreement with the value $g = 0.59\pm 0.07\pm 0.01$ obtained in Ref. \cite{Isola:2003fh}.

The couplings between the intermediate mesons $\mathcal{R}=\{\mathcal{S},\,f_2(1270)\}$ with $\mathcal{S}=\{f_0(500),\,f_0(980)\}$ and the charmed meson pair can be described by the following Lagrangians \cite{Meng:2008dd,Bai:2022cfz,Chen:2011qx,Chen:2015bma}:
\begin{align}
	\mathcal{L}_{\mathcal{R}\mathcal{D}\mathcal{D}} &= g_{\mathcal{S}\mathcal{D}\mathcal{D}}\mathcal{S}\mathcal{D}\mathcal{D}^\dagger-g_{\mathcal{S}\mathcal{D}^\ast\mathcal{D}^\ast}\mathcal{S}\mathcal{D}^\ast_\mu\mathcal{D}^{\ast\mu\dagger} \nonumber\\
	&+ g_{f_2\mathcal{D}^\ast\mathcal{D}^\ast}f_{2}^{\mu\nu}\mathcal{D}^\ast_\mu\mathcal{D}^{\ast\dagger}_\nu\,.
\end{align}
Here the coupling constants for the $\mathcal{S}=\{f_0(500),\,f_0(980)\}$ are
\begin{subequations}
	\begin{align}
	g_{\mathcal{S}\mathcal{D}\mathcal{D}}=am_Dg_\pi\,,\\
	g_{\mathcal{S}\mathcal{D}^\ast\mathcal{D}^\ast}=am_{D^\ast}g_\pi\,,
\end{align}
\end{subequations}
where $a=1/\sqrt{6}$ and $1/\sqrt{3}$ for the $f_0(500)$ and $f_0(980)$, respectively, and $g_\pi=3.73$ \cite{Bai:2022cfz}. In this work, we take $M_{f_0(500)}=449~\mathrm{MeV}$, $\Gamma_{\mathrm{tot}}[f_0(500)]=550~\mathrm{MeV}$; $M_{f_0(980)}=993~\mathrm{MeV}$, $\Gamma_{\mathrm{tot}}[f_0(980)]=61.3~\mathrm{MeV}$ \cite{Bai:2022cfz}. It is noted that, for the interaction between the $f_2(1270)$ and the charmed meson pair, we consider only the $S$-wave coupling while the higher-wave couplings are neglected. The coupling constant of the $f_2(1270)$ to a pair of charmed mesons is unknown. We shall make an estimation later. 

The decay of the scalar meson $\mathcal{S}$ into two pions is described by 
\begin{equation}
	\mathcal{L}_{\mathcal{S}\pi\pi} = g_{\mathcal{S}\pi\pi}\mathcal{S}\pi\pi\,,
\end{equation}
where $g_{\mathcal{S}\pi\pi} = 3.25~\mathrm{GeV}$ for the $ f_0(500)$ and $g_{\mathcal{S}\pi\pi} = 1.13~\mathrm{GeV}$ for the $ f_0(980)$ \cite{Bai:2022cfz}. Furthermore, the interaction of the $f_2(1270)$ with the $\pi^+\pi^- $ reads
\begin{equation}
	\mathcal{L}_{f_2\pi\pi} = g_{f_2\pi\pi}f_2^{\mu\nu}\partial_\mu\pi^+\partial_\nu\pi^-\,.
\end{equation}
Using the PDG data $\mathcal{B}(f_2\to\pi\pi)=84.3\%$ and $\Gamma_\mathrm{t}(f_2) = 186.6~\mathrm{MeV}$ \cite{ParticleDataGroup:2024cfk}, we find $g_{f_2\pi\pi}=14.64~\mathrm{GeV^{-1}}$ under the assumption of $\Gamma(f_2\to\pi^+\pi^-)=2\Gamma(f_2\to\pi^0\pi^0)$. 

Before proceeding, we provide here the estimation of the coupling constant $g_{f_2 \mathcal{D}^\ast\mathcal{D}^\ast}$.
%of the $f_2$ to the vector charmed meson pair.
In the heavy quark limit ($m_{c(b)}\to \infty$), the couplings of the $f_2$ to the charmed mesons and to the bottom ones appear equal. Fitting the Belle data for the process $\Upsilon(5S)\to\Upsilon(1S)\pi^+\pi^-$ \cite{Belle:2007xek} yields $g_{f_2B^\ast B^\ast}g_{f_2\pi\pi} = 12.36\pm 20.11$ \cite{Chen:2011qx}. As a consequence, the coupling $g_{f_2\mathcal{D}^\ast\mathcal{D}^\ast} $, if the masses of the $b$ and $c$ quarks both approach infinity, is $0.84~\mathrm{GeV}$. When the flavor SU(4) is an exact symmetry, the $g_{f_2\mathcal{D}^\ast\mathcal{D}^\ast}/g_{\chi_{c2}\mathcal{D}^\ast\mathcal{D}^\ast}=1$ accordingly. Using the formulas in Ref. \cite{Liu:2024ogo}, we find $g_{\chi_{c2}\mathcal{D}^\ast\mathcal{D}^\ast}\approx 11.21~\mathrm{GeV} $, and, hence, $g_{f_2\mathcal{D}^\ast\mathcal{D}^\ast}\approx 11.21~\mathrm{GeV} $ under the SU(4) symmetry. Based on the two estimations above, we take a midpoint value $g_{f_2\mathcal{D}^\ast\mathcal{D}^\ast} = 6.0~\mathrm{GeV} $ in our calculations.

The effective four-body interactions among the charmonium $J/\psi$, charmed mesons and the pions are described by the following Lagrangian \cite{Oh:2000qr,Lin:1999ad,Chen:2011xk}
\begin{align}
	\mathcal{L}_{J \mathcal{D}\mathcal{D}\pi} &=- \ii g_{J
	\mathcal{D}\mathcal{D}\pi}\epsilon_{\mu\nu\alpha\beta}J^\mu \partial^\nu \mathcal{D}\partial^\alpha\pi\partial^\beta \mathcal{D}^\dagger\nonumber\\
	&+g_{J\mathcal{D}\mathcal{D}^\ast \pi} J^\mu (\mathcal{D}\pi \mathcal{D}^{\ast\dagger}_\mu + \mathcal{D}^\ast_\mu\pi\mathcal{D}^\dagger)\nonumber\\
	&-\ii g_{J\mathcal{D}^\ast\mathcal{D}^\ast\pi}\epsilon_{\mu\nu\alpha\beta}J^\mu \mathcal{D}^{\ast\nu}\partial^\alpha \pi \mathcal{D}^{\ast\dagger \beta}\nonumber\\
	&-\ii h_{J\mathcal{D}^\ast\mathcal{D}^\ast\pi}\epsilon_{\mu\nu\alpha\beta}\partial^\mu J^\nu \mathcal{D}^{\ast\alpha}\pi\mathcal{D}^{\ast\dagger\beta}\,,
\end{align}
where $\pi = \vec{\tau}\cdot\vec{\pi}$ with $\tau_i$ being the Pauli matrices and $\vec{\pi}=(\pi_1,\,\pi_2,\,\pi_3)$.
%being associated with the pion isospin triplets by the following relations
%\begin{subequations}
%	\begin{align}
%	\pi^+ &= \frac{1}{\sqrt{2}}(\pi_1-\ii\pi_2),\\
%	\pi^-&= \frac{1}{\sqrt{2}}(\pi_1+\ii\pi_2),\\
%	\pi^0 &= \pi_3.
%\end{align}
%\end{subequations}
Following Ref. \cite{Oh:2000qr}, we adopt the four-point coupling constants obtained under the framework of SU(4) symmetry:
\begin{subequations}
	\begin{align}
		g_{J\mathcal{D}\mathcal{D}^\ast\pi}&=33.92, \\
		g_{J\mathcal{D}^\ast\mathcal{D}^\ast\pi} &= h_{J\mathcal{D}^\ast\mathcal{D}^\ast\pi} =38.19~\mathrm{GeV^{-1}},\\
		g_{J \mathcal{D}\mathcal{D}\pi} &=16.0~\mathrm{GeV^{-3}}.
	\end{align}
\end{subequations}

For the box diagrams, the general expression of the amplitudes can be described by 
\begin{equation}\label{eq:boxamp}
	\mathcal{M}_\mathrm{box}=\int\frac{\mathrm{d}^4q}{(2\pi)^4}\frac{V_\psi V_J V_{\pi^+}V_{\pi^-}}{P_1P_2P_3P} \mathcal{F}(q_3,M_3)\mathcal{F}(q,M)\,.
\end{equation}
For the triangle loops involving four-body interactions (hereafter referred as FB loops), the amplitudes have the following form:
\begin{equation}\label{eq:triFDamp}
	\mathcal{M}_{\mathrm{FB}}=\int\frac{\mathrm{d}^4q}{(2\pi)^4}\frac{V_\psi V_{J\pi} V_{\pi}}{P_1P_2P}\mathcal{F}^2(q,M)\,,
\end{equation}
while for the other triangle loops with the mesons $f_0(500)$, $f_0(980)$, and $f_2(1270)$ (called $\mathcal{R}$ loops for short) the amplitudes can be written as
\begin{equation}\label{eq:triRamp}
	\mathcal{M}_{\mathcal{R}} =\int\frac{\mathrm{d}^4q}{(2\pi)^4}\frac{V_\psi V_{J} V_{\mathcal{R}} V_{\pi\pi}}{P_1P_2P}\frac{\mathcal{N}\mathcal{F}^2(q,M)}{q_0^2-M_\mathcal{R}^2+\ii M_\mathcal{R}\Gamma_{\mathcal{R}}}\,.
\end{equation}
In Eqs. \eqref{eq:boxamp}--\eqref{eq:triRamp}, the $V$'s are the vertices related to the mesons specified by the index; the $P_1$, $P_2$, $P_3$, and $P$ stand for the propagators of the intermediate charmed mesons with momentum $q_1$, $q_2$, $q_3$, and $q$, respectively (see Fig. \ref{fig:feyndiags}); the width effect of the intermediate state $\mathcal{R}$ is considered. The numerator 
	\begin{align*}
		\mathcal{N} &= 1 ~~ \text{for $f_0(500)$ and $f_0(980)$},\\
		\mathcal{N} &= \frac{1}{2}(\bar{g}^{\mu\alpha}\bar{g}^{\nu\beta}+\bar{g}^{\mu\beta}\bar{g}^{\nu\alpha})-\frac{1}{3}\bar{g}^{\mu\nu}\bar{g}^{\alpha\beta}~~ \text{for $f_2(1270)$}
	\end{align*}
with $\bar{g}^{\mu\nu}=(g^{\mu\nu} - q_0^\mu q_0^\nu/M_{f_2}^2)$.
Here, it should be pointed out that, for the triangle $\mathcal{R}$ loops, we introduce two parameters $\phi_1$ and $\phi_2$ to describe the phase difference of the $f_0(980)$ and $f_2(1270)$ with respect to the $f_0(500)$. That is to say, the amplitude $\mathcal{M}_{\mathcal{R}}$ due to the $\mathcal{R}$ loops in Figs. \ref{fig:feyndiags}(o)--1(r) is of the form
\begin{equation}\label{eq:phi12def}
	\mathcal{M}_\mathcal{R} = \mathcal{M}_{f_0(500)} + \ee^{\ii\phi_1}\mathcal{M}_{f_0(980)} + \ee^{\ii\phi_2} \mathcal{M}_{f_2(1270)}\,.
\end{equation}
To describe the off-shell effects of the intermediate charmed mesons and the structure effect of the interactions vertices, a monopole form factor is introduced~\cite{Locher:1993cc,Li:1996yn,Cheng:2004ru}:
\begin{equation}
	\mathcal{F}(q,\,M) = \frac{M^2-\Lambda^2}{q^2-\Lambda^2}\,.
\end{equation} 
Here $\Lambda = M + \alpha \Lambda_{\mathrm{QCD}}$ with $\Lambda_{\mathrm{QCD}} = 0.22~\mathrm{GeV}$. The parameter $\alpha$ depends on the given processes, but is usually expected to be around unity~\cite{Cheng:2004ru}.

According to the mixed wave function of the $Y(4500)$ in Eq. \eqref{eq:wavefunction}, the total amplitude of the decays for the $Y(4500)\to\pi^+\pi^- J/\psi$ is written as
\begin{equation}
	\mathcal{M}_{\mathrm{tot}} = -\mathcal{M}_S\sin\theta + \mathcal{M}_D\cos\theta\,,
\end{equation}
where $\mathcal{M}_S$ and $\mathcal{M}_D$ are the amplitudes due to the pure $\psi(5S)$ and $\psi(4\,{}^3D_1)$ contributions, respectively, of which the proportion is described by the mixing angle $\theta$. It is recalled that we adopt $\theta=30^\circ$ in this work.

Once the amplitude $\mathcal{M}_\mathrm{tot}$ is obtained, the differential partial decay width is then evaluated by \cite{ParticleDataGroup:2024cfk}
\begin{equation}
	\frac{\mathrm{d}\Gamma}{\mathrm{d}m_{\pi\pi}^2\mathrm{d}m_{J/\psi\pi}^2} = \frac{1}{3} \frac{1}{(2\pi)^3}\frac{1}{32m^3} \sum_{\mathrm{spin}}\left| \mathcal{M}_\mathrm{tot}\right| ^2 \,.
\end{equation}
Here the symbol $\sum_{\mathrm{spin}}$ means the summation over the spins of the initial $Y(4500)$ and final $J/\psi$. Additionally, the $m_{\pi\pi}$ and $m_{J/\psi\pi}$ stand for the invariant mass of the two pions and of the $J/\psi$ and $\pi$, respectively, while $m$ is the $Y(4500)$ mass. The calculations were conducted with the assistance of the FeynCalc \cite{Mertig:1990an,Shtabovenko:2020gxv} and LoopTools \cite{Hahn:1998yk} packages.

\section{Numerical Results and Discussion}\label{sec:results}

\begin{table}
	\caption{The coupling constants used to evaluate the decay process $Y(4500)\to\pi^+\pi^- J/\psi$. The couplings for the four-body interactions are taken from Ref. \cite{Oh:2000qr}, the couplings related to the $f_0(500)$ and $f_0(980)$ are from Ref. \cite{Bai:2022cfz}, and the estimations for the others based on experimental data or theoretical predictions are discussed in the text and the Appendix.}
	\label{tab:couplingconstants}
	\begin{ruledtabular}
		\begin{tabular}{ll|ll}
			Coupling & Value & Coupling & Value\\
			\colrule
			$g_{\psi \mathcal{D}\mathcal{D}}$  & $0.37$  & $g_{\psi_1 \mathcal{D}\mathcal{D}}$  & $-0.72$ \\
			$g_{\psi \mathcal{D}\mathcal{D}^\ast}$ & $0.041~ \mathrm{GeV^{-1}}$ & $g_{\psi_1\mathcal{D}\mathcal{D}^\ast}$  & $0.065~ \mathrm{GeV^{-1}}$ \\
			$g_{\psi \mathcal{D}^\ast\mathcal{D}^\ast}$ & $0.38$ & $g_{\psi_1 \mathcal{D}^\ast\mathcal{D}^\ast}$& $0.20$ \\   \colrule
			$g_{J\mathcal{D}\mathcal{D}}$&$7.44$&$g_{J\mathcal{D}\mathcal{D}\pi}$&$16.0~\mathrm{GeV^{-3}}$ \\
			$g_{J\mathcal{D}\mathcal{D}^\ast}$&$2.49~\mathrm{GeV^{-1}}$&$g_{J\mathcal{D}\mathcal{D}^\ast\pi}$&$33.92$\\
			$g_{\mathcal{D}^\ast\mathcal{D}^\ast}$&$8.00$ &$g(h)_{J\mathcal{D}^\ast\mathcal{D}^\ast\pi}$ &$38.19~\mathrm{GeV^{-1}}$   \\
			\colrule
			$g_{f_0(500)\mathcal{D}\mathcal{D}}$& $2.84~\mathrm{GeV}$ &$g_{f_0(980)\mathcal{D}\mathcal{D}} $ &$4.02~\mathrm{GeV}$\\
			$g_{f_0(500)\mathcal{D}^\ast\mathcal{D}^\ast}$& $3.06~\mathrm{GeV}$ &$g_{f_0(980)\mathcal{D}^\ast\mathcal{D}^\ast} $ &$4.33~\mathrm{GeV}$\\
			$g_{f_0(500)\pi\pi}$&$3.25~\mathrm{GeV}$&$g_{f_0(980)\pi\pi}$&$1.13~\mathrm{GeV}$\\
			$g_{f_2\mathcal{D}^\ast\mathcal{D}^\ast}$&$6.0~\mathrm{GeV}$&$g_{f_2\pi\pi}$&$14.64~\mathrm{GeV^{-1}} $\\
			\colrule
			$g_{\mathcal{D}\mathcal{D}^\ast\pi}$&$17.31$&$g_{\mathcal{D}^\ast\mathcal{D}^\ast\pi}$&$8.94~\mathrm{GeV^{-1}}$
		\end{tabular}
	\end{ruledtabular}
\end{table}

The following calculations are devoted to the analysis of
the partial decay widths contributed from different kinds of Feynman diagrams. The calculations were
performed with the coupling constants summarized in Table \ref{tab:couplingconstants}. To obtain more theoretical information about the decay process $Y(4500)\to\pi^+\pi^- J/\psi$, we evaluated the invariant mass spectra of the $\pi^+\pi^-$ and $\pi^+ J/\psi$ via those three kinds of loops depicted in Fig. \ref{fig:feyndiags}. These spectra can be directly compared with the future BESIII and Belle II experiments. In the absence of experimental data, we are incapable of limiting the superposition among these three kinds of loops. As a consequence, in the following, we first describe their contributions separately.

\begin{figure*}
	\centering
	\includegraphics[width=.96\linewidth]{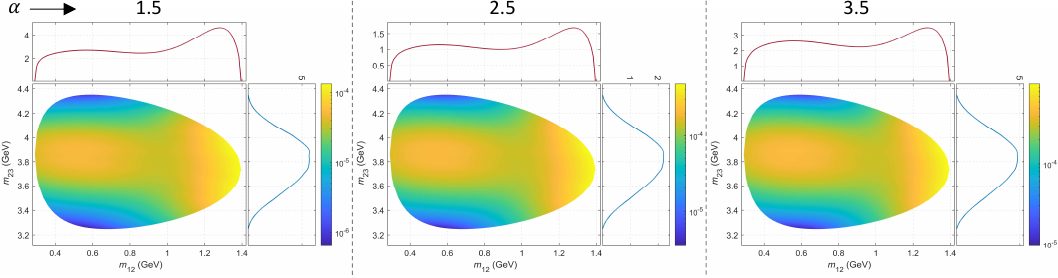}
	\caption{ The Dalitz plots and distributions of the invariant mass of the $\pi^+\pi^-$ ($m_{12}$) and $\pi^+ J/\psi$ ($m_{23}$) due to the box diagrams in Figs. \ref{fig:feyndiags}(a)--1(j) at $\alpha=1.5,\,2.5$, and $3.5$. The solid lines are the corresponding spectra projected onto the $m_{12}$ (red) and $m_{23}$ axis, namely, $(\mathrm{d} \Gamma / \mathrm{d} m_{12})$ and $(\mathrm{d} \Gamma / \mathrm{d} m_{23})$. For graphical reasons, the projected spectra for $\alpha=1.5$ are increased by $10^5$, while for $\alpha=2.5$ and $3.5$ they are increased by $10^4$. The color bars are not the same.}
	\label{fig:boxdgdm}
\end{figure*}

In Fig. \ref{fig:boxdgdm}, two-dimensional distributions of the invariant mass of the $\pi^+\pi^-$ and $\pi^+ J/\psi$ pairs due to the box loops [Figs. \ref{fig:feyndiags}(a)--1(j)] are shown for three different model parameters $\alpha=1.5$, $2.5$, and $3.5$. Moreover, the corresponding projected line spectra are also attached marginally. Roughly speaking, the spectral pattern are nearly model($\alpha$) independent. 
%The invariant mass distribution of the $\pi^+\pi^-$ ($m_{12}$) seems to show signals of the mesons $f_0(500)$ and $f_2(1270)$. With increasing the model parameter $\alpha$, the signal of the $f_0(500)$, if it really exists in this process, is relatively enhanced. Under the present parameters, no obvious structure in the $\pi^+ J/\psi$ mass spectrum is indicated. 
%
The scattered events in the Dalitz plot are concentrated in the kinematic regions $m_{23}=3.6$--$4$ GeV and $m_{12}=1$--$1.4$ GeV. As shown in Figs. \ref{fig:feyndiags}(a)--1(j), the $J/\psi\pi$ final states originate from the $D^{(*)}\bar{D}^{(*)}\to J/\psi\pi$ scattering processes mediated by the t-channel charmed meson exchange. Therefore, in the range of $m_{23}=3.6$--$4$, i.e., the mass threshold of $D^{(*)}\bar{D}^{(*)}$, the enhancement of the interaction between $D^{(*)}\bar{D}^{(*)}$ and $J/\psi\pi$ leads to the event clustering. The partial decay width contributed from the box loops shows strong dependence on the $\alpha$, which increases from $30$ to $600~\mathrm{keV}$ as $\alpha$ is varied from $1.5$ to $5.0$.

\begin{figure*}
	\centering
	\includegraphics[width=.96\linewidth]{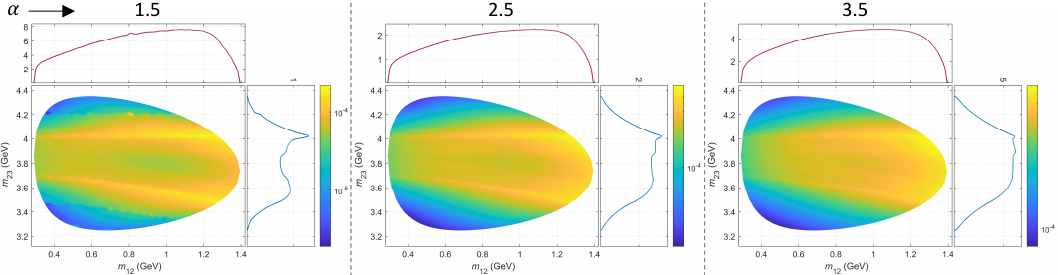}
	\caption{As in Fig. \ref{fig:boxdgdm}, but these data are obtained using the triangle loops shown in Figs. \ref{fig:feyndiags}(k)--1(n).	For graphical reasons, $(\mathrm{d}\Gamma/\mathrm{d}m_{12})\times 10^5$ and $(\mathrm{d}\Gamma/\mathrm{d}m_{23})\times 10^4$ for $\alpha=1.5$, while $(\mathrm{d}\Gamma/\mathrm{d}m_{12})\times 10^4$ and $(\mathrm{d}\Gamma/\mathrm{d}m_{23})\times 10^4$ for $\alpha=2.5$ and $3.5$.}
	\label{fig:fourbodydgdm}
\end{figure*}

Next, in Fig. \ref{fig:fourbodydgdm}, the results contributed from the FB loops [Figs. \ref{fig:feyndiags}(k)--1(n)] are compared for the three model parameters $\alpha =1.5$, $2.5$, and $3.5$. Here, the selection of $\alpha$'s is the same as that for the box loop case so that the calculated data shown before in Fig. \ref{fig:boxdgdm} can be included in the comparison. It is seen that the invariant mass of the $\pi^+\pi^-$ is nearly independent of the model parameter. However, the $\pi^+ J/\psi$ mass distribution exhibits a noticeable change with the model parameter.
%There exists a narrow peak at $m_{23}=4.02~\mathrm{GeV}$ at $\alpha=1.5$, but with increasing $\alpha$, this peak would be weakened and a small bump arises near $m_{23}=3.89~\mathrm{GeV}$. This two structures in the line spectrum of the $\pi^+ J/\psi$ invariant mass might correspond to the exotic states $Z_c(3900)$ and $X(4020)$ \cite{ParticleDataGroup:2024cfk}.
%
In the first graph in Fig.~\ref{fig:fourbodydgdm}, two clear band distributions are shown at $\alpha=1.5$. The horizontal band is located at $m_{23}=4.02$ GeV with a significant peak in the $m_{23}$ invariant mass spectra, while the oblique band appears near $m_{23}=3.6$ GeV as a lower broad bump. Similar to the analysis of $Y(4260\to J/\psi\pi\pi)$ in Ref.~\cite{Wang:2013cya}, the oblique band and lower bump originate from the interference between the $J/\psi \pi^+$ and $J/\psi \pi^-$ rescattering from $D^{*}\bar{D}^{*}$ and are a reflection of the narrow structure at 3.9 GeV. These two structures in the mass spectrum of the $\pi^+ J/\psi$ invariant mass might correspond to the exotic states $Z_c(3900)$ and $Z_c(4020)$ \cite{ParticleDataGroup:2024cfk}. From Fig.~\ref{fig:fourbodydgdm}, these two structures gradually diminish with increasing $\alpha$ and manifest as two narrow peaks at $\alpha=3.5$.
It is noticed that the shape of the $m_{\pi^+ J/\psi}$ distribution at $\alpha=1.5$ is similar to the experimental data of the $Y(4230)\to\pi^+\pi^- J/\psi$ \cite{BESIII:2013ris}. Additionally, the invariant mass spectrum of the $\pi^+ J/\psi$ obtained by the present model is similar to that through the initial single chiral particle emission mechanism proposed by Chen and Liu \cite{Chen:2013coa,Chen:2013wca,Chen:2011xk}. The partial decay width through this kind of loops is about $65~\mathrm{keV}$ at $\alpha =1.5$ and can reach up to $960~\mathrm{keV}$ at $\alpha=5.0$.

\begin{figure*}
	\centering
	\includegraphics[width=.99\linewidth]{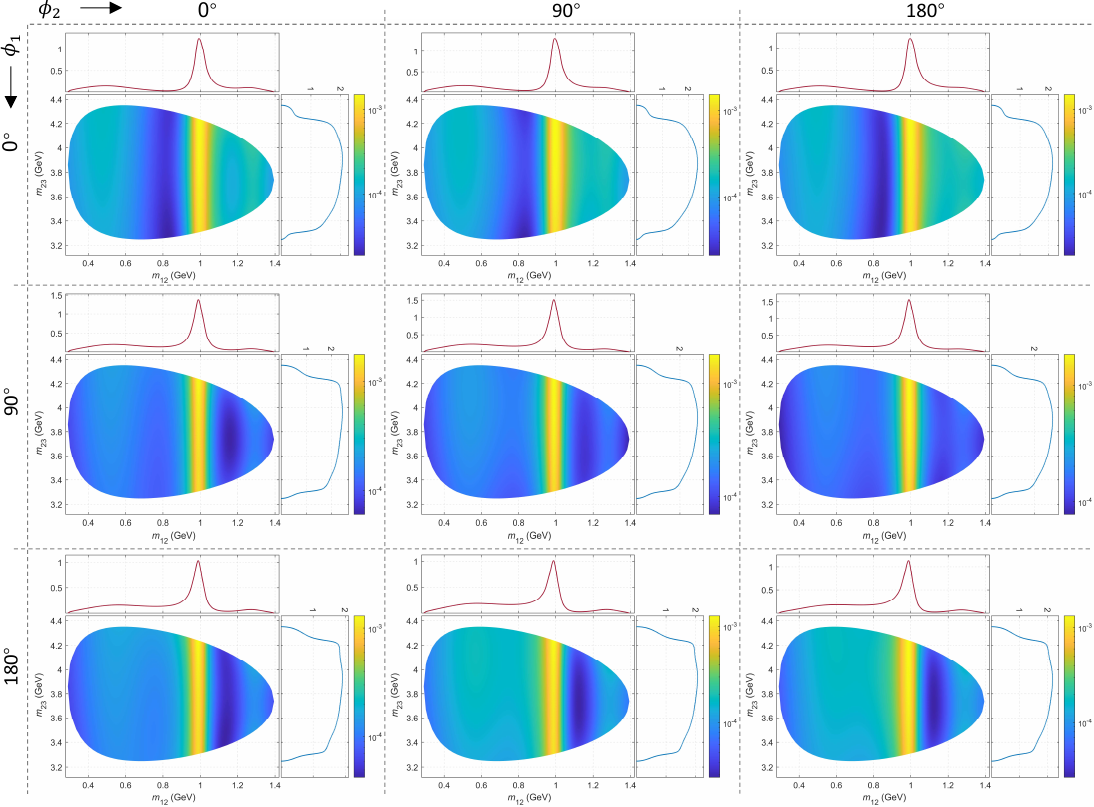}
	\caption{Distributions of the invariant mass of the $\pi^+\pi^-$ ($m_{12}$) and $\pi^+ J/\psi$ ($m_{23}$) for different combinations of the $\phi_1$ and $\phi_2$ (in units of degree) at $\alpha=3.5$, obtained using the triangle $\mathcal{R}$ loops (o)--(r) in Fig. \ref{fig:feyndiags}. The solid lines are the corresponding spectra projected onto the $m_{12}$ (red) and $m_{23}$ axis, namely $(\mathrm{d} \Gamma / \mathrm{d} m_{12})\times 10^{3}$ and $(\mathrm{d} \Gamma / \mathrm{d} m_{23})\times 10^{4}$. The color bars for different combinations are not the same.}
	\label{fig:trigdgdm}
\end{figure*}

 Figure \ref{fig:trigdgdm} shows the invariant mass distributions obtained by means of the $\mathcal{R}$ diagrams [namely, Figs. \ref{fig:feyndiags}(o)-(r)]. In this case, the spectra of the $\pi^+\pi^-$ and $\pi^+ J/\psi$ are found to be completely independent of the model parameter $\alpha$. Hence, we take the results at $\alpha=3.5$ for illustration. To indicate the influence of the phases $\phi_1$ and $\phi_2$ [defined in Eq. \eqref{eq:phi12def}], we choose three special values of $0^\circ$, $90^\circ$, and $180^\circ$. On the whole, the spectra are weakly dependent on the phase angles $\phi_1$ and $\phi_2$. This finding indicates that the interference effect among the $f_0(500)$, $f_0(980)$ and $f_2(1270)$ contributions, if their relevant parameters are those we use here, is of minor importance. 
%In the line spectrum of the $\pi^+\pi^-$, there are three peaks or bulges near the $f_0(500)$, $f_0(980)$, and $f_2(1270)$ mass, just as it should be, since in the $\mathcal{R}$ loops the two pions are assumed to be produced directly via theses three intermediate mesons [see Figs. \ref{fig:feyndiags}(o)--(r)]. 
%
The Dalitz plots exhibit a distinct band at $m_{12}=0.98$ GeV, regardless of the $\phi_1$ and $\phi_2$ values. Meanwhile, a significant peak appears at $0.98$ GeV in the $\pi^+\pi^-$ invariant mass spectra. Given that the $Y(4500)$ is regarded as a mixed state of the $\psi(5S)$ and $\psi(4\,{}^3D_1)$, with a dominant 
$D$-wave component, the charmed meson loop contributions via the $f_0(500/980)$ are strongly suppressed in the heavy quark limit. This suppression arises from the fact that the angular momenta of the $\psi(4\,{}^3D_1)$ and $J/\psi$ are $L=2$ and $L=0$, respectively. Similarly, the $\psi(5S)$ state in the wave function of the $Y(4500)$ also suppresses the charmed meson loop contributions via the $f_2(1270)$. One should notice that the broad widths of the $f_0(500)/f_2(1270)$ amplify the suppression of associated loop contributions and result in two small bumps around $m_{12}=0.5$ and $1.27$ GeV in the line shape. In addition, for both charmonium~\cite{BES:2006eer,Wang:2015xsa} and bottomonium~\cite{CLEO:2007rbi,Meng:2007tk} dipion transitions, the dipion systems are predominantly $S$-wave dominated, highlighting the essential role of scalar resonances in these processes. As a result, only a single peak or band located at $m_{12}=0.98$ GeV is found for the triangle $\mathcal{R}$ loops, corresponding to the $f_0(980)$.  
At $\alpha=3.5$, the $\mathcal{R}$ loop mechanism predicts the partial decay width to be of the order of $200~\mathrm{keV}$, reflecting the suppression effect discussed above. In Table \ref{tab:Rwidth} the partial decay widths due to the $\mathcal{R}$ loops for different combinations of phases $\phi_1$ and $\phi_2$ are listed. Despite the $\alpha$ independence of the invariant mass distributions, the partial width depends strongly on the model parameter $\alpha$. According to the calculated results, it can be concluded that the phase influence in the case of $\mathcal{R}$ loops is of minor importance. 

\begin{table}
	\caption{Partial decay widths (in units of keV) due to the $\mathcal{R}$ loops for different combinations of phases $\phi_1$ and $\phi_2$. Here, the model parameters $\alpha =1.5$, $2.5$, and $3.5$ are chosen.}\label{tab:Rwidth}
	\begin{tabular}{@{}c|c|c|c|c|c|c|c|c|c@{}}
		\hline\hline
		& \multicolumn{3}{c|}{$\phi_2=0^\circ$} & \multicolumn{3}{c|}{$\phi_2=90^\circ$} & \multicolumn{3}{c}{$\phi_2=180^\circ$} \\ \cline{2-10} 
		&
		\multicolumn{1}{c|}{$\alpha=1.5$} &
		\multicolumn{1}{c|}{$\alpha=2.5$} &
		$\alpha=3.5$ &
		\multicolumn{1}{c|}{$\alpha=1.5$} &
		\multicolumn{1}{c|}{$\alpha=2.5$} &
		$\alpha=3.5$ &
		\multicolumn{1}{c|}{$\alpha=1.5$} &
		\multicolumn{1}{c|}{$\alpha=2.5$} &
		$\alpha=3.5$  \\ \hline
		\multicolumn{1}{c|}{$\phi_1=0^\circ$}   & 10.7       & 60.7       & 175.4       & 12.0        & 67.7       & 195.7       & 11.1        & 62.3        & 180.0       \\
		\multicolumn{1}{c|}{$\phi_1=90^\circ$}  & 13.6       & 76.5       & 221.3       & 15.3        & 86.1       & 249.2       & 15.3        & 86.3        & 250.0       \\
		\multicolumn{1}{c|}{$\phi_1=180^\circ$} & 10.8       & 60.7       & 175.6       & 11.5        & 64.8       & 187.6       & 12.0        & 67.6        & 195.6  \\ \hline\hline    
	\end{tabular}
\end{table}

\begin{figure}
	\centering
	\includegraphics[width=0.48\linewidth]{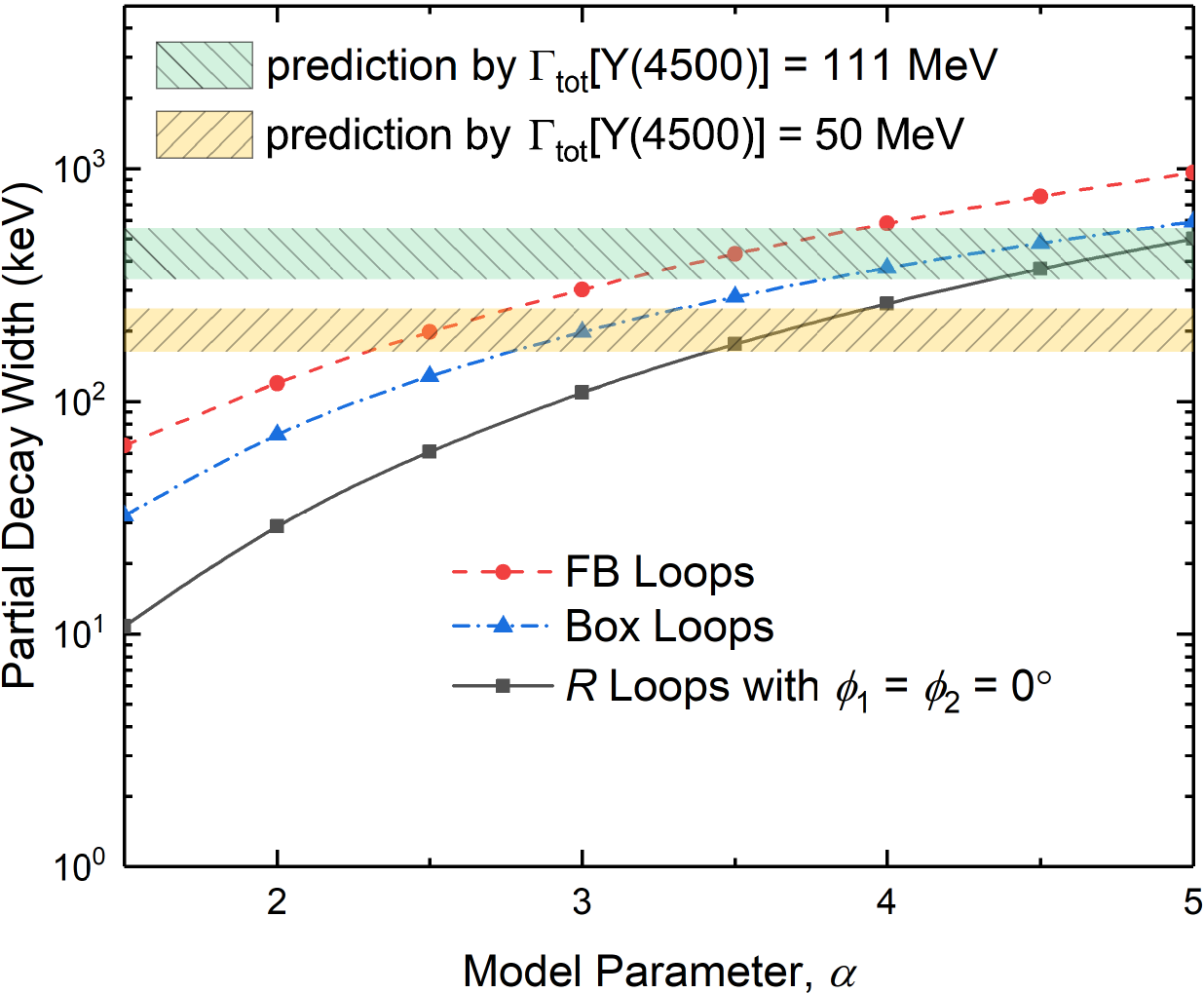}
	\caption{Partial decay width as a function of the model parameter $\alpha$. The closed circles, triangles, and squares represent results contributed the FB, box, and $\mathcal{R}$ loops, respectively. The calculations for $\mathcal{R}$ loops were conducted at phases $\phi_1=\phi_2=0^\circ$. The two bands depict the estimations using the BESIII experiments \cite{BESIII:2018iop} and the theoretical predictions for the $Y(4500)\to K^+K^- J/\psi$ \cite{Wang:2022jxj} (see the text).}
	\label{fig:widthvsalpha}
\end{figure}

From the invariant mass distributions in Figs. \ref{fig:boxdgdm}-\ref{fig:trigdgdm}, we found that the FB and box diagrams play a role as background terms with respect to the $\mathcal{R}$ loops diagrams for the $\pi\pi$. As for $J/\psi\pi$, the $\mathcal{R}$ loops and box diagrams play a role as background terms with respect to the FB diagrams. The predicted line shapes are helpful for experimentally searching the decay $Y(4500)\to J/\psi \pi\pi$.
In Fig. \ref{fig:widthvsalpha}, we plot the partial decay widths contributed from the three kinds of loops as a function of the model parameter $\alpha$. Although patterns of the invariant mass distributions are hardly dependent on the model parameter $\alpha$ as mentioned above, the partial decay widths increase distinctly with increasing $\alpha$. It is seen that under the present conditions we adopted, the FB loops show the most importance among the three kinds of loops. In the range $\alpha\in [1.5,\,5]$, the decay widths through the FB, box, and $\mathcal{R}$ loops are, respectively,
\begin{subequations}
	\begin{align}
		\Gamma_{\mathrm{FB}} &= (64.6\text{--}960.2)~\mathrm{keV}\,,\\
		\Gamma_{\mathrm{box}} &= (32.2\text{--} 591.1)~\mathrm{keV}\,,\\
		\Gamma_{\mathcal{R}} &= (10.8\text{--}499.4)~\mathrm{keV}\,.
	\end{align}
\end{subequations}

To characterize the interference effects among the three kinds of loop contributions, we introduce two other phase angles, $\chi_1$ and $\chi_2$. Consequently, the total contribution to the $Y(4500)\to\pi^+\pi^- J/\psi$ from the three kinds of loop diagrams shown in Fig. \ref{fig:feyndiags} could be expressed as
\begin{equation}
	\mathcal{M}_\mathrm{tot} = \mathcal{M}_{\mathrm{box}} +\ee^{\ii \chi_1}\mathcal{M}_{\mathrm{FB}}+\ee^{\ii \chi_2}\mathcal{M}_{\mathcal{R}}\,.
\end{equation}

\begin{figure*}
	\centering
	\includegraphics[width=.99\linewidth]{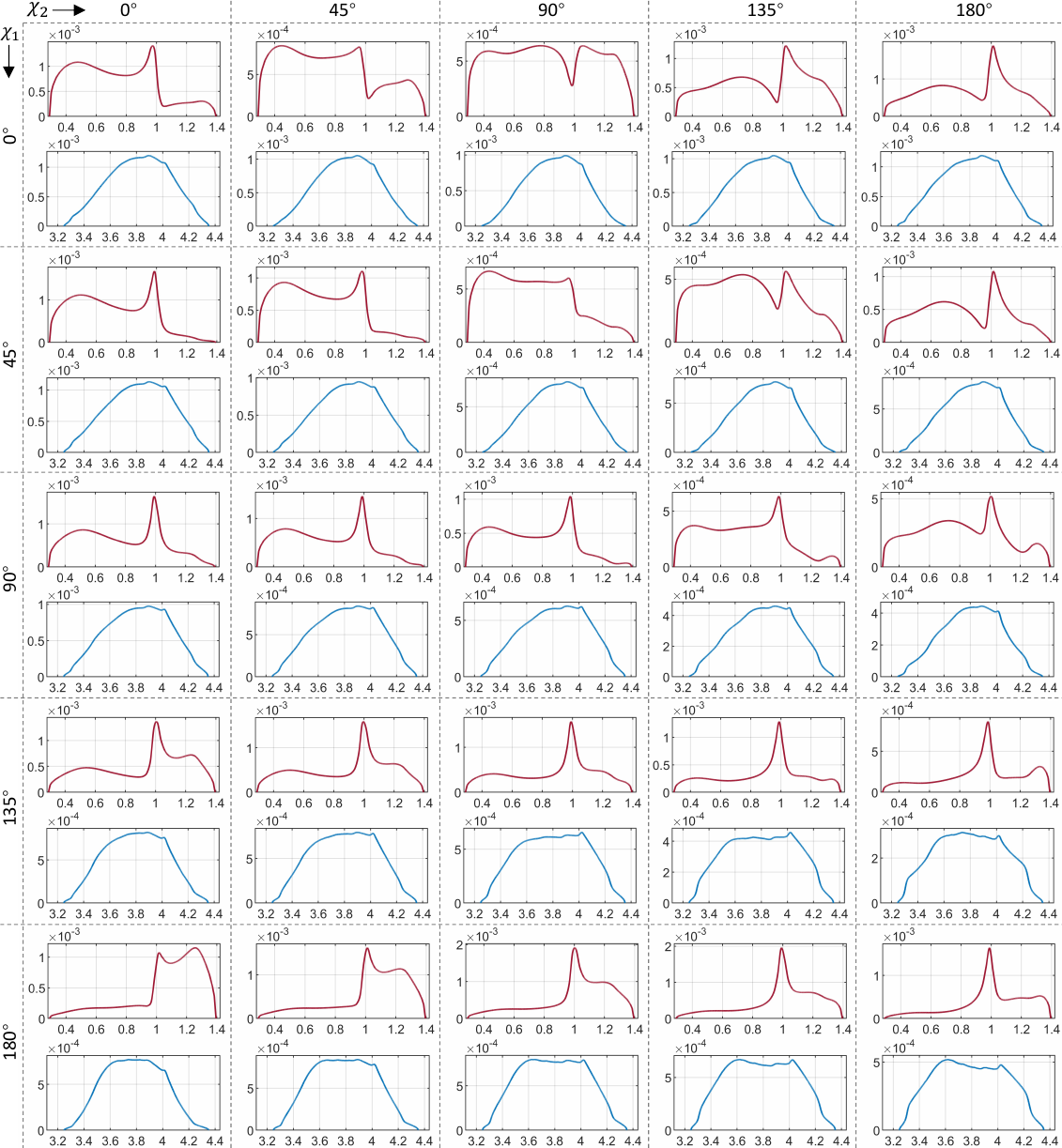}
	\caption{Distributions of the invariant mass of the $\pi^+\pi^-$ ($m_{12}$) and $\pi^+ J/\psi$ ($m_{23}$) due to the three kinds of loop Feynman diagrams depicted in Fig. \ref{fig:feyndiags}. The calculations were performed using $\alpha=3.0$ and $\phi_1=\phi_2=0^\circ$. The phase angles $\chi_1$ and $\chi_2$ were varied from $0^\circ$ to $180^\circ$ as indicated. The solid red line represents the $\mathrm{d}\Gamma/ \mathrm{d} m_{12}$ and the blue line is the $\mathrm{d}\Gamma/ \mathrm{d} m_{23}$. For graphical reasons, the $x$- and $y$-axis labels are omitted.}
	\label{fig:dgdmvsphasechi}
\end{figure*}

In the absence of experimental results, the $\chi_1$ and $\chi_2$ remain completely free. Notice that the patterns shown in Figs. \ref{fig:boxdgdm}--\ref{fig:trigdgdm} are nearly independent of the model parameter $\alpha$. Moreover, the phase angles $\phi_1$ and $\phi_2$ [see Eq. \eqref{eq:phi12def}], describing the interference of the $\mathcal{R}$ loops, also have little effect on the decay process (see Fig. \ref{fig:trigdgdm} and Table \ref{tab:Rwidth}). Hence, these calculations were conducted using $\alpha=3.0$ and $\phi_1=\phi_2=0^\circ$. In Fig. \ref{fig:dgdmvsphasechi}, the invariant mass distributions of the $\pi^+\pi^-$ and $\pi^+ J/\psi$ are presented for the $\chi_1$ and $\chi_2$ in the range of $0^\circ$--$ 180^\circ$. It is seen that the pattern of the $\pi^+\pi^-$ invariant mass distributions exhibits strong dependence on the $\chi_1$ and $\chi_2$. In contrast, for the $\pi^+ J/\psi$ system, the invariant-mass-distribution pattern shows a rather small sensitivity to the phase angle $\chi_2$. Moreover, it varies marginally with the $\chi_1$ when compared to that of the $\pi^+\pi^-$. The significant dependence of the $\pi^+\pi^-$ invariant mass distribution on the angles $\chi_1$ and $\chi_2$ is understandable, because the $\pi^+\pi^-$ invariant mass intensities (namely the $\mathrm{d}\Gamma/\mathrm{d}m_{12}$ shown in Figs. \ref{fig:boxdgdm} and \ref{fig:fourbodydgdm}) for the box and FB loops have nearly the same order of magnitude as the peak intensity near the $f_0(980)$ for the $\mathcal{R}$ loops, thereby leading to obvious interference effect.

\begin{figure*}
	\centering
	\includegraphics[width=0.48\linewidth]{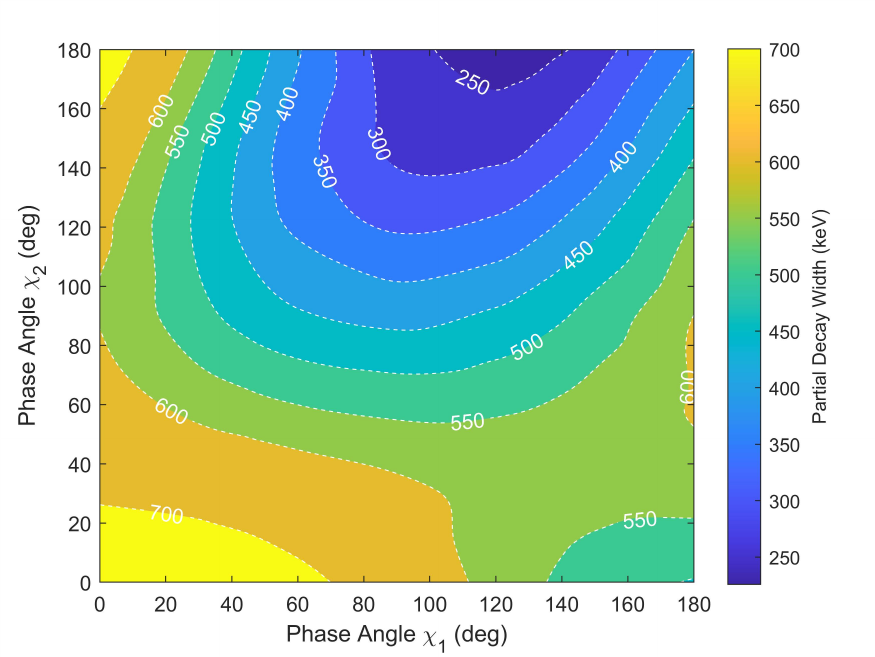}
	\caption{Partial decay width due to the three kinds of the loops in Fig. \ref{fig:feyndiags}. The $\alpha=3.0$ and $\phi_1=\phi_2=0^\circ$ were used.}
	\label{fig:widthvsphase1and2}
\end{figure*}

We plot in Fig. \ref{fig:widthvsphase1and2} the partial decay width of the $Y(4500)\to\pi^+\pi^- J/ \psi$ as a function of the $\chi_1$ and $\chi_2$. As seen, the width value varies significantly with the $\chi_1$ and $\chi_2$.
 Given the dependence of the partial decay widths on the model parameter $\alpha$, as illustrated in Fig. \ref{fig:widthvsalpha}, the foregoing predicted width ranges alone are insufficient to constrain the phase angles $\chi_1$ and $\chi_2$.
 %Since the partial decay widths also depend on the model parameter $\alpha$ as indicated in Fig. \ref{fig:widthvsalpha}, we cannot, only according to the foregoing predicted range of the widths, restrict the phase angles $\chi_1$ and $\chi_2$. 
 A precise measurement of the invariant mass distributions of the $\pi^+\pi^-$ and $\pi J/\psi$ would enable the determination of $\chi_1$ and $\chi_2$, thereby providing constraints on the model parameter $\alpha$.

If the $Y(4500)$ is the only contribution to the $\pi^+\pi^- J/\psi$ and $K^+ K^- J/\psi$, the BESIII experiments \cite{BESIII:2018iop} imply $\Gamma[Y(4500)\to K^+K^- J/\psi]/\Gamma[Y(4500)\to \pi^+\pi^- J/\psi]\approx 0.4$. In view of the theoretical prediction in Ref. \cite{Wang:2022jxj} that the $\Gamma[Y(4500)\to K^+K^- J/\psi] \approx (65\sim 100)~\mathrm{keV}$ when $\Gamma_{\mathrm{tot}}[Y(4500)]=50~\mathrm{MeV}$ is adopted, the partial width $\Gamma[Y(4500)\to \pi^+\pi^- J/\psi] \approx (162.5$--$250)~\mathrm{keV}$. However, if the experimental center value $\Gamma_{\mathrm{tot}}[Y(4500)]=111~\mathrm{MeV}$ is adopted, the corresponding decay width would be increased by approximately a factor of 2. These two estimations are indicated in Fig. \ref{fig:widthvsalpha} as the two bands. Unfortunately, without any experimental constraints, the parameters $\chi_1$ and $\chi_2$ are entirely free. Nevertheless, the predicted partial width of $Y(4500)\to \pi^+\pi^- J/\psi$, varying from 250 to 700 keV with $\alpha=3.0$ and $\phi_1=\phi_2=0^\circ$, is consistent with the estimation obtained by combining the measurement by BESIII Collaboration and theoretical prediction for the $Y(4500)\to K^+K^- J/\psi$.

\section{Summary}\label{sec:summary}
The dipionic transition of the newly observed charmoniumlike resonance $Y(4500)$ by the BESIII Collaboration to the $J/\psi$ was investigated using an effective Lagrangian approach. This work was motivated by BESIII observations that the Born cross section of the $e^+e^-\to\pi^+\pi^- J/\psi$ exhibits a small enhancement around $4.5~\mathrm{GeV}$ \cite{BESIII:2022qal}, hinting the well-constructed $Y(4500)$ in the $e^+e^-\to K^+ K^- J/\psi$ \cite{BESIII:2022joj}. In this study, the transition was assumed to occur via charmed meson loops, including box loops and two kinds of triangle loops. Similar to the treatment by Wang and Liu \cite{Wang:2022jxj}, the $Y(4500)$ was considered as a mixture of the $\psi(5S)$ and $\psi(4\,{}^3D_1)$.

Our calculations indicate that the patterns of the invariant mass distributions of $\pi^+\pi^-$ and $\pi^+ J/\psi$ are hardly dependent on the model, especially for those contributed from the box and $\mathcal{R}$ loops. Different loop mechanisms give different invariant mass spectra. The interference among the three different kinds of loops is also exhibited. We could determine the decay mechanism of the process $Y(4500)\to \pi^+\pi^- J/\psi$ by comparing the present calculated spectra with the future experimental measurements by BESIII or Belle II. The partial decay widths are found to be model dependent. Combining the BESIII experiments \cite{BESIII:2022joj} and the theoretical predictions \cite{Wang:2022jxj}, the width of $Y(4500)\to\pi^+\pi^- J/\psi$ was roughly estimated to between $160$ and $250~\mathrm{keV}$ or from $360$ to $550~\mathrm{keV}$. The different ranges arise from different total widths of the $Y(4500)$. Our results agree with the combined predictions. We hope that our present calculations would be tested by the future BESIII or Belle II experiments.

\begin{acknowledgments}\label{sec:acknowledgements}
We thank Xiao-Hai Liu at Tianjin University for some helpful discussion. Particular thanks are devoted to Jun-Zhang Wang at Peking University for some helpful information about the work \cite{Wang:2022jxj}. 
This work is partly supported by the National Natural Science Foundation of China under Grants No. 12475081, No. 12105153, and No. 12405093, and by the Natural Science Foundation of Shandong Province under Grants No. ZR2025MS04, No. ZR2021MA082, and No.ZR2022ZD26. It is also supported by Taishan Scholar Project of Shandong Province (Grant No.tsqn202103062) and partly by National Key Research and Development Program under Grant No. 2024YFA1610504.	
\end{acknowledgments}
		
%\onecolumngrid
\appendix
\section{Estimate of Coupling Constants}\label{app:couplings}
In this appendix, we perform the estimations of the coupling constants for the $Y(4500)$. As noted, we treat the $Y(4500)$ as the mixed state of the $\psi(5S)$ and $\psi(4D)$. According to the mixing scheme \cite{Wang:2019mhs}, 
%\begin{subequations}
%	\begin{align}
%		\tilde{\psi}(4500) &= -\tilde{\psi}(5S) \sin\theta  + \tilde{\psi}(4\,{}^3D_1) \cos\theta\,,\\
%		\tilde{\psi}(4415) &= \tilde{\psi}(5S) \cos\theta  + \tilde{\psi}(4\,{}^3D_1) \sin\theta\,,
%	\end{align}
%\end{subequations}
\begin{equation}\label{eq:matrix}
	\left( 
	\begin{array}{c}
		\tilde{\psi}(4415)  \\
		\tilde{Y}(4500) 
		\end{array}
	\right) 
	=\left( 
	\begin{array}{cc}
		\cos\theta& \sin\theta  \\
		-\sin\theta & \cos\theta
		\end{array}
		\right)
		\left( 
		\begin{array}{c}
		\tilde{\psi}(5S) \\
		\tilde{\psi}(4\,{}^3D_1)  
		\end{array}
		\right)
\end{equation}

To reasonably estimate the couplings, we could fit the partial width of the $Y(4500)$ and $\psi(4415)$ decaying into the open charmed meson pairs by the Lagrangians in Eq. \eqref{eq:LagSD}. Using the theoretical widths predicted in Ref. \cite{Wang:2019mhs}, we give
\begin{subequations}\label{eq:gpsis}
	\begin{align}
			&\left| g_{\psi\mathcal{D}\mathcal{D}}\sin\theta+g_{\psi_{1}\mathcal{D}\mathcal{D}}\cos\theta\right|  = 0.437\,,\\
		&\left| g_{\psi\mathcal{D}\mathcal{D}}\cos\theta-g_{\psi_{1}\mathcal{D}\mathcal{D}}\sin\theta\right|  = 0.678\,,\\
			&\left| g_{\psi\mathcal{D}\mathcal{D}^\ast}\sin\theta+g_{\psi_{1}\mathcal{D}\mathcal{D}^\ast}\cos\theta\right| =7.63\times 10^{-2}\,,\\
		&\left| g_{\psi\mathcal{D}\mathcal{D}^\ast}\cos\theta-g_{\psi_{1}\mathcal{D}\mathcal{D}^\ast}\sin\theta\right|  = 3.13\times 10^{-3}\,,\\
			&0.162 g_{\psi\mathcal{D}^\ast\mathcal{D}^\ast}^2\sin^2\theta-0.382g_{\psi\mathcal{D}^\ast\mathcal{D}^\ast}g_{\psi_{1}\mathcal{D}^\ast\mathcal{D}^\ast}\sin\theta\cos\theta\nonumber\\
			& + g_{\psi_{1}\mathcal{D}^\ast\mathcal{D}^\ast}^2\cos^2\theta = 2.38 \times 10^{-2}\,,\\
	&	g_{\psi\mathcal{D}^\ast\mathcal{D}^\ast}^2\cos^2\theta+2.44g_{\psi\mathcal{D}^\ast\mathcal{D}^\ast}g_{\psi_{1}\mathcal{D}^\ast\mathcal{D}^\ast}\sin\theta\cos\theta \nonumber\\
	&+6.06 g_{\psi_{1}\mathcal{D}^\ast\mathcal{D}^\ast}^2\sin^2\theta = 0.251\,.
	\end{align}
\end{subequations}
It is expected that the couplings $g_{\psi\mathcal{D}^{(\ast)}\mathcal{D}^{(\ast)}} $ for the $\psi(5S)$ with the open charmed mesons are subject to the relations in Eq. \eqref{eq:gJs}. Moreover, we assume the couplings $g_{\psi\mathcal{D}^{(\ast)}\mathcal{D}^{(\ast)}} $ to be positive. Hence, solving Eq.\eqref{eq:gpsis} with $\theta = 30^\circ$ gives 
\begin{subequations}
	\begin{align}
		g_{\psi\mathcal{D}\mathcal{D}} &= 0.37\\
		g_{\psi\mathcal{D}\mathcal{D}^\ast} &= 4.10\times 10^{-2}~\mathrm{GeV^{-1}}\\
		g_{\psi\mathcal{D}^\ast\mathcal{D}^\ast} &= 0.38\\
		g_{\psi_1\mathcal{D}\mathcal{D}} &= -0.72\\
		g_{\psi_1\mathcal{D}\mathcal{D}^\ast} &= 6.50 \times 10^{-2}~\mathrm{GeV^{-1}}\\
		g_{\psi_1\mathcal{D}^\ast\mathcal{D}^\ast} &= 0.20
	\end{align}
\end{subequations}
%\twocolumngrid
	% References using bib
	\bibliography{particlePhys.bib}

\begin{thebibliography}{100}%
\makeatletter
\providecommand \@ifxundefined [1]{%
 \@ifx{#1\undefined}
}%
\providecommand \@ifnum [1]{%
 \ifnum #1\expandafter \@firstoftwo
 \else \expandafter \@secondoftwo
 \fi
}%
\providecommand \@ifx [1]{%
 \ifx #1\expandafter \@firstoftwo
 \else \expandafter \@secondoftwo
 \fi
}%
\providecommand \natexlab [1]{#1}%
\providecommand \enquote  [1]{``#1''}%
\providecommand \bibnamefont  [1]{#1}%
\providecommand \bibfnamefont [1]{#1}%
\providecommand \citenamefont [1]{#1}%
\providecommand \href@noop [0]{\@secondoftwo}%
\providecommand \href [0]{\begingroup \@sanitize@url \@href}%
\providecommand \@href[1]{\@@startlink{#1}\@@href}%
\providecommand \@@href[1]{\endgroup#1\@@endlink}%
\providecommand \@sanitize@url [0]{\catcode `\\12\catcode `\$12\catcode
  `\&12\catcode `\#12\catcode `\^12\catcode `\_12\catcode `\%12\relax}%
\providecommand \@@startlink[1]{}%
\providecommand \@@endlink[0]{}%
\providecommand \url  [0]{\begingroup\@sanitize@url \@url }%
\providecommand \@url [1]{\endgroup\@href {#1}{\urlprefix }}%
\providecommand \urlprefix  [0]{URL }%
\providecommand \Eprint [0]{\href }%
\providecommand \doibase [0]{https://doi.org/}%
\providecommand \selectlanguage [0]{\@gobble}%
\providecommand \bibinfo  [0]{\@secondoftwo}%
\providecommand \bibfield  [0]{\@secondoftwo}%
\providecommand \translation [1]{[#1]}%
\providecommand \BibitemOpen [0]{}%
\providecommand \bibitemStop [0]{}%
\providecommand \bibitemNoStop [0]{.\EOS\space}%
\providecommand \EOS [0]{\spacefactor3000\relax}%
\providecommand \BibitemShut  [1]{\csname bibitem#1\endcsname}%
\let\auto@bib@innerbib\@empty
%</preamble>
\bibitem [{\citenamefont {Liu}\ and\ \citenamefont
  {Mitchell}(2023)}]{Liu:2023hhl}%
  \BibitemOpen
  \bibfield  {author} {\bibinfo {author} {\bibfnamefont {Z.}~\bibnamefont
  {Liu}}\ and\ \bibinfo {author} {\bibfnamefont {R.~E.}\ \bibnamefont
  {Mitchell}},\ }\bibfield  {title} {\bibinfo {title} {{New hadrons discovered
  at BESIII}},\ }\href {https://doi.org/10.1016/j.scib.2023.08.025} {\bibfield
  {journal} {\bibinfo  {journal} {Sci. Bull.}\ }\textbf {\bibinfo {volume}
  {68}},\ \bibinfo {pages} {2148} (\bibinfo {year} {2023})},\ \Eprint
  {https://arxiv.org/abs/2310.09465} {arXiv:2310.09465 [hep-ex]} \BibitemShut
  {NoStop}%
\bibitem [{\citenamefont {Gershon}(2022)}]{Gershon:2022xnn}%
  \BibitemOpen
  \bibfield  {author} {\bibinfo {author} {\bibfnamefont {T.}~\bibnamefont
  {Gershon}} (\bibinfo {collaboration} {LHCb}),\ }\href
  {https://doi.org/10.17181/CERN.7XZO.HPH7} {\bibinfo {title} {{Exotic hadron
  naming convention}}} (\bibinfo {year} {2022}),\ \Eprint
  {https://arxiv.org/abs/2206.15233} {arXiv:2206.15233 [hep-ex]} \BibitemShut
  {NoStop}%
\bibitem [{\citenamefont {Navas}\ \emph {et~al.}(2024)\citenamefont {Navas}
  \emph {et~al.}}]{ParticleDataGroup:2024cfk}%
  \BibitemOpen
  \bibfield  {author} {\bibinfo {author} {\bibfnamefont {S.}~\bibnamefont
  {Navas}} \emph {et~al.} (\bibinfo {collaboration} {Particle Data Group}),\
  }\bibfield  {title} {\bibinfo {title} {{Review of particle physics}},\ }\href
  {https://doi.org/10.1103/PhysRevD.110.030001} {\bibfield  {journal} {\bibinfo
   {journal} {Phys. Rev. D}\ }\textbf {\bibinfo {volume} {110}},\ \bibinfo
  {pages} {030001} (\bibinfo {year} {2024})}\BibitemShut {NoStop}%
\bibitem [{\citenamefont {Mo}\ \emph {et~al.}(2006)\citenamefont {Mo},
  \citenamefont {Li}, \citenamefont {Yuan}, \citenamefont {He}, \citenamefont
  {Hu}, \citenamefont {Hu}, \citenamefont {Wang},\ and\ \citenamefont
  {Wang}}]{Mo:2006ss}%
  \BibitemOpen
  \bibfield  {author} {\bibinfo {author} {\bibfnamefont {X.~H.}\ \bibnamefont
  {Mo}}, \bibinfo {author} {\bibfnamefont {G.}~\bibnamefont {Li}}, \bibinfo
  {author} {\bibfnamefont {C.~Z.}\ \bibnamefont {Yuan}}, \bibinfo {author}
  {\bibfnamefont {K.~L.}\ \bibnamefont {He}}, \bibinfo {author} {\bibfnamefont
  {H.~M.}\ \bibnamefont {Hu}}, \bibinfo {author} {\bibfnamefont {J.~H.}\
  \bibnamefont {Hu}}, \bibinfo {author} {\bibfnamefont {P.}~\bibnamefont
  {Wang}},\ and\ \bibinfo {author} {\bibfnamefont {Z.~Y.}\ \bibnamefont
  {Wang}},\ }\bibfield  {title} {\bibinfo {title} {{Determining the upper limit
  of $\Gamma_{ee}$ for the $Y(4260)$}},\ }\href
  {https://doi.org/10.1016/j.physletb.2006.07.060} {\bibfield  {journal}
  {\bibinfo  {journal} {Phys. Lett. B}\ }\textbf {\bibinfo {volume} {640}},\
  \bibinfo {pages} {182} (\bibinfo {year} {2006})},\ \Eprint
  {https://arxiv.org/abs/hep-ex/0603024} {arXiv:hep-ex/0603024} \BibitemShut
  {NoStop}%
\bibitem [{\citenamefont {Wang}\ \emph {et~al.}(2019)\citenamefont {Wang},
  \citenamefont {Chen}, \citenamefont {Liu},\ and\ \citenamefont
  {Matsuki}}]{Wang:2019mhs}%
  \BibitemOpen
  \bibfield  {author} {\bibinfo {author} {\bibfnamefont {J.-Z.}\ \bibnamefont
  {Wang}}, \bibinfo {author} {\bibfnamefont {D.-Y.}\ \bibnamefont {Chen}},
  \bibinfo {author} {\bibfnamefont {X.}~\bibnamefont {Liu}},\ and\ \bibinfo
  {author} {\bibfnamefont {T.}~\bibnamefont {Matsuki}},\ }\bibfield  {title}
  {\bibinfo {title} {Constructing {$J/\ensuremath{\psi}$} family with updated
  data of charmoniumlike {$Y$} states},\ }\href
  {https://doi.org/10.1103/PhysRevD.99.114003} {\bibfield  {journal} {\bibinfo
  {journal} {Phys. Rev. D}\ }\textbf {\bibinfo {volume} {99}},\ \bibinfo
  {pages} {114003} (\bibinfo {year} {2019})},\ \Eprint
  {https://arxiv.org/abs/1903.07115} {arXiv:1903.07115 [hep-ph]} \BibitemShut
  {NoStop}%
\bibitem [{\citenamefont {Wang}\ \emph {et~al.}(2018)\citenamefont {Wang},
  \citenamefont {Sun}, \citenamefont {Liu},\ and\ \citenamefont
  {Matsuki}}]{Wang:2018rjg}%
  \BibitemOpen
  \bibfield  {author} {\bibinfo {author} {\bibfnamefont {J.-Z.}\ \bibnamefont
  {Wang}}, \bibinfo {author} {\bibfnamefont {Z.-F.}\ \bibnamefont {Sun}},
  \bibinfo {author} {\bibfnamefont {X.}~\bibnamefont {Liu}},\ and\ \bibinfo
  {author} {\bibfnamefont {T.}~\bibnamefont {Matsuki}},\ }\bibfield  {title}
  {\bibinfo {title} {Higher bottomonium zoo},\ }\href
  {https://doi.org/10.1140/epjc/s10052-018-6372-1} {\bibfield  {journal}
  {\bibinfo  {journal} {Eur. Phys. J. C}\ }\textbf {\bibinfo {volume} {78}},\
  \bibinfo {pages} {915} (\bibinfo {year} {2018})},\ \Eprint
  {https://arxiv.org/abs/1802.04938} {arXiv:1802.04938 [hep-ph]} \BibitemShut
  {NoStop}%
\bibitem [{\citenamefont {Ablikim}\ \emph {et~al.}(2017)\citenamefont {Ablikim}
  \emph {et~al.}}]{BESIII:2016bnd}%
  \BibitemOpen
  \bibfield  {author} {\bibinfo {author} {\bibfnamefont {M.}~\bibnamefont
  {Ablikim}} \emph {et~al.} (\bibinfo {collaboration} {BESIII}),\ }\bibfield
  {title} {\bibinfo {title} {{Precise measurement of the $e^+e^-\to
  \pi^+\pi^-J/\psi$ cross section at center-of-mass energies from 3.77 to 4.60
  GeV}},\ }\href {https://doi.org/10.1103/PhysRevLett.118.092001} {\bibfield
  {journal} {\bibinfo  {journal} {Phys. Rev. Lett.}\ }\textbf {\bibinfo
  {volume} {118}},\ \bibinfo {pages} {092001} (\bibinfo {year} {2017})},\
  \Eprint {https://arxiv.org/abs/1611.01317} {arXiv:1611.01317 [hep-ex]}
  \BibitemShut {NoStop}%
\bibitem [{\citenamefont {Chen}\ \emph {et~al.}(2025)\citenamefont {Chen},
  \citenamefont {Chen}, \citenamefont {Guo}, \citenamefont {Ma}, \citenamefont
  {Shen}, \citenamefont {Shou}, \citenamefont {Shou}, \citenamefont {Wang},
  \citenamefont {Wu},\ and\ \citenamefont {Zou}}]{Chen:2024eaq}%
  \BibitemOpen
  \bibfield  {author} {\bibinfo {author} {\bibfnamefont {J.-H.}\ \bibnamefont
  {Chen}}, \bibinfo {author} {\bibfnamefont {J.}~\bibnamefont {Chen}}, \bibinfo
  {author} {\bibfnamefont {F.-K.}\ \bibnamefont {Guo}}, \bibinfo {author}
  {\bibfnamefont {Y.-G.}\ \bibnamefont {Ma}}, \bibinfo {author} {\bibfnamefont
  {C.-P.}\ \bibnamefont {Shen}}, \bibinfo {author} {\bibfnamefont {Q.-Y.}\
  \bibnamefont {Shou}}, \bibinfo {author} {\bibfnamefont {Q.}~\bibnamefont
  {Shou}}, \bibinfo {author} {\bibfnamefont {Q.}~\bibnamefont {Wang}}, \bibinfo
  {author} {\bibfnamefont {J.-J.}\ \bibnamefont {Wu}},\ and\ \bibinfo {author}
  {\bibfnamefont {B.-S.}\ \bibnamefont {Zou}},\ }\bibfield  {title} {\bibinfo
  {title} {{Production of exotic hadrons in pp and nuclear collisions}},\
  }\href {https://doi.org/10.1007/s41365-025-01664-w} {\bibfield  {journal}
  {\bibinfo  {journal} {Nucl. Sci. Tech.}\ }\textbf {\bibinfo {volume} {36}},\
  \bibinfo {pages} {55} (\bibinfo {year} {2025})},\ \Eprint
  {https://arxiv.org/abs/2411.18257} {arXiv:2411.18257 [hep-ph]} \BibitemShut
  {NoStop}%
\bibitem [{\citenamefont {Liu}\ \emph {et~al.}(2025)\citenamefont {Liu},
  \citenamefont {Pan}, \citenamefont {Liu}, \citenamefont {Wu}, \citenamefont
  {Lu},\ and\ \citenamefont {Geng}}]{Liu:2024uxn}%
  \BibitemOpen
  \bibfield  {author} {\bibinfo {author} {\bibfnamefont {M.-Z.}\ \bibnamefont
  {Liu}}, \bibinfo {author} {\bibfnamefont {Y.-W.}\ \bibnamefont {Pan}},
  \bibinfo {author} {\bibfnamefont {Z.-W.}\ \bibnamefont {Liu}}, \bibinfo
  {author} {\bibfnamefont {T.-W.}\ \bibnamefont {Wu}}, \bibinfo {author}
  {\bibfnamefont {J.-X.}\ \bibnamefont {Lu}},\ and\ \bibinfo {author}
  {\bibfnamefont {L.-S.}\ \bibnamefont {Geng}},\ }\bibfield  {title} {\bibinfo
  {title} {{Three ways to decipher the nature of exotic hadrons: Multiplets,
  three-body hadronic molecules, and correlation functions}},\ }\href
  {https://doi.org/10.1016/j.physrep.2024.12.001} {\bibfield  {journal}
  {\bibinfo  {journal} {Phys. Rept.}\ }\textbf {\bibinfo {volume} {1108}},\
  \bibinfo {pages} {1} (\bibinfo {year} {2025})},\ \Eprint
  {https://arxiv.org/abs/2404.06399} {arXiv:2404.06399 [hep-ph]} \BibitemShut
  {NoStop}%
\bibitem [{\citenamefont {Chen}\ \emph
  {et~al.}(2016{\natexlab{a}})\citenamefont {Chen}, \citenamefont {Chen},
  \citenamefont {Liu},\ and\ \citenamefont {Zhu}}]{Chen:2016qju}%
  \BibitemOpen
  \bibfield  {author} {\bibinfo {author} {\bibfnamefont {H.-X.}\ \bibnamefont
  {Chen}}, \bibinfo {author} {\bibfnamefont {W.}~\bibnamefont {Chen}}, \bibinfo
  {author} {\bibfnamefont {X.}~\bibnamefont {Liu}},\ and\ \bibinfo {author}
  {\bibfnamefont {S.-L.}\ \bibnamefont {Zhu}},\ }\bibfield  {title} {\bibinfo
  {title} {The hidden-charm pentaquark and tetraquark states},\ }\href
  {https://doi.org/10.1016/j.physrep.2016.05.004} {\bibfield  {journal}
  {\bibinfo  {journal} {Phys. Rept.}\ }\textbf {\bibinfo {volume} {639}},\
  \bibinfo {pages} {1} (\bibinfo {year} {2016}{\natexlab{a}})},\ \Eprint
  {https://arxiv.org/abs/1601.02092} {arXiv:1601.02092 [hep-ph]} \BibitemShut
  {NoStop}%
\bibitem [{\citenamefont {Chen}\ \emph {et~al.}(2022)\citenamefont {Chen},
  \citenamefont {Chen}, \citenamefont {Liu}, \citenamefont {Liu},\ and\
  \citenamefont {Zhu}}]{Chen:2022asf}%
  \BibitemOpen
  \bibfield  {author} {\bibinfo {author} {\bibfnamefont {H.-X.}\ \bibnamefont
  {Chen}}, \bibinfo {author} {\bibfnamefont {W.}~\bibnamefont {Chen}}, \bibinfo
  {author} {\bibfnamefont {X.}~\bibnamefont {Liu}}, \bibinfo {author}
  {\bibfnamefont {Y.-R.}\ \bibnamefont {Liu}},\ and\ \bibinfo {author}
  {\bibfnamefont {S.-L.}\ \bibnamefont {Zhu}},\ }\bibfield  {title} {\bibinfo
  {title} {An updated review of the new hadron states},\ }\href
  {https://doi.org/10.1088/1361-6633/aca3b6} {\bibfield  {journal} {\bibinfo
  {journal} {Rept. Prog. Phys.}\ }\textbf {\bibinfo {volume} {86}},\ \bibinfo
  {pages} {026201} (\bibinfo {year} {2022})},\ \Eprint
  {https://arxiv.org/abs/2204.02649} {arXiv:2204.02649 [hep-ph]} \BibitemShut
  {NoStop}%
\bibitem [{\citenamefont {Meng}\ \emph {et~al.}(2023)\citenamefont {Meng},
  \citenamefont {Wang}, \citenamefont {Wang},\ and\ \citenamefont
  {Zhu}}]{Meng:2022ozq}%
  \BibitemOpen
  \bibfield  {author} {\bibinfo {author} {\bibfnamefont {L.}~\bibnamefont
  {Meng}}, \bibinfo {author} {\bibfnamefont {B.}~\bibnamefont {Wang}}, \bibinfo
  {author} {\bibfnamefont {G.-J.}\ \bibnamefont {Wang}},\ and\ \bibinfo
  {author} {\bibfnamefont {S.-L.}\ \bibnamefont {Zhu}},\ }\bibfield  {title}
  {\bibinfo {title} {Chiral perturbation theory for heavy hadrons and chiral
  effective field theory for heavy hadronic molecules},\ }\href
  {https://doi.org/10.1016/j.physrep.2023.04.003} {\bibfield  {journal}
  {\bibinfo  {journal} {Phys. Rept.}\ }\textbf {\bibinfo {volume} {1019}},\
  \bibinfo {pages} {1} (\bibinfo {year} {2023})},\ \Eprint
  {https://arxiv.org/abs/2204.08716} {arXiv:2204.08716 [hep-ph]} \BibitemShut
  {NoStop}%
\bibitem [{\citenamefont {Ablikim}\ \emph {et~al.}(2020)\citenamefont {Ablikim}
  \emph {et~al.}}]{BESIII:2020nme}%
  \BibitemOpen
  \bibfield  {author} {\bibinfo {author} {\bibfnamefont {M.}~\bibnamefont
  {Ablikim}} \emph {et~al.} (\bibinfo {collaboration} {BESIII}),\ }\bibfield
  {title} {\bibinfo {title} {{Future Physics Programme of BESIII}},\ }\href
  {https://doi.org/10.1088/1674-1137/44/4/040001} {\bibfield  {journal}
  {\bibinfo  {journal} {Chin. Phys. C}\ }\textbf {\bibinfo {volume} {44}},\
  \bibinfo {pages} {040001} (\bibinfo {year} {2020})},\ \Eprint
  {https://arxiv.org/abs/1912.05983} {arXiv:1912.05983 [hep-ex]} \BibitemShut
  {NoStop}%
\bibitem [{\citenamefont {Guo}\ \emph {et~al.}(2020)\citenamefont {Guo},
  \citenamefont {Liu},\ and\ \citenamefont {Sakai}}]{Guo:2019twa}%
  \BibitemOpen
  \bibfield  {author} {\bibinfo {author} {\bibfnamefont {F.-K.}\ \bibnamefont
  {Guo}}, \bibinfo {author} {\bibfnamefont {X.-H.}\ \bibnamefont {Liu}},\ and\
  \bibinfo {author} {\bibfnamefont {S.}~\bibnamefont {Sakai}},\ }\bibfield
  {title} {\bibinfo {title} {Threshold cusps and triangle singularities in
  hadronic reactions},\ }\href {https://doi.org/10.1016/j.ppnp.2020.103757}
  {\bibfield  {journal} {\bibinfo  {journal} {Prog. Part. Nucl. Phys.}\
  }\textbf {\bibinfo {volume} {112}},\ \bibinfo {pages} {103757} (\bibinfo
  {year} {2020})},\ \Eprint {https://arxiv.org/abs/1912.07030}
  {arXiv:1912.07030 [hep-ph]} \BibitemShut {NoStop}%
\bibitem [{\citenamefont {Brambilla}\ \emph {et~al.}(2020)\citenamefont
  {Brambilla}, \citenamefont {Eidelman}, \citenamefont {Hanhart}, \citenamefont
  {Nefediev}, \citenamefont {Shen}, \citenamefont {Thomas}, \citenamefont
  {Vairo},\ and\ \citenamefont {Yuan}}]{Brambilla:2019esw}%
  \BibitemOpen
  \bibfield  {author} {\bibinfo {author} {\bibfnamefont {N.}~\bibnamefont
  {Brambilla}}, \bibinfo {author} {\bibfnamefont {S.}~\bibnamefont {Eidelman}},
  \bibinfo {author} {\bibfnamefont {C.}~\bibnamefont {Hanhart}}, \bibinfo
  {author} {\bibfnamefont {A.}~\bibnamefont {Nefediev}}, \bibinfo {author}
  {\bibfnamefont {C.-P.}\ \bibnamefont {Shen}}, \bibinfo {author}
  {\bibfnamefont {C.~E.}\ \bibnamefont {Thomas}}, \bibinfo {author}
  {\bibfnamefont {A.}~\bibnamefont {Vairo}},\ and\ \bibinfo {author}
  {\bibfnamefont {C.-Z.}\ \bibnamefont {Yuan}},\ }\bibfield  {title} {\bibinfo
  {title} {The xyz states: Experimental and theoretical status and
  perspectives},\ }\href {https://doi.org/10.1016/j.physrep.2020.05.001}
  {\bibfield  {journal} {\bibinfo  {journal} {Phys. Rept.}\ }\textbf {\bibinfo
  {volume} {873}},\ \bibinfo {pages} {1} (\bibinfo {year} {2020})},\ \Eprint
  {https://arxiv.org/abs/1907.07583} {arXiv:1907.07583 [hep-ex]} \BibitemShut
  {NoStop}%
\bibitem [{\citenamefont {Liu}\ \emph {et~al.}(2019)\citenamefont {Liu},
  \citenamefont {Chen}, \citenamefont {Chen}, \citenamefont {Liu},\ and\
  \citenamefont {Zhu}}]{Liu:2019zoy}%
  \BibitemOpen
  \bibfield  {author} {\bibinfo {author} {\bibfnamefont {Y.-R.}\ \bibnamefont
  {Liu}}, \bibinfo {author} {\bibfnamefont {H.-X.}\ \bibnamefont {Chen}},
  \bibinfo {author} {\bibfnamefont {W.}~\bibnamefont {Chen}}, \bibinfo {author}
  {\bibfnamefont {X.}~\bibnamefont {Liu}},\ and\ \bibinfo {author}
  {\bibfnamefont {S.-L.}\ \bibnamefont {Zhu}},\ }\bibfield  {title} {\bibinfo
  {title} {Pentaquark and tetraquark states},\ }\href
  {https://doi.org/10.1016/j.ppnp.2019.04.003} {\bibfield  {journal} {\bibinfo
  {journal} {Prog. Part. Nucl. Phys.}\ }\textbf {\bibinfo {volume} {107}},\
  \bibinfo {pages} {237} (\bibinfo {year} {2019})},\ \Eprint
  {https://arxiv.org/abs/1903.11976} {arXiv:1903.11976 [hep-ph]} \BibitemShut
  {NoStop}%
\bibitem [{\citenamefont {Guo}\ \emph {et~al.}(2018)\citenamefont {Guo},
  \citenamefont {Hanhart}, \citenamefont {Meißner}, \citenamefont {Wang},
  \citenamefont {Zhao},\ and\ \citenamefont {Zou}}]{Guo:2017jvc}%
  \BibitemOpen
  \bibfield  {author} {\bibinfo {author} {\bibfnamefont {F.-K.}\ \bibnamefont
  {Guo}}, \bibinfo {author} {\bibfnamefont {C.}~\bibnamefont {Hanhart}},
  \bibinfo {author} {\bibfnamefont {U.-G.}\ \bibnamefont {Meißner}}, \bibinfo
  {author} {\bibfnamefont {Q.}~\bibnamefont {Wang}}, \bibinfo {author}
  {\bibfnamefont {Q.}~\bibnamefont {Zhao}},\ and\ \bibinfo {author}
  {\bibfnamefont {B.-S.}\ \bibnamefont {Zou}},\ }\bibfield  {title} {\bibinfo
  {title} {Hadronic molecules},\ }\href
  {https://doi.org/10.1103/RevModPhys.90.015004} {\bibfield  {journal}
  {\bibinfo  {journal} {Rev. Mod. Phys.}\ }\textbf {\bibinfo {volume} {90}},\
  \bibinfo {pages} {015004} (\bibinfo {year} {2018})},\ \Eprint
  {https://arxiv.org/abs/1705.00141} {arXiv:1705.00141 [hep-ph]} \BibitemShut
  {NoStop}%
\bibitem [{\citenamefont {Liu}(2014{\natexlab{a}})}]{Liu:2013waa}%
  \BibitemOpen
  \bibfield  {author} {\bibinfo {author} {\bibfnamefont {X.}~\bibnamefont
  {Liu}},\ }\bibfield  {title} {\bibinfo {title} {{An overview of $XYZ$ new
  particles}},\ }\href {https://doi.org/10.1007/s11434-014-0407-2} {\bibfield
  {journal} {\bibinfo  {journal} {Chin. Sci. Bull.}\ }\textbf {\bibinfo
  {volume} {59}},\ \bibinfo {pages} {3815} (\bibinfo {year}
  {2014}{\natexlab{a}})},\ \Eprint {https://arxiv.org/abs/1312.7408}
  {arXiv:1312.7408 [hep-ph]} \BibitemShut {NoStop}%
\bibitem [{\citenamefont {Wang}(2025)}]{Wang:2025sic}%
  \BibitemOpen
  \bibfield  {author} {\bibinfo {author} {\bibfnamefont {Z.-G.}\ \bibnamefont
  {Wang}},\ }\href@noop {} {\bibinfo {title} {{Review of the QCD sum rules for
  exotic states}}} (\bibinfo {year} {2025}),\ \Eprint
  {https://arxiv.org/abs/2502.11351} {arXiv:2502.11351 [hep-ph]} \BibitemShut
  {NoStop}%
\bibitem [{\citenamefont {Aubert}\ \emph {et~al.}(2005)\citenamefont {Aubert}
  \emph {et~al.}}]{BaBar:2005hhc}%
  \BibitemOpen
  \bibfield  {author} {\bibinfo {author} {\bibfnamefont {B.}~\bibnamefont
  {Aubert}} \emph {et~al.} (\bibinfo {collaboration} {BaBar}),\ }\bibfield
  {title} {\bibinfo {title} {{Observation of a broad structure in the $\pi^+
  \pi^- J/\psi$ mass spectrum around 4.26-GeV/c$^2$}},\ }\href
  {https://doi.org/10.1103/PhysRevLett.95.142001} {\bibfield  {journal}
  {\bibinfo  {journal} {Phys. Rev. Lett.}\ }\textbf {\bibinfo {volume} {95}},\
  \bibinfo {pages} {142001} (\bibinfo {year} {2005})},\ \Eprint
  {https://arxiv.org/abs/hep-ex/0506081} {arXiv:hep-ex/0506081} \BibitemShut
  {NoStop}%
\bibitem [{\citenamefont {Coan}\ \emph {et~al.}(2006)\citenamefont {Coan},
  \citenamefont {Gao}, \citenamefont {Liu} \emph {et~al.}}]{CLEO:2006ike}%
  \BibitemOpen
  \bibfield  {author} {\bibinfo {author} {\bibfnamefont {T.}~\bibnamefont
  {Coan}}, \bibinfo {author} {\bibfnamefont {Y.}~\bibnamefont {Gao}}, \bibinfo
  {author} {\bibfnamefont {F.}~\bibnamefont {Liu}}, \emph {et~al.} (\bibinfo
  {collaboration} {CLEO}),\ }\bibfield  {title} {\bibinfo {title} {{Charmonium
  Decays of $Y(4260)$, $\ensuremath{\psi}(4160)$, and
  $\ensuremath{\psi}(4040)$}},\ }\href
  {https://doi.org/10.1103/PhysRevLett.96.162003} {\bibfield  {journal}
  {\bibinfo  {journal} {Phys. Rev. Lett.}\ }\textbf {\bibinfo {volume} {96}},\
  \bibinfo {pages} {162003} (\bibinfo {year} {2006})},\ \Eprint
  {https://arxiv.org/abs/hep-ex/0602034} {arXiv:hep-ex/0602034} \BibitemShut
  {NoStop}%
\bibitem [{\citenamefont {Yuan}\ \emph {et~al.}(2007)\citenamefont {Yuan} \emph
  {et~al.}}]{Belle:2007dxy}%
  \BibitemOpen
  \bibfield  {author} {\bibinfo {author} {\bibfnamefont {C.~Z.}\ \bibnamefont
  {Yuan}} \emph {et~al.} (\bibinfo {collaboration} {Belle}),\ }\bibfield
  {title} {\bibinfo {title} {{Measurement of $e^+e^-\to \pi^+\pi^- J/\psi$
  cross-section via initial state radiation at Belle}},\ }\href
  {https://doi.org/10.1103/PhysRevLett.99.182004} {\bibfield  {journal}
  {\bibinfo  {journal} {Phys. Rev. Lett.}\ }\textbf {\bibinfo {volume} {99}},\
  \bibinfo {pages} {182004} (\bibinfo {year} {2007})},\ \Eprint
  {https://arxiv.org/abs/0707.2541} {arXiv:0707.2541 [hep-ex]} \BibitemShut
  {NoStop}%
\bibitem [{\citenamefont {Ablikim}\ \emph
  {et~al.}(2022{\natexlab{a}})\citenamefont {Ablikim} \emph
  {et~al.}}]{BESIII:2022joj}%
  \BibitemOpen
  \bibfield  {author} {\bibinfo {author} {\bibfnamefont {M.}~\bibnamefont
  {Ablikim}} \emph {et~al.} (\bibinfo {collaboration} {BESIII}),\ }\bibfield
  {title} {\bibinfo {title} {{Observation of the $Y(4230)$ and a new structure
  in $e^+ e^-\to K^+ K^- J/\psi$}},\ }\href
  {https://doi.org/10.1088/1674-1137/ac945c} {\bibfield  {journal} {\bibinfo
  {journal} {Chin. Phys. C}\ }\textbf {\bibinfo {volume} {46}},\ \bibinfo
  {pages} {111002} (\bibinfo {year} {2022}{\natexlab{a}})},\ \Eprint
  {https://arxiv.org/abs/2204.07800} {arXiv:2204.07800 [hep-ex]} \BibitemShut
  {NoStop}%
\bibitem [{\citenamefont {Ablikim}\ \emph
  {et~al.}(2022{\natexlab{b}})\citenamefont {Ablikim}, \citenamefont {Achasov},
  \citenamefont {Adlarson} \emph {et~al.}}]{BESIII:2022qal}%
  \BibitemOpen
  \bibfield  {author} {\bibinfo {author} {\bibfnamefont {M.}~\bibnamefont
  {Ablikim}}, \bibinfo {author} {\bibfnamefont {M.~N.}\ \bibnamefont
  {Achasov}}, \bibinfo {author} {\bibfnamefont {P.~A.}\ \bibnamefont
  {Adlarson}}, \emph {et~al.} (\bibinfo {collaboration} {BESIII}),\ }\bibfield
  {title} {\bibinfo {title} {{Study of the resonance structures in $e^{+}e^{-}
  \rightarrow \pi^{+}\pi^{-}J/\psi$ process}},\ }\href
  {https://doi.org/10.1103/PhysRevD.106.072001} {\bibfield  {journal} {\bibinfo
   {journal} {Phys. Rev. D}\ }\textbf {\bibinfo {volume} {106}},\ \bibinfo
  {pages} {072001} (\bibinfo {year} {2022}{\natexlab{b}})},\ \Eprint
  {https://arxiv.org/abs/2206.08554} {arXiv:2206.08554 [hep-ex]} \BibitemShut
  {NoStop}%
\bibitem [{\citenamefont {Ablikim}\ \emph
  {et~al.}(2023{\natexlab{a}})\citenamefont {Ablikim}, \citenamefont {Achasov},
  \citenamefont {Adlarson} \emph {et~al.}}]{BESIII:2023cmv}%
  \BibitemOpen
  \bibfield  {author} {\bibinfo {author} {\bibfnamefont {M.}~\bibnamefont
  {Ablikim}}, \bibinfo {author} {\bibfnamefont {M.}~\bibnamefont {Achasov}},
  \bibinfo {author} {\bibfnamefont {P.}~\bibnamefont {Adlarson}}, \emph
  {et~al.} (\bibinfo {collaboration} {BESIII}),\ }\bibfield  {title} {\bibinfo
  {title} {{Observation of Three Charmoniumlike States with
  ${J}^{PC}={1}^{\ensuremath{-}\ensuremath{-}}$ in
  ${e}^{+}{e}^{\ensuremath{-}}\ensuremath{\rightarrow}{D}^{*0}{D}^{*\ensuremath{-}}{\ensuremath{\pi}}^{+}$}},\
  }\href {https://doi.org/10.1103/PhysRevLett.130.121901} {\bibfield  {journal}
  {\bibinfo  {journal} {Phys. Rev. Lett.}\ }\textbf {\bibinfo {volume} {130}},\
  \bibinfo {pages} {121901} (\bibinfo {year} {2023}{\natexlab{a}})},\ \Eprint
  {https://arxiv.org/abs/2301.07321} {arXiv:2301.07321 [hep-ex]} \BibitemShut
  {NoStop}%
\bibitem [{\citenamefont {Zhu}(2005)}]{Zhu:2005hp}%
  \BibitemOpen
  \bibfield  {author} {\bibinfo {author} {\bibfnamefont {S.-L.}\ \bibnamefont
  {Zhu}},\ }\bibfield  {title} {\bibinfo {title} {{The possible interpretations
  of $Y(4260)$}},\ }\href {https://doi.org/10.1016/j.physletb.2005.08.068}
  {\bibfield  {journal} {\bibinfo  {journal} {Phys. Lett. B}\ }\textbf
  {\bibinfo {volume} {625}},\ \bibinfo {pages} {212} (\bibinfo {year}
  {2005})},\ \Eprint {https://arxiv.org/abs/hep-ph/0507025}
  {arXiv:hep-ph/0507025} \BibitemShut {NoStop}%
\bibitem [{\citenamefont {Dubnička}\ \emph {et~al.}(2020)\citenamefont
  {Dubnička}, \citenamefont {Dubničková}, \citenamefont {Issadykov},
  \citenamefont {Ivanov},\ and\ \citenamefont {Liptaj}}]{Dubnicka:2020xoh}%
  \BibitemOpen
  \bibfield  {author} {\bibinfo {author} {\bibfnamefont {S.}~\bibnamefont
  {Dubnička}}, \bibinfo {author} {\bibfnamefont {A.}~\bibnamefont
  {Dubničková}}, \bibinfo {author} {\bibfnamefont {A.}~\bibnamefont
  {Issadykov}}, \bibinfo {author} {\bibfnamefont {M.}~\bibnamefont {Ivanov}},\
  and\ \bibinfo {author} {\bibfnamefont {A.}~\bibnamefont {Liptaj}},\
  }\bibfield  {title} {\bibinfo {title} {{$Y(4260)$ as a four-quark state}},\
  }\href {https://doi.org/10.1103/PhysRevD.101.094030} {\bibfield  {journal}
  {\bibinfo  {journal} {Phys. Rev. D}\ }\textbf {\bibinfo {volume} {101}},\
  \bibinfo {pages} {094030} (\bibinfo {year} {2020})},\ \Eprint
  {https://arxiv.org/abs/2003.04142} {arXiv:2003.04142 [hep-ph]} \BibitemShut
  {NoStop}%
\bibitem [{\citenamefont {Wang}(2019)}]{Wang:2018ejf}%
  \BibitemOpen
  \bibfield  {author} {\bibinfo {author} {\bibfnamefont {Z.-G.}\ \bibnamefont
  {Wang}},\ }\bibfield  {title} {\bibinfo {title} {{Analysis of the vector
  tetraquark states with P-waves between the diquarks and antidiquarks via the
  QCD sum rules}},\ }\href {https://doi.org/10.1140/epjc/s10052-019-6568-z}
  {\bibfield  {journal} {\bibinfo  {journal} {Eur. Phys. J. C}\ }\textbf
  {\bibinfo {volume} {79}},\ \bibinfo {pages} {29} (\bibinfo {year} {2019})},\
  \Eprint {https://arxiv.org/abs/1811.02726} {arXiv:1811.02726 [hep-ph]}
  \BibitemShut {NoStop}%
\bibitem [{\citenamefont {Wang}(2018)}]{Wang:2018ntv}%
  \BibitemOpen
  \bibfield  {author} {\bibinfo {author} {\bibfnamefont {Z.-G.}\ \bibnamefont
  {Wang}},\ }\bibfield  {title} {\bibinfo {title} {{Lowest vector tetraquark
  states: $Y(4260/4220)$ or $Z_c(4100)$}},\ }\href
  {https://doi.org/10.1140/epjc/s10052-018-6417-5} {\bibfield  {journal}
  {\bibinfo  {journal} {Eur. Phys. J. C}\ }\textbf {\bibinfo {volume} {78}},\
  \bibinfo {pages} {933} (\bibinfo {year} {2018})},\ \Eprint
  {https://arxiv.org/abs/1809.10299} {arXiv:1809.10299 [hep-ph]} \BibitemShut
  {NoStop}%
\bibitem [{\citenamefont {Chiu}\ and\ \citenamefont
  {Hsieh}(2006)}]{Chiu:2005ey}%
  \BibitemOpen
  \bibfield  {author} {\bibinfo {author} {\bibfnamefont {T.-W.}\ \bibnamefont
  {Chiu}}\ and\ \bibinfo {author} {\bibfnamefont {T.-H.}\ \bibnamefont {Hsieh}}
  (\bibinfo {collaboration} {TWQCD}),\ }\bibfield  {title} {\bibinfo {title}
  {{$Y(4260)$ on the lattice}},\ }\href
  {https://doi.org/10.1103/PhysRevD.73.094510} {\bibfield  {journal} {\bibinfo
  {journal} {Phys. Rev. D}\ }\textbf {\bibinfo {volume} {73}},\ \bibinfo
  {pages} {094510} (\bibinfo {year} {2006})},\ \Eprint
  {https://arxiv.org/abs/hep-lat/0512029} {arXiv:hep-lat/0512029} \BibitemShut
  {NoStop}%
\bibitem [{\citenamefont {Yuan}\ \emph {et~al.}(2006)\citenamefont {Yuan},
  \citenamefont {Wang},\ and\ \citenamefont {Mo}}]{Yuan:2005dr}%
  \BibitemOpen
  \bibfield  {author} {\bibinfo {author} {\bibfnamefont {C.~Z.}\ \bibnamefont
  {Yuan}}, \bibinfo {author} {\bibfnamefont {P.}~\bibnamefont {Wang}},\ and\
  \bibinfo {author} {\bibfnamefont {X.~H.}\ \bibnamefont {Mo}},\ }\bibfield
  {title} {\bibinfo {title} {{The $Y(4260)$ as an $\omega \chi_{c1}$ molecular
  state}},\ }\href {https://doi.org/10.1016/j.physletb.2006.01.031} {\bibfield
  {journal} {\bibinfo  {journal} {Phys. Lett. B}\ }\textbf {\bibinfo {volume}
  {634}},\ \bibinfo {pages} {399} (\bibinfo {year} {2006})},\ \Eprint
  {https://arxiv.org/abs/hep-ph/0511107} {arXiv:hep-ph/0511107} \BibitemShut
  {NoStop}%
\bibitem [{\citenamefont {Albuquerque}\ and\ \citenamefont
  {Nielsen}(2009)}]{Albuquerque:2008up}%
  \BibitemOpen
  \bibfield  {author} {\bibinfo {author} {\bibfnamefont {R.~M.}\ \bibnamefont
  {Albuquerque}}\ and\ \bibinfo {author} {\bibfnamefont {M.}~\bibnamefont
  {Nielsen}},\ }\bibfield  {title} {\bibinfo {title} {{QCD sum rules study of
  the $J^{PC}=1^{--}$ charmonium $Y$ mesons}},\ }\href
  {https://doi.org/10.1016/j.nuclphysa.2011.04.001} {\bibfield  {journal}
  {\bibinfo  {journal} {Nucl. Phys. A}\ }\textbf {\bibinfo {volume} {815}},\
  \bibinfo {pages} {53} (\bibinfo {year} {2009})},\ \Eprint
  {https://arxiv.org/abs/0804.4817} {arXiv:0804.4817 [hep-ph]} \BibitemShut
  {NoStop}%
\bibitem [{\citenamefont {Ding}(2009)}]{Ding:2008gr}%
  \BibitemOpen
  \bibfield  {author} {\bibinfo {author} {\bibfnamefont {G.-J.}\ \bibnamefont
  {Ding}},\ }\bibfield  {title} {\bibinfo {title} {{Are $Y(4260)$ and
  ${Z}_{2}^{+}(4250)$ ${D}_{1}D$ or ${D}_{0}{D}^{*}$ hadronic molecules?}},\
  }\href {https://doi.org/10.1103/PhysRevD.79.014001} {\bibfield  {journal}
  {\bibinfo  {journal} {Phys. Rev. D}\ }\textbf {\bibinfo {volume} {79}},\
  \bibinfo {pages} {014001} (\bibinfo {year} {2009})},\ \Eprint
  {https://arxiv.org/abs/0809.4818} {arXiv:0809.4818 [hep-ph]} \BibitemShut
  {NoStop}%
\bibitem [{\citenamefont {Close}\ \emph {et~al.}(2010)\citenamefont {Close},
  \citenamefont {Downum},\ and\ \citenamefont {Thomas}}]{Close:2010wq}%
  \BibitemOpen
  \bibfield  {author} {\bibinfo {author} {\bibfnamefont {F.}~\bibnamefont
  {Close}}, \bibinfo {author} {\bibfnamefont {C.}~\bibnamefont {Downum}},\ and\
  \bibinfo {author} {\bibfnamefont {C.~E.}\ \bibnamefont {Thomas}},\ }\bibfield
   {title} {\bibinfo {title} {{Novel Charmonium and Bottomonium Spectroscopies
  due to Deeply Bound Hadronic Molecules from Single Pion Exchange}},\ }\href
  {https://doi.org/10.1103/PhysRevD.81.074033} {\bibfield  {journal} {\bibinfo
  {journal} {Phys. Rev. D}\ }\textbf {\bibinfo {volume} {81}},\ \bibinfo
  {pages} {074033} (\bibinfo {year} {2010})},\ \Eprint
  {https://arxiv.org/abs/1001.2553} {arXiv:1001.2553 [hep-ph]} \BibitemShut
  {NoStop}%
\bibitem [{\citenamefont {Guo}\ \emph {et~al.}(2013)\citenamefont {Guo},
  \citenamefont {Hanhart}, \citenamefont {Meißner}, \citenamefont {Wang},\
  and\ \citenamefont {Zhao}}]{Guo:2013zbw}%
  \BibitemOpen
  \bibfield  {author} {\bibinfo {author} {\bibfnamefont {F.-K.}\ \bibnamefont
  {Guo}}, \bibinfo {author} {\bibfnamefont {C.}~\bibnamefont {Hanhart}},
  \bibinfo {author} {\bibfnamefont {U.-G.}\ \bibnamefont {Meißner}}, \bibinfo
  {author} {\bibfnamefont {Q.}~\bibnamefont {Wang}},\ and\ \bibinfo {author}
  {\bibfnamefont {Q.}~\bibnamefont {Zhao}},\ }\bibfield  {title} {\bibinfo
  {title} {{Production of the $X(3872)$ in charmonia radiative decays}},\
  }\href {https://doi.org/10.1016/j.physletb.2013.06.053} {\bibfield  {journal}
  {\bibinfo  {journal} {Phys. Lett. B}\ }\textbf {\bibinfo {volume} {725}},\
  \bibinfo {pages} {127} (\bibinfo {year} {2013})},\ \Eprint
  {https://arxiv.org/abs/1306.3096} {arXiv:1306.3096 [hep-ph]} \BibitemShut
  {NoStop}%
\bibitem [{\citenamefont {Li}\ and\ \citenamefont {Liu}(2013)}]{Li:2013yla}%
  \BibitemOpen
  \bibfield  {author} {\bibinfo {author} {\bibfnamefont {G.}~\bibnamefont
  {Li}}\ and\ \bibinfo {author} {\bibfnamefont {X.-H.}\ \bibnamefont {Liu}},\
  }\bibfield  {title} {\bibinfo {title} {{Investigating possible decay modes of
  $Y(4260)$ under ${D}_{1}(2420)\overline{D}+c.c.$ molecular state ansatz}},\
  }\href {https://doi.org/10.1103/PhysRevD.88.094008} {\bibfield  {journal}
  {\bibinfo  {journal} {Phys. Rev. D}\ }\textbf {\bibinfo {volume} {88}},\
  \bibinfo {pages} {094008} (\bibinfo {year} {2013})},\ \Eprint
  {https://arxiv.org/abs/1307.2622} {arXiv:1307.2622 [hep-ph]} \BibitemShut
  {NoStop}%
\bibitem [{\citenamefont {Liu}\ and\ \citenamefont {Li}(2013)}]{Liu:2013vfa}%
  \BibitemOpen
  \bibfield  {author} {\bibinfo {author} {\bibfnamefont {X.-H.}\ \bibnamefont
  {Liu}}\ and\ \bibinfo {author} {\bibfnamefont {G.}~\bibnamefont {Li}},\
  }\bibfield  {title} {\bibinfo {title} {{Exploring the threshold behavior and
  implications on the nature of $Y(4260)$ and $Z_c(3900)$}},\ }\href
  {https://doi.org/10.1103/PhysRevD.88.014013} {\bibfield  {journal} {\bibinfo
  {journal} {Phys. Rev. D}\ }\textbf {\bibinfo {volume} {88}},\ \bibinfo
  {pages} {014013} (\bibinfo {year} {2013})},\ \Eprint
  {https://arxiv.org/abs/1306.1384} {arXiv:1306.1384 [hep-ph]} \BibitemShut
  {NoStop}%
\bibitem [{\citenamefont {Wang}\ \emph {et~al.}(2013)\citenamefont {Wang},
  \citenamefont {Hanhart},\ and\ \citenamefont {Zhao}}]{Wang:2013cya}%
  \BibitemOpen
  \bibfield  {author} {\bibinfo {author} {\bibfnamefont {Q.}~\bibnamefont
  {Wang}}, \bibinfo {author} {\bibfnamefont {C.}~\bibnamefont {Hanhart}},\ and\
  \bibinfo {author} {\bibfnamefont {Q.}~\bibnamefont {Zhao}},\ }\bibfield
  {title} {\bibinfo {title} {{Decoding the riddle of $Y(4260)$ and
  $Z_c(3900)$}},\ }\href {https://doi.org/10.1103/PhysRevLett.111.132003}
  {\bibfield  {journal} {\bibinfo  {journal} {Phys. Rev. Lett.}\ }\textbf
  {\bibinfo {volume} {111}},\ \bibinfo {pages} {132003} (\bibinfo {year}
  {2013})},\ \Eprint {https://arxiv.org/abs/1303.6355} {arXiv:1303.6355
  [hep-ph]} \BibitemShut {NoStop}%
\bibitem [{\citenamefont {Cleven}\ \emph {et~al.}(2014)\citenamefont {Cleven},
  \citenamefont {Wang}, \citenamefont {Guo}, \citenamefont {Hanhart},
  \citenamefont {Meißner},\ and\ \citenamefont {Zhao}}]{Cleven:2013mka}%
  \BibitemOpen
  \bibfield  {author} {\bibinfo {author} {\bibfnamefont {M.}~\bibnamefont
  {Cleven}}, \bibinfo {author} {\bibfnamefont {Q.}~\bibnamefont {Wang}},
  \bibinfo {author} {\bibfnamefont {F.-K.}\ \bibnamefont {Guo}}, \bibinfo
  {author} {\bibfnamefont {C.}~\bibnamefont {Hanhart}}, \bibinfo {author}
  {\bibfnamefont {U.-G.}\ \bibnamefont {Meißner}},\ and\ \bibinfo {author}
  {\bibfnamefont {Q.}~\bibnamefont {Zhao}},\ }\bibfield  {title} {\bibinfo
  {title} {{$Y(4260)$ as the first $S$-wave open charm vector molecular
  state?}},\ }\href {https://doi.org/10.1103/PhysRevD.90.074039} {\bibfield
  {journal} {\bibinfo  {journal} {Phys. Rev. D}\ }\textbf {\bibinfo {volume}
  {90}},\ \bibinfo {pages} {074039} (\bibinfo {year} {2014})},\ \Eprint
  {https://arxiv.org/abs/1310.2190} {arXiv:1310.2190 [hep-ph]} \BibitemShut
  {NoStop}%
\bibitem [{\citenamefont {Peng}\ \emph {et~al.}(2023)\citenamefont {Peng},
  \citenamefont {Yan}, \citenamefont {Sánchez},\ and\ \citenamefont
  {Valderrama}}]{Peng:2022nrj}%
  \BibitemOpen
  \bibfield  {author} {\bibinfo {author} {\bibfnamefont {F.-Z.}\ \bibnamefont
  {Peng}}, \bibinfo {author} {\bibfnamefont {M.-J.}\ \bibnamefont {Yan}},
  \bibinfo {author} {\bibfnamefont {M.~S.}\ \bibnamefont {Sánchez}},\ and\
  \bibinfo {author} {\bibfnamefont {M.~P.}\ \bibnamefont {Valderrama}},\
  }\bibfield  {title} {\bibinfo {title} {{Light- and heavy-quark symmetries and
  the $Y(4230)$, $Y(4360)$, $Y(4500)$, $Y(4620)$ and $X(4630)$ resonances}},\
  }\href {https://doi.org/10.1103/PhysRevD.107.016001} {\bibfield  {journal}
  {\bibinfo  {journal} {Phys. Rev. D}\ }\textbf {\bibinfo {volume} {107}},\
  \bibinfo {pages} {016001} (\bibinfo {year} {2023})},\ \Eprint
  {https://arxiv.org/abs/2205.13590} {arXiv:2205.13590 [hep-ph]} \BibitemShut
  {NoStop}%
\bibitem [{\citenamefont {Dong}\ \emph {et~al.}(2020)\citenamefont {Dong},
  \citenamefont {Lin},\ and\ \citenamefont {Zou}}]{Dong:2019ofp}%
  \BibitemOpen
  \bibfield  {author} {\bibinfo {author} {\bibfnamefont {X.-K.}\ \bibnamefont
  {Dong}}, \bibinfo {author} {\bibfnamefont {Y.-H.}\ \bibnamefont {Lin}},\ and\
  \bibinfo {author} {\bibfnamefont {B.-S.}\ \bibnamefont {Zou}},\ }\bibfield
  {title} {\bibinfo {title} {{Prediction of an exotic state around 4240 MeV
  with $J^{PC}=1^{-+}$ as C-parity partner of $Y(4260)$ in molecular
  picture}},\ }\href {https://doi.org/10.1103/PhysRevD.101.076003} {\bibfield
  {journal} {\bibinfo  {journal} {Phys. Rev. D}\ }\textbf {\bibinfo {volume}
  {101}},\ \bibinfo {pages} {076003} (\bibinfo {year} {2020})},\ \Eprint
  {https://arxiv.org/abs/1910.14455} {arXiv:1910.14455 [hep-ph]} \BibitemShut
  {NoStop}%
\bibitem [{\citenamefont {Cleven}\ and\ \citenamefont
  {Zhao}(2017)}]{Cleven:2016qbn}%
  \BibitemOpen
  \bibfield  {author} {\bibinfo {author} {\bibfnamefont {M.}~\bibnamefont
  {Cleven}}\ and\ \bibinfo {author} {\bibfnamefont {Q.}~\bibnamefont {Zhao}},\
  }\bibfield  {title} {\bibinfo {title} {{Cross section line shape of
  $e^+e^-\to\chi_{c0}\omega$ around the $Y(4260)$ mass region}},\ }\href
  {https://doi.org/10.1016/j.physletb.2017.02.041} {\bibfield  {journal}
  {\bibinfo  {journal} {Phys. Lett. B}\ }\textbf {\bibinfo {volume} {768}},\
  \bibinfo {pages} {52} (\bibinfo {year} {2017})},\ \Eprint
  {https://arxiv.org/abs/1611.04408} {arXiv:1611.04408 [hep-ph]} \BibitemShut
  {NoStop}%
\bibitem [{\citenamefont {Li}\ \emph {et~al.}(2015)\citenamefont {Li},
  \citenamefont {An}, \citenamefont {Li}, \citenamefont {Liu}, \citenamefont
  {Zhang},\ and\ \citenamefont {Zhou}}]{Li:2014gxa}%
  \BibitemOpen
  \bibfield  {author} {\bibinfo {author} {\bibfnamefont {G.}~\bibnamefont
  {Li}}, \bibinfo {author} {\bibfnamefont {C.-S.}\ \bibnamefont {An}}, \bibinfo
  {author} {\bibfnamefont {P.-Y.}\ \bibnamefont {Li}}, \bibinfo {author}
  {\bibfnamefont {D.}~\bibnamefont {Liu}}, \bibinfo {author} {\bibfnamefont
  {X.}~\bibnamefont {Zhang}},\ and\ \bibinfo {author} {\bibfnamefont
  {Z.}~\bibnamefont {Zhou}},\ }\bibfield  {title} {\bibinfo {title}
  {{Investigations on the charmless decays of $Y(4260)$*}},\ }\href
  {https://doi.org/10.1088/1674-1137/39/6/063102} {\bibfield  {journal}
  {\bibinfo  {journal} {Chin. Phys. C}\ }\textbf {\bibinfo {volume} {39}},\
  \bibinfo {pages} {063102} (\bibinfo {year} {2015})},\ \Eprint
  {https://arxiv.org/abs/1412.3221} {arXiv:1412.3221 [hep-ph]} \BibitemShut
  {NoStop}%
\bibitem [{\citenamefont {Qiao}(2006)}]{Qiao:2005av}%
  \BibitemOpen
  \bibfield  {author} {\bibinfo {author} {\bibfnamefont {C.-F.}\ \bibnamefont
  {Qiao}},\ }\bibfield  {title} {\bibinfo {title} {{One explanation for the
  exotic state $Y(4260)$}},\ }\href
  {https://doi.org/10.1016/j.physletb.2006.06.038} {\bibfield  {journal}
  {\bibinfo  {journal} {Phys. Lett. B}\ }\textbf {\bibinfo {volume} {639}},\
  \bibinfo {pages} {263} (\bibinfo {year} {2006})},\ \Eprint
  {https://arxiv.org/abs/hep-ph/0510228} {arXiv:hep-ph/0510228} \BibitemShut
  {NoStop}%
\bibitem [{\citenamefont {Chen}\ \emph {et~al.}(2019)\citenamefont {Chen},
  \citenamefont {Dai}, \citenamefont {Guo},\ and\ \citenamefont
  {Kubis}}]{Chen:2019mgp}%
  \BibitemOpen
  \bibfield  {author} {\bibinfo {author} {\bibfnamefont {Y.-H.}\ \bibnamefont
  {Chen}}, \bibinfo {author} {\bibfnamefont {L.-Y.}\ \bibnamefont {Dai}},
  \bibinfo {author} {\bibfnamefont {F.-K.}\ \bibnamefont {Guo}},\ and\ \bibinfo
  {author} {\bibfnamefont {B.}~\bibnamefont {Kubis}},\ }\bibfield  {title}
  {\bibinfo {title} {{Nature of the $Y\mathbf{(}4260\mathbf{)}$: A light-quark
  perspective}},\ }\href {https://doi.org/10.1103/PhysRevD.99.074016}
  {\bibfield  {journal} {\bibinfo  {journal} {Phys. Rev. D}\ }\textbf {\bibinfo
  {volume} {99}},\ \bibinfo {pages} {074016} (\bibinfo {year} {2019})},\
  \Eprint {https://arxiv.org/abs/1902.10957} {arXiv:1902.10957 [hep-ph]}
  \BibitemShut {NoStop}%
\bibitem [{\citenamefont {Fu}\ and\ \citenamefont {Jiang}(2019)}]{Fu:2018yxq}%
  \BibitemOpen
  \bibfield  {author} {\bibinfo {author} {\bibfnamefont {H.-f.}\ \bibnamefont
  {Fu}}\ and\ \bibinfo {author} {\bibfnamefont {L.}~\bibnamefont {Jiang}},\
  }\bibfield  {title} {\bibinfo {title} {{Coupled-Channel-Induced $S-D$ mixing
  of Charmonia and Testing Possible Assignments for $Y(4260)$ and $Y(4360)$}},\
  }\href {https://doi.org/10.1140/epjc/s10052-019-6976-0} {\bibfield  {journal}
  {\bibinfo  {journal} {Eur. Phys. J. C}\ }\textbf {\bibinfo {volume} {79}},\
  \bibinfo {pages} {460} (\bibinfo {year} {2019})},\ \Eprint
  {https://arxiv.org/abs/1812.00179} {arXiv:1812.00179 [hep-ph]} \BibitemShut
  {NoStop}%
\bibitem [{\citenamefont {Chen}\ \emph
  {et~al.}(2011{\natexlab{a}})\citenamefont {Chen}, \citenamefont {He},\ and\
  \citenamefont {Liu}}]{Chen:2010nv}%
  \BibitemOpen
  \bibfield  {author} {\bibinfo {author} {\bibfnamefont {D.-Y.}\ \bibnamefont
  {Chen}}, \bibinfo {author} {\bibfnamefont {J.}~\bibnamefont {He}},\ and\
  \bibinfo {author} {\bibfnamefont {X.}~\bibnamefont {Liu}},\ }\bibfield
  {title} {\bibinfo {title} {{Nonresonant explanation for the $Y(4260)$
  structure observed in the
  ${e}^{+}{e}^{\ensuremath{-}}\ensuremath{\rightarrow}J/\ensuremath{\psi}{\ensuremath{\pi}}^{+}{\ensuremath{\pi}}^{\ensuremath{-}}$
  process}},\ }\href {https://doi.org/10.1103/PhysRevD.83.054021} {\bibfield
  {journal} {\bibinfo  {journal} {Phys. Rev. D}\ }\textbf {\bibinfo {volume}
  {83}},\ \bibinfo {pages} {054021} (\bibinfo {year} {2011}{\natexlab{a}})},\
  \Eprint {https://arxiv.org/abs/1012.5362} {arXiv:1012.5362 [hep-ph]}
  \BibitemShut {NoStop}%
\bibitem [{\citenamefont {Liu}(2014{\natexlab{b}})}]{Liu:2014spa}%
  \BibitemOpen
  \bibfield  {author} {\bibinfo {author} {\bibfnamefont {X.-H.}\ \bibnamefont
  {Liu}},\ }\bibfield  {title} {\bibinfo {title} {{Influence of threshold
  effects induced by charmed meson rescattering}},\ }\href
  {https://doi.org/10.1103/PhysRevD.90.074004} {\bibfield  {journal} {\bibinfo
  {journal} {Phys. Rev. D}\ }\textbf {\bibinfo {volume} {90}},\ \bibinfo
  {pages} {074004} (\bibinfo {year} {2014}{\natexlab{b}})},\ \Eprint
  {https://arxiv.org/abs/1403.2818} {arXiv:1403.2818 [hep-ph]} \BibitemShut
  {NoStop}%
\bibitem [{\citenamefont {Ablikim}\ \emph
  {et~al.}(2023{\natexlab{b}})\citenamefont {Ablikim}, \citenamefont {Achasov},
  \citenamefont {Adlarson} \emph {et~al.}}]{BESIII:2023wqy}%
  \BibitemOpen
  \bibfield  {author} {\bibinfo {author} {\bibfnamefont {M.}~\bibnamefont
  {Ablikim}}, \bibinfo {author} {\bibfnamefont {M.}~\bibnamefont {Achasov}},
  \bibinfo {author} {\bibfnamefont {P.}~\bibnamefont {Adlarson}}, \emph
  {et~al.} (\bibinfo {collaboration} {BESIII}),\ }\bibfield  {title} {\bibinfo
  {title} {{Observation of a Vector Charmoniumlike State at $4.7\text{ }\text{
  }\mathrm{GeV}/{c}^{2}$ and Search for ${Z}_{cs}$ in
  ${e}^{+}{e}^{\ensuremath{-}}\ensuremath{\rightarrow}{K}^{+}{K}^{\ensuremath{-}}J/\ensuremath{\psi}$}},\
  }\href {https://doi.org/10.1103/PhysRevLett.131.211902} {\bibfield  {journal}
  {\bibinfo  {journal} {Phys. Rev. Lett.}\ }\textbf {\bibinfo {volume} {131}},\
  \bibinfo {pages} {211902} (\bibinfo {year} {2023}{\natexlab{b}})},\ \Eprint
  {https://arxiv.org/abs/2308.15362} {arXiv:2308.15362 [hep-ex]} \BibitemShut
  {NoStop}%
\bibitem [{\citenamefont {Ablikim}\ \emph
  {et~al.}(2023{\natexlab{c}})\citenamefont {Ablikim} \emph
  {et~al.}}]{BESIII:2023wsc}%
  \BibitemOpen
  \bibfield  {author} {\bibinfo {author} {\bibfnamefont {M.}~\bibnamefont
  {Ablikim}} \emph {et~al.} (\bibinfo {collaboration} {BESIII}),\ }\bibfield
  {title} {\bibinfo {title} {{Precise Measurement of the $e^+e^-\to
  D_s^{*+}D_s^{*-}$ Cross Sections at Center-of-Mass Energies from Threshold to
  4.95~GeV}},\ }\href {https://doi.org/10.1103/PhysRevLett.131.151903}
  {\bibfield  {journal} {\bibinfo  {journal} {Phys. Rev. Lett.}\ }\textbf
  {\bibinfo {volume} {131}},\ \bibinfo {pages} {151903} (\bibinfo {year}
  {2023}{\natexlab{c}})},\ \Eprint {https://arxiv.org/abs/2305.10789}
  {arXiv:2305.10789 [hep-ex]} \BibitemShut {NoStop}%
\bibitem [{\citenamefont {Ablikim}\ \emph {et~al.}(2024)\citenamefont
  {Ablikim}, \citenamefont {Achasov}, \citenamefont {Adlarson} \emph
  {et~al.}}]{BESIII:2024jzg}%
  \BibitemOpen
  \bibfield  {author} {\bibinfo {author} {\bibfnamefont {M.}~\bibnamefont
  {Ablikim}}, \bibinfo {author} {\bibfnamefont {M.~N.}\ \bibnamefont
  {Achasov}}, \bibinfo {author} {\bibfnamefont {P.~A.}\ \bibnamefont
  {Adlarson}}, \emph {et~al.} (\bibinfo {collaboration} {BESIII}),\ }\bibfield
  {title} {\bibinfo {title} {Observation of structures in the processes
  $e^+e^-\rightarrow\omega\chi_{c1}$ and $\omega\chi_{c2}$},\ }\href
  {https://doi.org/10.1103/PhysRevLett.132.161901} {\bibfield  {journal}
  {\bibinfo  {journal} {Phys. Rev. Lett.}\ }\textbf {\bibinfo {volume} {132}},\
  \bibinfo {pages} {161901} (\bibinfo {year} {2024})},\ \Eprint
  {https://arxiv.org/abs/2401.14720} {arXiv:2401.14720 [hep-ex]} \BibitemShut
  {NoStop}%
\bibitem [{\citenamefont {Ablikim}\ \emph {et~al.}(2025)\citenamefont {Ablikim}
  \emph {et~al.}}]{BESIII:2025bce}%
  \BibitemOpen
  \bibfield  {author} {\bibinfo {author} {\bibfnamefont {M.}~\bibnamefont
  {Ablikim}} \emph {et~al.} (\bibinfo {collaboration} {BESIII Collaboration}),\
  }\bibfield  {title} {\bibinfo {title} {Observation of three resonant
  structures in the cross section of
  ${e}^{+}{e}^{\ensuremath{-}}\ensuremath{\rightarrow}{\ensuremath{\pi}}^{+}{\ensuremath{\pi}}^{\ensuremath{-}}{h}_{c}$},\
  }\href {https://doi.org/10.1103/ljnf-4jfr} {\bibfield  {journal} {\bibinfo
  {journal} {Phys. Rev. Lett.}\ }\textbf {\bibinfo {volume} {135}},\ \bibinfo
  {pages} {071901} (\bibinfo {year} {2025})}\BibitemShut {NoStop}%
\bibitem [{\citenamefont {Deng}\ \emph {et~al.}(2024)\citenamefont {Deng},
  \citenamefont {Ni}, \citenamefont {Li},\ and\ \citenamefont
  {Zhong}}]{Deng:2023mza}%
  \BibitemOpen
  \bibfield  {author} {\bibinfo {author} {\bibfnamefont {Q.}~\bibnamefont
  {Deng}}, \bibinfo {author} {\bibfnamefont {R.-H.}\ \bibnamefont {Ni}},
  \bibinfo {author} {\bibfnamefont {Q.}~\bibnamefont {Li}},\ and\ \bibinfo
  {author} {\bibfnamefont {X.-H.}\ \bibnamefont {Zhong}},\ }\bibfield  {title}
  {\bibinfo {title} {{Charmonia in an unquenched quark model}},\ }\href
  {https://doi.org/10.1103/PhysRevD.110.056034} {\bibfield  {journal} {\bibinfo
   {journal} {Phys. Rev. D}\ }\textbf {\bibinfo {volume} {110}},\ \bibinfo
  {pages} {056034} (\bibinfo {year} {2024})},\ \Eprint
  {https://arxiv.org/abs/2312.10296} {arXiv:2312.10296 [hep-ph]} \BibitemShut
  {NoStop}%
\bibitem [{\citenamefont {Dong}\ \emph {et~al.}(2021)\citenamefont {Dong},
  \citenamefont {Guo},\ and\ \citenamefont {Zou}}]{Dong:2021juy}%
  \BibitemOpen
  \bibfield  {author} {\bibinfo {author} {\bibfnamefont {X.-K.}\ \bibnamefont
  {Dong}}, \bibinfo {author} {\bibfnamefont {F.-K.}\ \bibnamefont {Guo}},\ and\
  \bibinfo {author} {\bibfnamefont {B.-S.}\ \bibnamefont {Zou}},\ }\bibfield
  {title} {\bibinfo {title} {A survey of heavy-antiheavy hadronic molecules},\
  }\href {https://doi.org/10.13725/j.cnki.pip.2021.02.001} {\bibfield
  {journal} {\bibinfo  {journal} {Progr. Phys.}\ }\textbf {\bibinfo {volume}
  {41}},\ \bibinfo {pages} {65} (\bibinfo {year} {2021})},\ \Eprint
  {https://arxiv.org/abs/2101.01021} {arXiv:2101.01021 [hep-ph]} \BibitemShut
  {NoStop}%
\bibitem [{\citenamefont {Wang}(2023)}]{Wang:2023hsc}%
  \BibitemOpen
  \bibfield  {author} {\bibinfo {author} {\bibfnamefont {Z.-G.}\ \bibnamefont
  {Wang}},\ }\bibfield  {title} {\bibinfo {title} {{Analysis of the decay
  $Y(4500)\to D^*D^-\pi $ with the light-cone QCD sum rules}},\ }\href
  {https://doi.org/10.1016/j.nuclphysb.2023.116265} {\bibfield  {journal}
  {\bibinfo  {journal} {Nucl. Phys. B}\ }\textbf {\bibinfo {volume} {993}},\
  \bibinfo {pages} {116265} (\bibinfo {year} {2023})},\ \Eprint
  {https://arxiv.org/abs/2304.14153} {arXiv:2304.14153 [hep-ph]} \BibitemShut
  {NoStop}%
\bibitem [{\citenamefont {Wang}(2024)}]{Wang:2024qqa}%
  \BibitemOpen
  \bibfield  {author} {\bibinfo {author} {\bibfnamefont {Z.-G.}\ \bibnamefont
  {Wang}},\ }\bibfield  {title} {\bibinfo {title} {{Strong decays of the vector
  tetraquark states with the masses about $4.5\,\rm{GeV}$ via the QCD sum
  rules}},\ }\href {https://doi.org/10.1016/j.nuclphysb.2024.116580} {\bibfield
   {journal} {\bibinfo  {journal} {Nucl. Phys. B}\ }\textbf {\bibinfo {volume}
  {1005}},\ \bibinfo {pages} {116580} (\bibinfo {year} {2024})},\ \Eprint
  {https://arxiv.org/abs/2404.05328} {arXiv:2404.05328 [hep-ph]} \BibitemShut
  {NoStop}%
\bibitem [{\citenamefont {Wang}\ and\ \citenamefont
  {Liu}(2023)}]{Wang:2022jxj}%
  \BibitemOpen
  \bibfield  {author} {\bibinfo {author} {\bibfnamefont {J.-Z.}\ \bibnamefont
  {Wang}}\ and\ \bibinfo {author} {\bibfnamefont {X.}~\bibnamefont {Liu}},\
  }\bibfield  {title} {\bibinfo {title} {{Confirming the existence of a new
  higher charmonium $\ensuremath{\psi}(4500)$ by the newly released data of
  ${e}^{+}{e}^{\ensuremath{-}}\ensuremath{\rightarrow}{K}^{+}{K}^{\ensuremath{-}}J/\ensuremath{\psi}$}},\
  }\href {https://doi.org/10.1103/PhysRevD.107.054016} {\bibfield  {journal}
  {\bibinfo  {journal} {Phys. Rev. D}\ }\textbf {\bibinfo {volume} {107}},\
  \bibinfo {pages} {054016} (\bibinfo {year} {2023})},\ \Eprint
  {https://arxiv.org/abs/2212.13512} {arXiv:2212.13512 [hep-ph]} \BibitemShut
  {NoStop}%
\bibitem [{\citenamefont {Ablikim}\ \emph {et~al.}(2018)\citenamefont
  {Ablikim}, \citenamefont {Achasov}, \citenamefont {Ahmed} \emph
  {et~al.}}]{BESIII:2018iop}%
  \BibitemOpen
  \bibfield  {author} {\bibinfo {author} {\bibfnamefont {M.}~\bibnamefont
  {Ablikim}}, \bibinfo {author} {\bibfnamefont {M.~N.}\ \bibnamefont
  {Achasov}}, \bibinfo {author} {\bibfnamefont {S.~A.~N.}\ \bibnamefont
  {Ahmed}}, \emph {et~al.} (\bibinfo {collaboration} {BESIII}),\ }\bibfield
  {title} {\bibinfo {title} {{Measurement of
  ${e}^{+}{e}^{\ensuremath{-}}\ensuremath{\rightarrow}K\overline{K}J/\ensuremath{\psi}$
  cross sections at center-of-mass energies from 4.189 to 4.600 GeV}},\ }\href
  {https://doi.org/10.1103/PhysRevD.97.071101} {\bibfield  {journal} {\bibinfo
  {journal} {Phys. Rev. D}\ }\textbf {\bibinfo {volume} {97}},\ \bibinfo
  {pages} {071101} (\bibinfo {year} {2018})},\ \Eprint
  {https://arxiv.org/abs/1802.01216} {arXiv:1802.01216 [hep-ex]} \BibitemShut
  {NoStop}%
\bibitem [{\citenamefont {Yan}(1980)}]{Yan:1980uh}%
  \BibitemOpen
  \bibfield  {author} {\bibinfo {author} {\bibfnamefont {T.-M.}\ \bibnamefont
  {Yan}},\ }\bibfield  {title} {\bibinfo {title} {{Hadronic Transitions Between
  Heavy Quark States in Quantum Chromodynamics}},\ }\href
  {https://doi.org/10.1103/PhysRevD.22.1652} {\bibfield  {journal} {\bibinfo
  {journal} {Phys. Rev. D}\ }\textbf {\bibinfo {volume} {22}},\ \bibinfo
  {pages} {1652} (\bibinfo {year} {1980})}\BibitemShut {NoStop}%
\bibitem [{\citenamefont {Kuang}\ and\ \citenamefont
  {Yan}(1981)}]{Kuang:1981se}%
  \BibitemOpen
  \bibfield  {author} {\bibinfo {author} {\bibfnamefont {Y.-P.}\ \bibnamefont
  {Kuang}}\ and\ \bibinfo {author} {\bibfnamefont {T.-M.}\ \bibnamefont
  {Yan}},\ }\bibfield  {title} {\bibinfo {title} {{Predictions for Hadronic
  Transitions in the $b\bar{b}$ System}},\ }\href
  {https://doi.org/10.1103/PhysRevD.24.2874} {\bibfield  {journal} {\bibinfo
  {journal} {Phys. Rev. D}\ }\textbf {\bibinfo {volume} {24}},\ \bibinfo
  {pages} {2874} (\bibinfo {year} {1981})}\BibitemShut {NoStop}%
\bibitem [{\citenamefont {Wang}\ \emph {et~al.}(2016)\citenamefont {Wang},
  \citenamefont {Xu}, \citenamefont {Liu}, \citenamefont {Chen}, \citenamefont
  {Coito},\ and\ \citenamefont {Eichten}}]{Wang:2015xsa}%
  \BibitemOpen
  \bibfield  {author} {\bibinfo {author} {\bibfnamefont {B.}~\bibnamefont
  {Wang}}, \bibinfo {author} {\bibfnamefont {H.}~\bibnamefont {Xu}}, \bibinfo
  {author} {\bibfnamefont {X.}~\bibnamefont {Liu}}, \bibinfo {author}
  {\bibfnamefont {D.-Y.}\ \bibnamefont {Chen}}, \bibinfo {author}
  {\bibfnamefont {S.}~\bibnamefont {Coito}},\ and\ \bibinfo {author}
  {\bibfnamefont {E.}~\bibnamefont {Eichten}},\ }\bibfield  {title} {\bibinfo
  {title} {{Using $X(3823)\to J/\psi \pi^+\pi^-$ to identify coupled-channel
  effects}},\ }\href {https://doi.org/10.1007/s11467-016-0564-7} {\bibfield
  {journal} {\bibinfo  {journal} {Front. Phys. (Beijing)}\ }\textbf {\bibinfo
  {volume} {11}},\ \bibinfo {pages} {111402} (\bibinfo {year} {2016})},\
  \Eprint {https://arxiv.org/abs/1507.07985} {arXiv:1507.07985 [hep-ph]}
  \BibitemShut {NoStop}%
\bibitem [{\citenamefont {Jia}\ \emph {et~al.}(2024)\citenamefont {Jia},
  \citenamefont {Zhang}, \citenamefont {Qin},\ and\ \citenamefont
  {Li}}]{Jia:2023pud}%
  \BibitemOpen
  \bibfield  {author} {\bibinfo {author} {\bibfnamefont {Z.-S.}\ \bibnamefont
  {Jia}}, \bibinfo {author} {\bibfnamefont {Z.-H.}\ \bibnamefont {Zhang}},
  \bibinfo {author} {\bibfnamefont {W.-H.}\ \bibnamefont {Qin}},\ and\ \bibinfo
  {author} {\bibfnamefont {G.}~\bibnamefont {Li}},\ }\bibfield  {title}
  {\bibinfo {title} {{Hunting for $X_b$ via hidden bottomonium decays $X_b\to
  \pi\pi\chi_{bJ}$}},\ }\href {https://doi.org/10.1103/PhysRevD.109.034017}
  {\bibfield  {journal} {\bibinfo  {journal} {Phys. Rev. D}\ }\textbf {\bibinfo
  {volume} {109}},\ \bibinfo {pages} {034017} (\bibinfo {year} {2024})},\
  \Eprint {https://arxiv.org/abs/2311.15527} {arXiv:2311.15527 [hep-ph]}
  \BibitemShut {NoStop}%
\bibitem [{\citenamefont {Liu}\ \emph {et~al.}(2024{\natexlab{a}})\citenamefont
  {Liu}, \citenamefont {Wang}, \citenamefont {Jia}, \citenamefont {Li},
  \citenamefont {Liu},\ and\ \citenamefont {Xie}}]{Liu:2024ogo}%
  \BibitemOpen
  \bibfield  {author} {\bibinfo {author} {\bibfnamefont {S.-D.}\ \bibnamefont
  {Liu}}, \bibinfo {author} {\bibfnamefont {F.}~\bibnamefont {Wang}}, \bibinfo
  {author} {\bibfnamefont {Z.-S.}\ \bibnamefont {Jia}}, \bibinfo {author}
  {\bibfnamefont {G.}~\bibnamefont {Li}}, \bibinfo {author} {\bibfnamefont
  {X.-H.}\ \bibnamefont {Liu}},\ and\ \bibinfo {author} {\bibfnamefont {J.-J.}\
  \bibnamefont {Xie}},\ }\bibfield  {title} {\bibinfo {title} {{Pionic
  transitions of the spin-2 partner of $X(3872)$ to
  ${\ensuremath{\chi}}_{cJ}$}},\ }\href
  {https://doi.org/10.1103/PhysRevD.110.054048} {\bibfield  {journal} {\bibinfo
   {journal} {Phys. Rev. D}\ }\textbf {\bibinfo {volume} {110}},\ \bibinfo
  {pages} {054048} (\bibinfo {year} {2024}{\natexlab{a}})},\ \Eprint
  {https://arxiv.org/abs/2406.01874} {arXiv:2406.01874 [hep-ph]} \BibitemShut
  {NoStop}%
\bibitem [{\citenamefont {Chen}(2020)}]{Chen:2019gfp}%
  \BibitemOpen
  \bibfield  {author} {\bibinfo {author} {\bibfnamefont {Y.-H.}\ \bibnamefont
  {Chen}},\ }\bibfield  {title} {\bibinfo {title} {{Predictions of
  $\Upsilon(4S) \to h_b (1P,2P) \pi^+\pi^-$ transitions*}},\ }\href
  {https://doi.org/10.1088/1674-1137/44/2/023103} {\bibfield  {journal}
  {\bibinfo  {journal} {Chin. Phys. C}\ }\textbf {\bibinfo {volume} {44}},\
  \bibinfo {pages} {023103} (\bibinfo {year} {2020})},\ \Eprint
  {https://arxiv.org/abs/1907.05547} {arXiv:1907.05547 [hep-ph]} \BibitemShut
  {NoStop}%
\bibitem [{\citenamefont {Chen}\ \emph {et~al.}(2017)\citenamefont {Chen},
  \citenamefont {Cleven}, \citenamefont {Daub}, \citenamefont {Guo},
  \citenamefont {Hanhart}, \citenamefont {Kubis}, \citenamefont {Meißner},\
  and\ \citenamefont {Zou}}]{Chen:2016mjn}%
  \BibitemOpen
  \bibfield  {author} {\bibinfo {author} {\bibfnamefont {Y.-H.}\ \bibnamefont
  {Chen}}, \bibinfo {author} {\bibfnamefont {M.}~\bibnamefont {Cleven}},
  \bibinfo {author} {\bibfnamefont {J.~T.}\ \bibnamefont {Daub}}, \bibinfo
  {author} {\bibfnamefont {F.-K.}\ \bibnamefont {Guo}}, \bibinfo {author}
  {\bibfnamefont {C.}~\bibnamefont {Hanhart}}, \bibinfo {author} {\bibfnamefont
  {B.}~\bibnamefont {Kubis}}, \bibinfo {author} {\bibfnamefont {U.-G.}\
  \bibnamefont {Meißner}},\ and\ \bibinfo {author} {\bibfnamefont {B.-S.}\
  \bibnamefont {Zou}},\ }\bibfield  {title} {\bibinfo {title} {{Effects of
  ${Z}_{b}$ states and bottom meson loops on
  $\mathrm{\ensuremath{\Upsilon}}(4S)\ensuremath{\rightarrow}\mathrm{\ensuremath{\Upsilon}}(1S,2S){\ensuremath{\pi}}^{+}{\ensuremath{\pi}}^{\ensuremath{-}}$
  transitions}},\ }\href {https://doi.org/10.1103/PhysRevD.95.034022}
  {\bibfield  {journal} {\bibinfo  {journal} {Phys. Rev. D}\ }\textbf {\bibinfo
  {volume} {95}},\ \bibinfo {pages} {034022} (\bibinfo {year} {2017})},\
  \Eprint {https://arxiv.org/abs/1611.00913} {arXiv:1611.00913 [hep-ph]}
  \BibitemShut {NoStop}%
\bibitem [{\citenamefont {Badalian}\ \emph {et~al.}(2009)\citenamefont
  {Badalian}, \citenamefont {Bakker},\ and\ \citenamefont
  {Danilkin}}]{Badalian:2008dv}%
  \BibitemOpen
  \bibfield  {author} {\bibinfo {author} {\bibfnamefont {A.~M.}\ \bibnamefont
  {Badalian}}, \bibinfo {author} {\bibfnamefont {B.~L.~G.}\ \bibnamefont
  {Bakker}},\ and\ \bibinfo {author} {\bibfnamefont {I.~V.}\ \bibnamefont
  {Danilkin}},\ }\bibfield  {title} {\bibinfo {title} {{The $S$-$D$ mixing and
  di-electron widths of higher charmonium $1^{--}$ states}},\ }\href
  {https://doi.org/10.1134/S1063778809040085} {\bibfield  {journal} {\bibinfo
  {journal} {Phys. Atom. Nucl.}\ }\textbf {\bibinfo {volume} {72}},\ \bibinfo
  {pages} {638} (\bibinfo {year} {2009})},\ \Eprint
  {https://arxiv.org/abs/0805.2291} {arXiv:0805.2291 [hep-ph]} \BibitemShut
  {NoStop}%
\bibitem [{\citenamefont {Barnes}\ \emph {et~al.}(2005)\citenamefont {Barnes},
  \citenamefont {Godfrey},\ and\ \citenamefont {Swanson}}]{Barnes:2005pb}%
  \BibitemOpen
  \bibfield  {author} {\bibinfo {author} {\bibfnamefont {T.}~\bibnamefont
  {Barnes}}, \bibinfo {author} {\bibfnamefont {S.}~\bibnamefont {Godfrey}},\
  and\ \bibinfo {author} {\bibfnamefont {E.~S.}\ \bibnamefont {Swanson}},\
  }\bibfield  {title} {\bibinfo {title} {Higher charmonia},\ }\href
  {https://doi.org/10.1103/PhysRevD.72.054026} {\bibfield  {journal} {\bibinfo
  {journal} {Phys. Rev. D}\ }\textbf {\bibinfo {volume} {72}},\ \bibinfo
  {pages} {054026} (\bibinfo {year} {2005})},\ \Eprint
  {https://arxiv.org/abs/hep-ph/0505002} {arXiv:hep-ph/0505002} \BibitemShut
  {NoStop}%
\bibitem [{\citenamefont {Ebert}\ \emph {et~al.}(2011)\citenamefont {Ebert},
  \citenamefont {Faustov},\ and\ \citenamefont {Galkin}}]{Ebert:2011jc}%
  \BibitemOpen
  \bibfield  {author} {\bibinfo {author} {\bibfnamefont {D.}~\bibnamefont
  {Ebert}}, \bibinfo {author} {\bibfnamefont {R.~N.}\ \bibnamefont {Faustov}},\
  and\ \bibinfo {author} {\bibfnamefont {V.~O.}\ \bibnamefont {Galkin}},\
  }\bibfield  {title} {\bibinfo {title} {Spectroscopy and regge trajectories of
  heavy quarkonia and bcmesons},\ }\href
  {https://doi.org/10.1140/epjc/s10052-011-1825-9} {\bibfield  {journal}
  {\bibinfo  {journal} {Eur. Phys. J. C}\ }\textbf {\bibinfo {volume} {71}},\
  \bibinfo {pages} {1825} (\bibinfo {year} {2011})},\ \Eprint
  {https://arxiv.org/abs/1111.0454} {arXiv:1111.0454 [hep-ph]} \BibitemShut
  {NoStop}%
\bibitem [{\citenamefont {Gui}\ \emph {et~al.}(2018)\citenamefont {Gui},
  \citenamefont {Lu}, \citenamefont {Lü}, \citenamefont {Zhong},\ and\
  \citenamefont {Zhao}}]{Gui:2018rvv}%
  \BibitemOpen
  \bibfield  {author} {\bibinfo {author} {\bibfnamefont {L.-C.}\ \bibnamefont
  {Gui}}, \bibinfo {author} {\bibfnamefont {L.-S.}\ \bibnamefont {Lu}},
  \bibinfo {author} {\bibfnamefont {Q.-F.}\ \bibnamefont {Lü}}, \bibinfo
  {author} {\bibfnamefont {X.-H.}\ \bibnamefont {Zhong}},\ and\ \bibinfo
  {author} {\bibfnamefont {Q.}~\bibnamefont {Zhao}},\ }\bibfield  {title}
  {\bibinfo {title} {Strong decays of higher charmonium states into open-charm
  meson pairs},\ }\href {https://doi.org/10.1103/PhysRevD.98.016010} {\bibfield
   {journal} {\bibinfo  {journal} {Phys. Rev. D}\ }\textbf {\bibinfo {volume}
  {98}},\ \bibinfo {pages} {016010} (\bibinfo {year} {2018})},\ \Eprint
  {https://arxiv.org/abs/1801.08791} {arXiv:1801.08791 [hep-ph]} \BibitemShut
  {NoStop}%
\bibitem [{\citenamefont {Deng}\ \emph {et~al.}(2017)\citenamefont {Deng},
  \citenamefont {Liu}, \citenamefont {Gui},\ and\ \citenamefont
  {Zhong}}]{Deng:2016stx}%
  \BibitemOpen
  \bibfield  {author} {\bibinfo {author} {\bibfnamefont {W.-J.}\ \bibnamefont
  {Deng}}, \bibinfo {author} {\bibfnamefont {H.}~\bibnamefont {Liu}}, \bibinfo
  {author} {\bibfnamefont {L.-C.}\ \bibnamefont {Gui}},\ and\ \bibinfo {author}
  {\bibfnamefont {X.-H.}\ \bibnamefont {Zhong}},\ }\bibfield  {title} {\bibinfo
  {title} {Charmonium spectrum and electromagnetic transitions with higher
  multipole contributions},\ }\href
  {https://doi.org/10.1103/PhysRevD.95.034026} {\bibfield  {journal} {\bibinfo
  {journal} {Phys. Rev. D}\ }\textbf {\bibinfo {volume} {95}},\ \bibinfo
  {pages} {034026} (\bibinfo {year} {2017})},\ \Eprint
  {https://arxiv.org/abs/1608.00287} {arXiv:1608.00287 [hep-ph]} \BibitemShut
  {NoStop}%
\bibitem [{\citenamefont {Chen}\ and\ \citenamefont {Liu}(2011)}]{Chen:2011xk}%
  \BibitemOpen
  \bibfield  {author} {\bibinfo {author} {\bibfnamefont {D.-Y.}\ \bibnamefont
  {Chen}}\ and\ \bibinfo {author} {\bibfnamefont {X.}~\bibnamefont {Liu}},\
  }\bibfield  {title} {\bibinfo {title} {Predicted charged charmoniumlike
  structures in the hidden-charm dipion decay of higher charmonia},\ }\href
  {https://doi.org/10.1103/PhysRevD.84.034032} {\bibfield  {journal} {\bibinfo
  {journal} {Phys. Rev. D}\ }\textbf {\bibinfo {volume} {84}},\ \bibinfo
  {pages} {034032} (\bibinfo {year} {2011})},\ \Eprint
  {https://arxiv.org/abs/1106.5290} {arXiv:1106.5290 [hep-ph]} \BibitemShut
  {NoStop}%
\bibitem [{\citenamefont {Chen}\ \emph
  {et~al.}(2013{\natexlab{a}})\citenamefont {Chen}, \citenamefont {Liu},\ and\
  \citenamefont {Matsuki}}]{Chen:2013wca}%
  \BibitemOpen
  \bibfield  {author} {\bibinfo {author} {\bibfnamefont {D.-Y.}\ \bibnamefont
  {Chen}}, \bibinfo {author} {\bibfnamefont {X.}~\bibnamefont {Liu}},\ and\
  \bibinfo {author} {\bibfnamefont {T.}~\bibnamefont {Matsuki}},\ }\bibfield
  {title} {\bibinfo {title} {Predictions of charged charmoniumlike structures
  with hidden-charm and open-strange channels},\ }\href
  {https://doi.org/10.1103/PhysRevLett.110.232001} {\bibfield  {journal}
  {\bibinfo  {journal} {Phys. Rev. Lett.}\ }\textbf {\bibinfo {volume} {110}},\
  \bibinfo {pages} {232001} (\bibinfo {year} {2013}{\natexlab{a}})},\ \Eprint
  {https://arxiv.org/abs/1303.6842} {arXiv:1303.6842 [hep-ph]} \BibitemShut
  {NoStop}%
\bibitem [{\citenamefont {Anwar}\ \emph {et~al.}(2017)\citenamefont {Anwar},
  \citenamefont {Lu},\ and\ \citenamefont {Zou}}]{Anwar:2016mxo}%
  \BibitemOpen
  \bibfield  {author} {\bibinfo {author} {\bibfnamefont {M.~N.}\ \bibnamefont
  {Anwar}}, \bibinfo {author} {\bibfnamefont {Y.}~\bibnamefont {Lu}},\ and\
  \bibinfo {author} {\bibfnamefont {B.-S.}\ \bibnamefont {Zou}},\ }\bibfield
  {title} {\bibinfo {title} {Modeling charmonium-$\ensuremath{\eta}$ decays of
  ${J}^{PC}={1}^{\ensuremath{-}\ensuremath{-}}$ higher charmonia},\ }\href
  {https://doi.org/10.1103/PhysRevD.95.114031} {\bibfield  {journal} {\bibinfo
  {journal} {Phys. Rev. D}\ }\textbf {\bibinfo {volume} {95}},\ \bibinfo
  {pages} {114031} (\bibinfo {year} {2017})},\ \Eprint
  {https://arxiv.org/abs/1612.05396} {arXiv:1612.05396 [hep-ph]} \BibitemShut
  {NoStop}%
\bibitem [{\citenamefont {Casalbuoni}\ \emph {et~al.}(1997)\citenamefont
  {Casalbuoni}, \citenamefont {Deandrea}, \citenamefont {Di~Bartolomeo},
  \citenamefont {Gatto}, \citenamefont {Feruglio},\ and\ \citenamefont
  {Nardulli}}]{Casalbuoni:1996pg}%
  \BibitemOpen
  \bibfield  {author} {\bibinfo {author} {\bibfnamefont {R.}~\bibnamefont
  {Casalbuoni}}, \bibinfo {author} {\bibfnamefont {A.}~\bibnamefont
  {Deandrea}}, \bibinfo {author} {\bibfnamefont {N.}~\bibnamefont
  {Di~Bartolomeo}}, \bibinfo {author} {\bibfnamefont {R.}~\bibnamefont
  {Gatto}}, \bibinfo {author} {\bibfnamefont {F.}~\bibnamefont {Feruglio}},\
  and\ \bibinfo {author} {\bibfnamefont {G.}~\bibnamefont {Nardulli}},\
  }\bibfield  {title} {\bibinfo {title} {Phenomenology of heavy meson chiral
  lagrangians},\ }\href {https://doi.org/10.1016/S0370-1573(96)00027-0}
  {\bibfield  {journal} {\bibinfo  {journal} {Phys. Rept.}\ }\textbf {\bibinfo
  {volume} {281}},\ \bibinfo {pages} {145} (\bibinfo {year} {1997})},\ \Eprint
  {https://arxiv.org/abs/hep-ph/9605342} {arXiv:hep-ph/9605342} \BibitemShut
  {NoStop}%
\bibitem [{\citenamefont {Colangelo}\ \emph {et~al.}(2004)\citenamefont
  {Colangelo}, \citenamefont {De~Fazio},\ and\ \citenamefont
  {Pham}}]{Colangelo:2003sa}%
  \BibitemOpen
  \bibfield  {author} {\bibinfo {author} {\bibfnamefont {P.}~\bibnamefont
  {Colangelo}}, \bibinfo {author} {\bibfnamefont {F.}~\bibnamefont
  {De~Fazio}},\ and\ \bibinfo {author} {\bibfnamefont {T.}~\bibnamefont
  {Pham}},\ }\bibfield  {title} {\bibinfo {title} {Nonfactorizable
  contributions in b decays to charmonium: The case of
  ${B}^{\ensuremath{-}}\ensuremath{\rightarrow}{K}^{\ensuremath{-}}{h}_{c}$},\
  }\href {https://doi.org/10.1103/PhysRevD.69.054023} {\bibfield  {journal}
  {\bibinfo  {journal} {Phys. Rev. D}\ }\textbf {\bibinfo {volume} {69}},\
  \bibinfo {pages} {054023} (\bibinfo {year} {2004})},\ \Eprint
  {https://arxiv.org/abs/hep-ph/0310084} {arXiv:hep-ph/0310084} \BibitemShut
  {NoStop}%
\bibitem [{\citenamefont {Deandrea}\ \emph {et~al.}(2003)\citenamefont
  {Deandrea}, \citenamefont {Nardulli},\ and\ \citenamefont
  {Polosa}}]{Deandrea:2003pv}%
  \BibitemOpen
  \bibfield  {author} {\bibinfo {author} {\bibfnamefont {A.}~\bibnamefont
  {Deandrea}}, \bibinfo {author} {\bibfnamefont {G.}~\bibnamefont {Nardulli}},\
  and\ \bibinfo {author} {\bibfnamefont {A.~D.}\ \bibnamefont {Polosa}},\
  }\bibfield  {title} {\bibinfo {title} {{$J/\ensuremath{\psi}$} couplings to
  charmed resonances and to $\ensuremath{\pi}$},\ }\href
  {https://doi.org/10.1103/PhysRevD.68.034002} {\bibfield  {journal} {\bibinfo
  {journal} {Phys. Rev. D}\ }\textbf {\bibinfo {volume} {68}},\ \bibinfo
  {pages} {034002} (\bibinfo {year} {2003})},\ \Eprint
  {https://arxiv.org/abs/hep-ph/0302273} {arXiv:hep-ph/0302273} \BibitemShut
  {NoStop}%
\bibitem [{\citenamefont {Badalian}\ \emph {et~al.}(2010)\citenamefont
  {Badalian}, \citenamefont {Bakker},\ and\ \citenamefont
  {Danilkin}}]{Badalian:2009bu}%
  \BibitemOpen
  \bibfield  {author} {\bibinfo {author} {\bibfnamefont {A.~M.}\ \bibnamefont
  {Badalian}}, \bibinfo {author} {\bibfnamefont {B.~L.~G.}\ \bibnamefont
  {Bakker}},\ and\ \bibinfo {author} {\bibfnamefont {I.~V.}\ \bibnamefont
  {Danilkin}},\ }\bibfield  {title} {\bibinfo {title} {Dielectron widths of the
  {$S$-}, {$D$-}vector bottomonium states},\ }\href
  {https://doi.org/10.1134/S1063778810010163} {\bibfield  {journal} {\bibinfo
  {journal} {Phys. Atom. Nucl.}\ }\textbf {\bibinfo {volume} {73}},\ \bibinfo
  {pages} {138} (\bibinfo {year} {2010})},\ \Eprint
  {https://arxiv.org/abs/0903.3643} {arXiv:0903.3643 [hep-ph]} \BibitemShut
  {NoStop}%
\bibitem [{\citenamefont {Li}\ \emph {et~al.}(2013)\citenamefont {Li},
  \citenamefont {Shao}, \citenamefont {Zhao},\ and\ \citenamefont
  {Zhao}}]{Li:2012as}%
  \BibitemOpen
  \bibfield  {author} {\bibinfo {author} {\bibfnamefont {G.}~\bibnamefont
  {Li}}, \bibinfo {author} {\bibfnamefont {F.-l.}\ \bibnamefont {Shao}},
  \bibinfo {author} {\bibfnamefont {C.-W.}\ \bibnamefont {Zhao}},\ and\
  \bibinfo {author} {\bibfnamefont {Q.}~\bibnamefont {Zhao}},\ }\bibfield
  {title} {\bibinfo {title}
  {${Z}_{b}/{Z}_{b}^{\ensuremath{'}}\ensuremath{\rightarrow}\ensuremath{\Upsilon}\ensuremath{\pi}$
  and ${h}_{b}\ensuremath{\pi}$ decays in intermediate meson loops model},\
  }\href {https://doi.org/10.1103/PhysRevD.87.034020} {\bibfield  {journal}
  {\bibinfo  {journal} {Phys. Rev. D}\ }\textbf {\bibinfo {volume} {87}},\
  \bibinfo {pages} {034020} (\bibinfo {year} {2013})},\ \Eprint
  {https://arxiv.org/abs/1212.3784} {arXiv:1212.3784 [hep-ph]} \BibitemShut
  {NoStop}%
\bibitem [{\citenamefont {Liu}\ \emph {et~al.}(2024{\natexlab{b}})\citenamefont
  {Liu}, \citenamefont {Cai}, \citenamefont {Jia}, \citenamefont {Li},\ and\
  \citenamefont {Xie}}]{Liu:2023gtx}%
  \BibitemOpen
  \bibfield  {author} {\bibinfo {author} {\bibfnamefont {S.}~\bibnamefont
  {Liu}}, \bibinfo {author} {\bibfnamefont {Z.}~\bibnamefont {Cai}}, \bibinfo
  {author} {\bibfnamefont {Z.}~\bibnamefont {Jia}}, \bibinfo {author}
  {\bibfnamefont {G.}~\bibnamefont {Li}},\ and\ \bibinfo {author}
  {\bibfnamefont {J.}~\bibnamefont {Xie}},\ }\bibfield  {title} {\bibinfo
  {title} {Hidden-bottom hadronic transitions of {$\Upsilon(10753)$}},\ }\href
  {https://doi.org/10.1103/PhysRevD.109.014039} {\bibfield  {journal} {\bibinfo
   {journal} {Phys. Rev. D}\ }\textbf {\bibinfo {volume} {109}},\ \bibinfo
  {pages} {014039} (\bibinfo {year} {2024}{\natexlab{b}})},\ \Eprint
  {https://arxiv.org/abs/2312.02761} {arXiv:2312.02761 [hep-ph]} \BibitemShut
  {NoStop}%
\bibitem [{\citenamefont {Wang}\ \emph {et~al.}(2022)\citenamefont {Wang},
  \citenamefont {Wu}, \citenamefont {Li}, \citenamefont {Qin}, \citenamefont
  {Liu}, \citenamefont {An},\ and\ \citenamefont {Xie}}]{Wang:2022qxe}%
  \BibitemOpen
  \bibfield  {author} {\bibinfo {author} {\bibfnamefont {Y.}~\bibnamefont
  {Wang}}, \bibinfo {author} {\bibfnamefont {Q.}~\bibnamefont {Wu}}, \bibinfo
  {author} {\bibfnamefont {G.}~\bibnamefont {Li}}, \bibinfo {author}
  {\bibfnamefont {W.-H.}\ \bibnamefont {Qin}}, \bibinfo {author} {\bibfnamefont
  {X.-H.}\ \bibnamefont {Liu}}, \bibinfo {author} {\bibfnamefont {C.-S.}\
  \bibnamefont {An}},\ and\ \bibinfo {author} {\bibfnamefont {J.-J.}\
  \bibnamefont {Xie}},\ }\bibfield  {title} {\bibinfo {title} {{Investigations
  of charmless decays of $X(3872)$ via intermediate meson loops}},\ }\href
  {https://doi.org/10.1103/PhysRevD.106.074015} {\bibfield  {journal} {\bibinfo
   {journal} {Phys. Rev. D}\ }\textbf {\bibinfo {volume} {106}},\ \bibinfo
  {pages} {074015} (\bibinfo {year} {2022})},\ \Eprint
  {https://arxiv.org/abs/2209.12206} {arXiv:2209.12206 [hep-ph]} \BibitemShut
  {NoStop}%
\bibitem [{\citenamefont {Wu}\ \emph {et~al.}(2021)\citenamefont {Wu},
  \citenamefont {Chen},\ and\ \citenamefont {Matsuki}}]{Wu:2021udi}%
  \BibitemOpen
  \bibfield  {author} {\bibinfo {author} {\bibfnamefont {Q.}~\bibnamefont
  {Wu}}, \bibinfo {author} {\bibfnamefont {D.-Y.}\ \bibnamefont {Chen}},\ and\
  \bibinfo {author} {\bibfnamefont {T.}~\bibnamefont {Matsuki}},\ }\bibfield
  {title} {\bibinfo {title} {{A phenomenological analysis on isospin-violating
  decay of $X(3872)$}},\ }\href
  {https://doi.org/10.1140/epjc/s10052-021-08984-2} {\bibfield  {journal}
  {\bibinfo  {journal} {Eur. Phys. J. C}\ }\textbf {\bibinfo {volume} {81}},\
  \bibinfo {pages} {193} (\bibinfo {year} {2021})},\ \Eprint
  {https://arxiv.org/abs/2102.08637} {arXiv:2102.08637 [hep-ph]} \BibitemShut
  {NoStop}%
\bibitem [{\citenamefont {Isola}\ \emph {et~al.}(2003)\citenamefont {Isola},
  \citenamefont {Ladisa}, \citenamefont {Nardulli},\ and\ \citenamefont
  {Santorelli}}]{Isola:2003fh}%
  \BibitemOpen
  \bibfield  {author} {\bibinfo {author} {\bibfnamefont {C.}~\bibnamefont
  {Isola}}, \bibinfo {author} {\bibfnamefont {M.}~\bibnamefont {Ladisa}},
  \bibinfo {author} {\bibfnamefont {G.}~\bibnamefont {Nardulli}},\ and\
  \bibinfo {author} {\bibfnamefont {P.}~\bibnamefont {Santorelli}},\ }\bibfield
   {title} {\bibinfo {title} {Charming penguin contributions in {$B\to K^\ast
  \pi, K(\rho,\,\omega,\,\phi)$} decays},\ }\href
  {https://doi.org/10.1103/PhysRevD.68.114001} {\bibfield  {journal} {\bibinfo
  {journal} {Phys. Rev. D}\ }\textbf {\bibinfo {volume} {68}},\ \bibinfo
  {pages} {114001} (\bibinfo {year} {2003})},\ \Eprint
  {https://arxiv.org/abs/hep-ph/0307367} {arXiv:hep-ph/0307367} \BibitemShut
  {NoStop}%
\bibitem [{\citenamefont {Meng}\ and\ \citenamefont
  {Chao}(2008{\natexlab{a}})}]{Meng:2008dd}%
  \BibitemOpen
  \bibfield  {author} {\bibinfo {author} {\bibfnamefont {C.}~\bibnamefont
  {Meng}}\ and\ \bibinfo {author} {\bibfnamefont {K.-T.}\ \bibnamefont
  {Chao}},\ }\bibfield  {title} {\bibinfo {title} {{Peak shifts due to
  ${B}^{(*)}\ensuremath{-}{\overline{B}}^{(*)}$ rescattering in
  $\ensuremath{\Upsilon}(5S)$ dipion transitions}},\ }\href
  {https://doi.org/10.1103/PhysRevD.78.034022} {\bibfield  {journal} {\bibinfo
  {journal} {Phys. Rev. D}\ }\textbf {\bibinfo {volume} {78}},\ \bibinfo
  {pages} {034022} (\bibinfo {year} {2008}{\natexlab{a}})},\ \Eprint
  {https://arxiv.org/abs/0805.0143} {arXiv:0805.0143 [hep-ph]} \BibitemShut
  {NoStop}%
\bibitem [{\citenamefont {Bai}\ \emph {et~al.}(2022)\citenamefont {Bai},
  \citenamefont {Li}, \citenamefont {Huang}, \citenamefont {Liu},\ and\
  \citenamefont {Matsuki}}]{Bai:2022cfz}%
  \BibitemOpen
  \bibfield  {author} {\bibinfo {author} {\bibfnamefont {Z.-Y.}\ \bibnamefont
  {Bai}}, \bibinfo {author} {\bibfnamefont {Y.-S.}\ \bibnamefont {Li}},
  \bibinfo {author} {\bibfnamefont {Q.}~\bibnamefont {Huang}}, \bibinfo
  {author} {\bibfnamefont {X.}~\bibnamefont {Liu}},\ and\ \bibinfo {author}
  {\bibfnamefont {T.}~\bibnamefont {Matsuki}},\ }\bibfield  {title} {\bibinfo
  {title}
  {$\mathrm{\ensuremath{\Upsilon}}(10753)\ensuremath{\rightarrow}\mathrm{\ensuremath{\Upsilon}}(\mathrm{n}\mathrm{S}){\ensuremath{\pi}}^{+}{\ensuremath{\pi}}^{\ensuremath{-}}$
  decays induced by hadronic loop mechanism},\ }\href
  {https://doi.org/10.1103/PhysRevD.105.074007} {\bibfield  {journal} {\bibinfo
   {journal} {Phys. Rev. D}\ }\textbf {\bibinfo {volume} {105}},\ \bibinfo
  {pages} {074007} (\bibinfo {year} {2022})},\ \Eprint
  {https://arxiv.org/abs/2201.12715} {arXiv:2201.12715 [hep-ph]} \BibitemShut
  {NoStop}%
\bibitem [{\citenamefont {Chen}\ \emph
  {et~al.}(2011{\natexlab{b}})\citenamefont {Chen}, \citenamefont {He},
  \citenamefont {Li},\ and\ \citenamefont {Liu}}]{Chen:2011qx}%
  \BibitemOpen
  \bibfield  {author} {\bibinfo {author} {\bibfnamefont {D.-Y.}\ \bibnamefont
  {Chen}}, \bibinfo {author} {\bibfnamefont {J.}~\bibnamefont {He}}, \bibinfo
  {author} {\bibfnamefont {X.-Q.}\ \bibnamefont {Li}},\ and\ \bibinfo {author}
  {\bibfnamefont {X.}~\bibnamefont {Liu}},\ }\bibfield  {title} {\bibinfo
  {title} {{Dipion invariant mass distribution of the anomalous
  $\ensuremath{\Upsilon}(1S){\ensuremath{\pi}}^{+}{\ensuremath{\pi}}^{\ensuremath{-}}$
  and
  $\ensuremath{\Upsilon}(2S){\ensuremath{\pi}}^{+}{\ensuremath{\pi}}^{\ensuremath{-}}$
  production near the peak of $\ensuremath{\Upsilon}(10860)$}},\ }\href
  {https://doi.org/10.1103/PhysRevD.84.074006} {\bibfield  {journal} {\bibinfo
  {journal} {Phys. Rev. D}\ }\textbf {\bibinfo {volume} {84}},\ \bibinfo
  {pages} {074006} (\bibinfo {year} {2011}{\natexlab{b}})},\ \Eprint
  {https://arxiv.org/abs/1105.1672} {arXiv:1105.1672 [hep-ph]} \BibitemShut
  {NoStop}%
\bibitem [{\citenamefont {Chen}\ \emph
  {et~al.}(2016{\natexlab{b}})\citenamefont {Chen}, \citenamefont {Liu},\ and\
  \citenamefont {Matsuki}}]{Chen:2015bma}%
  \BibitemOpen
  \bibfield  {author} {\bibinfo {author} {\bibfnamefont {D.-Y.}\ \bibnamefont
  {Chen}}, \bibinfo {author} {\bibfnamefont {X.}~\bibnamefont {Liu}},\ and\
  \bibinfo {author} {\bibfnamefont {T.}~\bibnamefont {Matsuki}},\ }\bibfield
  {title} {\bibinfo {title} {{Search for missing $\ensuremath{\psi}(4S)$ in the
  ${e}^{+}{e}^{\ensuremath{-}}\ensuremath{\rightarrow}{\ensuremath{\pi}}^{+}{\ensuremath{\pi}}^{\ensuremath{-}}\ensuremath{\psi}(2S)$
  process}},\ }\href {https://doi.org/10.1103/PhysRevD.93.034028} {\bibfield
  {journal} {\bibinfo  {journal} {Phys. Rev. D}\ }\textbf {\bibinfo {volume}
  {93}},\ \bibinfo {pages} {034028} (\bibinfo {year} {2016}{\natexlab{b}})},\
  \Eprint {https://arxiv.org/abs/1509.00736} {arXiv:1509.00736 [hep-ph]}
  \BibitemShut {NoStop}%
\bibitem [{\citenamefont {Chen}\ \emph {et~al.}(2008)\citenamefont {Chen} \emph
  {et~al.}}]{Belle:2007xek}%
  \BibitemOpen
  \bibfield  {author} {\bibinfo {author} {\bibfnamefont {K.~F.}\ \bibnamefont
  {Chen}} \emph {et~al.} (\bibinfo {collaboration} {Belle}),\ }\bibfield
  {title} {\bibinfo {title} {{Observation of anomalous $\Upsilon(1S) \pi^+
  \pi^-$ and $\Upsilon(2S) \pi^+ \pi^-$ production near the $\Upsilon(5S)$
  resonance}},\ }\href {https://doi.org/10.1103/PhysRevLett.100.112001}
  {\bibfield  {journal} {\bibinfo  {journal} {Phys. Rev. Lett.}\ }\textbf
  {\bibinfo {volume} {100}},\ \bibinfo {pages} {112001} (\bibinfo {year}
  {2008})},\ \Eprint {https://arxiv.org/abs/0710.2577} {arXiv:0710.2577
  [hep-ex]} \BibitemShut {NoStop}%
\bibitem [{\citenamefont {Oh}\ \emph {et~al.}(2001)\citenamefont {Oh},
  \citenamefont {Song},\ and\ \citenamefont {Lee}}]{Oh:2000qr}%
  \BibitemOpen
  \bibfield  {author} {\bibinfo {author} {\bibfnamefont {Y.}~\bibnamefont
  {Oh}}, \bibinfo {author} {\bibfnamefont {T.}~\bibnamefont {Song}},\ and\
  \bibinfo {author} {\bibfnamefont {S.~H.}\ \bibnamefont {Lee}},\ }\bibfield
  {title} {\bibinfo {title} {{$J/\ensuremath{\psi}$ absorption by
  $\ensuremath{\pi}$ and $\ensuremath{\rho}$ mesons in a meson exchange model
  with anomalous parity interactions}},\ }\href
  {https://doi.org/10.1103/PhysRevC.63.034901} {\bibfield  {journal} {\bibinfo
  {journal} {Phys. Rev. C}\ }\textbf {\bibinfo {volume} {63}},\ \bibinfo
  {pages} {034901} (\bibinfo {year} {2001})},\ \Eprint
  {https://arxiv.org/abs/nucl-th/0010064} {arXiv:nucl-th/0010064} \BibitemShut
  {NoStop}%
\bibitem [{\citenamefont {Lin}\ and\ \citenamefont {Ko}(2000)}]{Lin:1999ad}%
  \BibitemOpen
  \bibfield  {author} {\bibinfo {author} {\bibfnamefont {Z.}~\bibnamefont
  {Lin}}\ and\ \bibinfo {author} {\bibfnamefont {C.~M.}\ \bibnamefont {Ko}},\
  }\bibfield  {title} {\bibinfo {title} {{Model for $J/\ensuremath{\psi}$
  absorption in hadronic matter}},\ }\href
  {https://doi.org/10.1103/PhysRevC.62.034903} {\bibfield  {journal} {\bibinfo
  {journal} {Phys. Rev. C}\ }\textbf {\bibinfo {volume} {62}},\ \bibinfo
  {pages} {034903} (\bibinfo {year} {2000})},\ \Eprint
  {https://arxiv.org/abs/nucl-th/9912046} {arXiv:nucl-th/9912046} \BibitemShut
  {NoStop}%
\bibitem [{\citenamefont {Locher}\ \emph {et~al.}(1994)\citenamefont {Locher},
  \citenamefont {Lu},\ and\ \citenamefont {Zou}}]{Locher:1993cc}%
  \BibitemOpen
  \bibfield  {author} {\bibinfo {author} {\bibfnamefont {M.~P.}\ \bibnamefont
  {Locher}}, \bibinfo {author} {\bibfnamefont {Y.}~\bibnamefont {Lu}},\ and\
  \bibinfo {author} {\bibfnamefont {B.~S.}\ \bibnamefont {Zou}},\ }\bibfield
  {title} {\bibinfo {title} {{Rates for the reactions$\bar pp \to \pi \phi $
  and $\gamma \phi$}},\ }\href {https://doi.org/10.1007/BF01289796} {\bibfield
  {journal} {\bibinfo  {journal} {Z. Phys. A}\ }\textbf {\bibinfo {volume}
  {347}},\ \bibinfo {pages} {281} (\bibinfo {year} {1994})},\ \Eprint
  {https://arxiv.org/abs/nucl-th/9311021} {arXiv:nucl-th/9311021} \BibitemShut
  {NoStop}%
\bibitem [{\citenamefont {Li}\ \emph {et~al.}(1997)\citenamefont {Li},
  \citenamefont {Bugg},\ and\ \citenamefont {Zou}}]{Li:1996yn}%
  \BibitemOpen
  \bibfield  {author} {\bibinfo {author} {\bibfnamefont {X.-Q.}\ \bibnamefont
  {Li}}, \bibinfo {author} {\bibfnamefont {D.~V.}\ \bibnamefont {Bugg}},\ and\
  \bibinfo {author} {\bibfnamefont {B.-S.}\ \bibnamefont {Zou}},\ }\bibfield
  {title} {\bibinfo {title} {{Possible explanation of the
  ``$\ensuremath{\rho}\ensuremath{\pi}$ puzzle'' in $J/\ensuremath{\psi}$,
  ${\ensuremath{\psi}}^{\ensuremath{'}}$ decays}},\ }\href
  {https://doi.org/10.1103/PhysRevD.55.1421} {\bibfield  {journal} {\bibinfo
  {journal} {Phys. Rev. D}\ }\textbf {\bibinfo {volume} {55}},\ \bibinfo
  {pages} {1421} (\bibinfo {year} {1997})}\BibitemShut {NoStop}%
\bibitem [{\citenamefont {Cheng}\ \emph {et~al.}(2005)\citenamefont {Cheng},
  \citenamefont {Chua},\ and\ \citenamefont {Soni}}]{Cheng:2004ru}%
  \BibitemOpen
  \bibfield  {author} {\bibinfo {author} {\bibfnamefont {H.-Y.}\ \bibnamefont
  {Cheng}}, \bibinfo {author} {\bibfnamefont {C.-K.}\ \bibnamefont {Chua}},\
  and\ \bibinfo {author} {\bibfnamefont {A.}~\bibnamefont {Soni}},\ }\bibfield
  {title} {\bibinfo {title} {{Final state interactions in hadronic $B$
  decays}},\ }\href {https://doi.org/10.1103/PhysRevD.71.014030} {\bibfield
  {journal} {\bibinfo  {journal} {Phys. Rev. D}\ }\textbf {\bibinfo {volume}
  {71}},\ \bibinfo {pages} {014030} (\bibinfo {year} {2005})},\ \Eprint
  {https://arxiv.org/abs/hep-ph/0409317} {arXiv:hep-ph/0409317} \BibitemShut
  {NoStop}%
\bibitem [{\citenamefont {Mertig}\ \emph {et~al.}(1991)\citenamefont {Mertig},
  \citenamefont {Böhm},\ and\ \citenamefont {Denner}}]{Mertig:1990an}%
  \BibitemOpen
  \bibfield  {author} {\bibinfo {author} {\bibfnamefont {R.}~\bibnamefont
  {Mertig}}, \bibinfo {author} {\bibfnamefont {M.}~\bibnamefont {Böhm}},\ and\
  \bibinfo {author} {\bibfnamefont {A.}~\bibnamefont {Denner}},\ }\bibfield
  {title} {\bibinfo {title} {Feyn calc - computer-algebraic calculation of
  feynman amplitudes},\ }\href {https://doi.org/10.1016/0010-4655(91)90130-D}
  {\bibfield  {journal} {\bibinfo  {journal} {Comput. Phys. Commun.}\ }\textbf
  {\bibinfo {volume} {64}},\ \bibinfo {pages} {345} (\bibinfo {year}
  {1991})}\BibitemShut {NoStop}%
\bibitem [{\citenamefont {Shtabovenko}\ \emph {et~al.}(2020)\citenamefont
  {Shtabovenko}, \citenamefont {Mertig},\ and\ \citenamefont
  {Orellana}}]{Shtabovenko:2020gxv}%
  \BibitemOpen
  \bibfield  {author} {\bibinfo {author} {\bibfnamefont {V.}~\bibnamefont
  {Shtabovenko}}, \bibinfo {author} {\bibfnamefont {R.}~\bibnamefont
  {Mertig}},\ and\ \bibinfo {author} {\bibfnamefont {F.}~\bibnamefont
  {Orellana}},\ }\bibfield  {title} {\bibinfo {title} {Feyncalc 9.3: New
  features and improvements},\ }\href
  {https://doi.org/10.1016/j.cpc.2020.107478} {\bibfield  {journal} {\bibinfo
  {journal} {Comput. Phys. Commun.}\ }\textbf {\bibinfo {volume} {256}},\
  \bibinfo {pages} {107478} (\bibinfo {year} {2020})},\ \Eprint
  {https://arxiv.org/abs/2001.04407} {arXiv:2001.04407 [hep-ph]} \BibitemShut
  {NoStop}%
\bibitem [{\citenamefont {Hahn}\ and\ \citenamefont
  {Perez-Victoria}(1999)}]{Hahn:1998yk}%
  \BibitemOpen
  \bibfield  {author} {\bibinfo {author} {\bibfnamefont {T.}~\bibnamefont
  {Hahn}}\ and\ \bibinfo {author} {\bibfnamefont {M.}~\bibnamefont
  {Perez-Victoria}},\ }\bibfield  {title} {\bibinfo {title} {Automatized
  one-loop calculations in 4 and d dimensions},\ }\href
  {https://doi.org/10.1016/S0010-4655(98)00173-8} {\bibfield  {journal}
  {\bibinfo  {journal} {Comput. Phys. Commun.}\ }\textbf {\bibinfo {volume}
  {118}},\ \bibinfo {pages} {153} (\bibinfo {year} {1999})},\ \Eprint
  {https://arxiv.org/abs/hep-ph/9807565} {arXiv:hep-ph/9807565} \BibitemShut
  {NoStop}%
\bibitem [{\citenamefont {Ablikim}\ \emph {et~al.}(2013)\citenamefont
  {Ablikim}, \citenamefont {Achasov}, \citenamefont {Ai} \emph
  {et~al.}}]{BESIII:2013ris}%
  \BibitemOpen
  \bibfield  {author} {\bibinfo {author} {\bibfnamefont {M.}~\bibnamefont
  {Ablikim}}, \bibinfo {author} {\bibfnamefont {M.}~\bibnamefont {Achasov}},
  \bibinfo {author} {\bibfnamefont {X.}~\bibnamefont {Ai}}, \emph {et~al.}
  (\bibinfo {collaboration} {BESIII}),\ }\bibfield  {title} {\bibinfo {title}
  {{Observation of a Charged Charmoniumlike Structure in
  ${e}^{\mathbf{+}}{e}^{\mathbf{\ensuremath{-}}}\ensuremath{\rightarrow}{\ensuremath{\pi}}^{\mathbf{+}}{\ensuremath{\pi}}^{\mathbf{\ensuremath{-}}}J/\ensuremath{\psi}$
  at $\sqrt{s}\mathbf{=}4.26\text{ }\text{ }\mathrm{GeV}$}},\ }\href
  {https://doi.org/10.1103/PhysRevLett.110.252001} {\bibfield  {journal}
  {\bibinfo  {journal} {Phys. Rev. Lett.}\ }\textbf {\bibinfo {volume} {110}},\
  \bibinfo {pages} {252001} (\bibinfo {year} {2013})},\ \Eprint
  {https://arxiv.org/abs/1303.5949} {arXiv:1303.5949 [hep-ex]} \BibitemShut
  {NoStop}%
\bibitem [{\citenamefont {Chen}\ \emph
  {et~al.}(2013{\natexlab{b}})\citenamefont {Chen}, \citenamefont {Liu},\ and\
  \citenamefont {Matsuki}}]{Chen:2013coa}%
  \BibitemOpen
  \bibfield  {author} {\bibinfo {author} {\bibfnamefont {D.-Y.}\ \bibnamefont
  {Chen}}, \bibinfo {author} {\bibfnamefont {X.}~\bibnamefont {Liu}},\ and\
  \bibinfo {author} {\bibfnamefont {T.}~\bibnamefont {Matsuki}},\ }\bibfield
  {title} {\bibinfo {title} {Reproducing the ${Z}_{c}(3900)$ structure through
  the initial-single-pion-emission mechanism},\ }\href
  {https://doi.org/10.1103/PhysRevD.88.036008} {\bibfield  {journal} {\bibinfo
  {journal} {Phys. Rev. D}\ }\textbf {\bibinfo {volume} {88}},\ \bibinfo
  {pages} {036008} (\bibinfo {year} {2013}{\natexlab{b}})},\ \Eprint
  {https://arxiv.org/abs/1304.5845} {arXiv:1304.5845 [hep-ph]} \BibitemShut
  {NoStop}%
\bibitem [{\citenamefont {Ablikim}\ \emph {et~al.}(2007)\citenamefont {Ablikim}
  \emph {et~al.}}]{BES:2006eer}%
  \BibitemOpen
  \bibfield  {author} {\bibinfo {author} {\bibfnamefont {M.}~\bibnamefont
  {Ablikim}} \emph {et~al.} (\bibinfo {collaboration} {BES}),\ }\bibfield
  {title} {\bibinfo {title} {{Production of sigma in $\psi(2S) \to \pi^+\pi^-
  J/\psi$}},\ }\href {https://doi.org/10.1016/j.physletb.2006.11.056}
  {\bibfield  {journal} {\bibinfo  {journal} {Phys. Lett. B}\ }\textbf
  {\bibinfo {volume} {645}},\ \bibinfo {pages} {19} (\bibinfo {year} {2007})},\
  \Eprint {https://arxiv.org/abs/hep-ex/0610023} {arXiv:hep-ex/0610023}
  \BibitemShut {NoStop}%
\bibitem [{\citenamefont {Cronin-Hennessy}\ \emph {et~al.}(2007)\citenamefont
  {Cronin-Hennessy} \emph {et~al.}}]{CLEO:2007rbi}%
  \BibitemOpen
  \bibfield  {author} {\bibinfo {author} {\bibfnamefont {D.}~\bibnamefont
  {Cronin-Hennessy}} \emph {et~al.} (\bibinfo {collaboration} {CLEO}),\
  }\bibfield  {title} {\bibinfo {title} {{Study of di-pion transitions among
  $\Upsilon(3S)$, $\Upsilon(2S)$, and $\Upsilon(1S)$ states}},\ }\href
  {https://doi.org/10.1103/PhysRevD.76.072001} {\bibfield  {journal} {\bibinfo
  {journal} {Phys. Rev. D}\ }\textbf {\bibinfo {volume} {76}},\ \bibinfo
  {pages} {072001} (\bibinfo {year} {2007})},\ \Eprint
  {https://arxiv.org/abs/0706.2317} {arXiv:0706.2317 [hep-ex]} \BibitemShut
  {NoStop}%
\bibitem [{\citenamefont {Meng}\ and\ \citenamefont
  {Chao}(2008{\natexlab{b}})}]{Meng:2007tk}%
  \BibitemOpen
  \bibfield  {author} {\bibinfo {author} {\bibfnamefont {C.}~\bibnamefont
  {Meng}}\ and\ \bibinfo {author} {\bibfnamefont {K.-T.}\ \bibnamefont
  {Chao}},\ }\bibfield  {title} {\bibinfo {title} {{Scalar resonance
  contributions to the dipion transition rates of $\Upsilon(4S,5S)$ in the
  re-scattering model}},\ }\href {https://doi.org/10.1103/PhysRevD.77.074003}
  {\bibfield  {journal} {\bibinfo  {journal} {Phys. Rev. D}\ }\textbf {\bibinfo
  {volume} {77}},\ \bibinfo {pages} {074003} (\bibinfo {year}
  {2008}{\natexlab{b}})},\ \Eprint {https://arxiv.org/abs/0712.3595}
  {arXiv:0712.3595 [hep-ph]} \BibitemShut {NoStop}%
\end{thebibliography}%
\end{document}